\documentclass[aip,jcp,reprint,groupedaddress,floatfix]{revtex4-1}
\pdfoutput=1

\usepackage{siunitx} 

\usepackage{amsmath} 
\usepackage{amssymb}
\usepackage{mathrsfs}
\usepackage{mathtools}
\usepackage{stmaryrd}
\usepackage{esint}

\usepackage[shortlabels]{enumitem}
\usepackage{graphicx} 

\usepackage{multirow}
\usepackage{booktabs} 
\usepackage{hyperref} 

\newcommand\abs[1]{\left|#1\right|}

\begin{document}


\title{Perturbative density functional methods for cholesteric liquid crystals}
\date{\today}
\author{Maxime M. C. Tortora}
\author{Jonathan P. K. Doye}
\affiliation{Physical and Theoretical Chemistry Laboratory, University of Oxford, South Parks Road, Oxford, OX1 3QZ, United Kingdom}



\begin{abstract}

We introduce a comprehensive numerical framework to generically infer the emergent macroscopic properties of uniaxial nematic and cholesteric phases from that of their microscopic constituent mesogens. This approach, based on the full numerical resolution of the Poniewierski-Stecki equations in the weak chirality limit, may expediently handle a wide range of particle models through the use of Monte-Carlo sampling for all virial-type integrals. Its predictions in terms of equilibrium cholesteric structures are found to be in excellent agreement with previous full-functional descriptions, thereby demonstrating the quantitative validity of the perturbative treatment of chirality for pitch lengths as short as a few dozen particle diameters. Furthermore, the use of the full angle-dependent virial coefficients in the Onsager-Parsons-Lee formalism increases its numerical efficiency by several orders of magnitude over that of these previous methods. The comparison of our results with numerical simulations however reveals some shortcomings of the Parsons-Lee approximation for systems of strongly non-convex particles, notwithstanding the accurate inclusion of their full effective molecular volume. Further potential limitations of our theory in terms of phase symmetry assumptions are also examined, and prospective directions for future improvements discussed.

\end{abstract}


\maketitle 


\section{Introduction} \label{sec:Introduction}

Liquid crystals (LCs) are a fascinating example of self-assembling soft-matter systems, characterised by the singular coexistence of long-range order and fluidity. These organised phases typically arise from a specific set of structural and chemical features displayed by their constituent particles,\cite{deGennes} the most notable of which being shape anisotropy. In other words, these particles are elongated along one or more directions, and therefore generally possess both a long and a short axis --- respectively defined as the directions of maximum and minimum extent of their molecular backbone.
\par
A remarkable consequence of this microscopic anisotropy is the wealth of symmetry-breaking transitions involved in LC phase-ordering kinetics, as not only the positions, but also the orientations of the molecules may display long-range correlations.\cite{deGennes} Therefore, LCs are characterised by a degree of orientational order, potentially associated with a form of positional organisation --- leading to an impressive variety of macroscopic structures.\cite{Goodby-1} Understanding the link between this range of phase symmetries and the underlying properties of their constituent particles has remained a long-standing challenge of soft-condensed matter theory.\cite{Wilson}
\par
The simplest example of LC assembly is the so-called \textit{uniaxial nematic phase}, in which one of the molecular axes points on average in a particular direction called the \textit{nematic director}, denoted by $\mathbf{n}$, while retaining fluid-like positional disorder. The resulting phases may then be qualified as \textit{prolate} or \textit{oblate}, respectively referring to the local orientational order of the particle long or short axes. This structural anisotropy bestows upon nematic LCs a strong optical birefringence,\cite{Wu} much like regular uniaxial crystals, along with unique electrical\cite{Freedericksz} and rheological\cite{Doi} properties. In the absence of external aligning forces, the director generally fluctuates throughout the sample, and is therefore usually referred to as a \textit{director field} $\mathbf{n}(\mathbf{r})$, with $\mathbf{r}$ the spatial coordinate vector. 
\par
\begin{figure}[htpb]
  \includegraphics[width=\columnwidth]{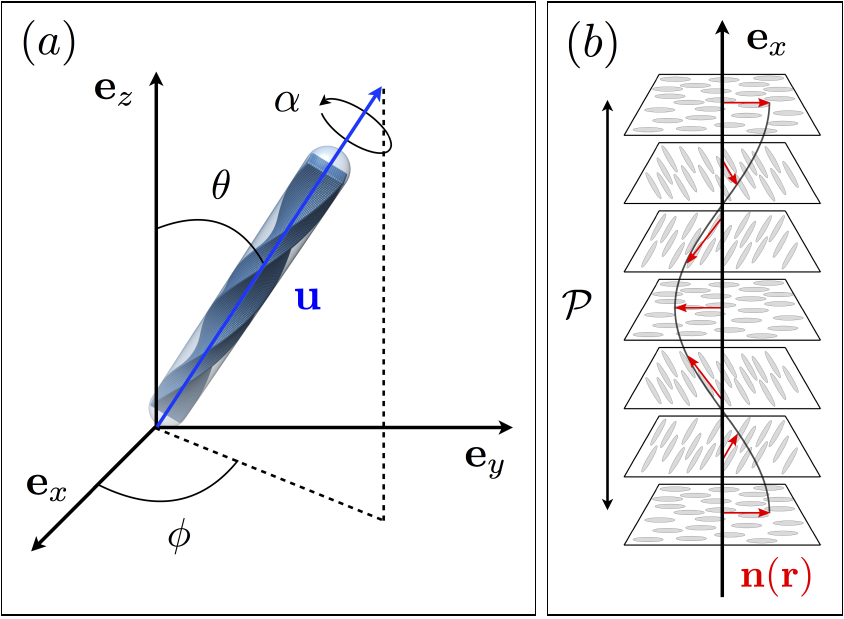}
  \caption{\label{fig1}Particle and phase chirality in cholesterics. $(a)$ Example of a calamitic (i.e.~rodlike) chiral mesogen with long axis $\mathbf{u}$. $(b)$ Sketch of a prolate cholesteric phase of axis $\mathbf{e}_x$, director field $\mathbf{n}$ and pitch $\mathcal{P}$.}
\end{figure}

An intriguing special case of this spatial modulation corresponds to the \textit{cholesteric phase}, typically formed by systems of chiral particles, in which $\mathbf{n}(\mathbf{r})$ periodically rotates along an axis normal to the local director, as illustrated in Fig.~\ref{fig1}. Hence, cholesteric LCs are commonly described in terms of a single lengthscale $\mathcal{P}$ --- termed the \textit{cholesteric pitch} --- defined as the spatial period of the director rotation, and fully characterising their macroscopic structure.

As the earliest LC phase to be identified experimentally in 1888,\cite{Reinitzer} cholesterics boast a remarkably long history along with a rich array of applications; the most notable example indubitably lies in the \textit{twisted nematic cell}, which single-handedly triggered the advent of modern LC display technology.\cite{Dunmur} However, an impressive variety of further uses of cholesteric systems have been proposed in recent years, ranging from physical\cite{Moreira} and chemical\cite{Mujahid} sensors to surface-morphing applications,\cite{Liu} including the design of tunable-wavelength lasers\cite{Finkelmann} and core-sheath fibers for smart textiles.\cite{Enz}
\par
Their original observation in cholesterol derivatives marked the discovery of the \textit{thermotropic} class of LCs, in which phase transitions are governed by temperature variations. However, \textit{lyotropic} cholesteric phases, in which transitions are mostly ruled by mesogen concentration, are also commonly observed in many bio-colloidal\cite{Robinson, Livolant, Dogic-1, Uematsu, Hiltrop, Sato-2, Werbowyj, Revol} and polymer\cite{Aharoni, Sato-1, Watanabe} solutions. A fundamental feature of both systems is the high sensitivity of the pitch $\mathcal{P}$ to the chemical structure and thermodynamic state of their constituent particles, which provides the basis for many of the above applications; the nature of this intricate dependence has been intensively investigated experimentally over the last decades in a number of different contexts.\cite{Yevdokimov, vanWinkle, Stanley, Dogic-2, Grelet-1, DuPre, Dong, Miller}
\par
Conversely, theoretical and numerical descriptions of the cholesteric phase have long been hampered by both the complexity of the microscopic chiral interactions and the multi-scale nature of the problem, as cholesteric pitches measured experimentally are usually several orders of magnitudes larger than typical particle sizes.\cite{deGennes} This large spatial periodicity renders direct atomistic simulations largely impractical, due to both the substantial size of the simulation boxes required and the unsuitability of standard periodic boundary conditions.\cite{Frenkel-3} Furthermore, the quick escalation of computational costs with increasing particle aspect ratios largely restricts their applicability to systems of short, highly chiral particles --- in stark contrast with the properties of many experimentally-relevant cholesteric mesogens.\cite{vanWinkle, Livolant, Stanley, Dogic-1, Dogic-2, Grelet-1, Usov, Wang}
\par
The majority of theoretical attempts to overcome these limitations have thus chosen to follow in the footsteps of Onsager,\cite{Onsager} whose seminal microscopic description of the nematic state provides a natural framework to investigate the self-assembly of highly anisotropic particles. This original approach, which prefigures the modern formulation of classical density functional theory (DFT),\cite{Evans-1} has been successfully used to predict the isotropic-nematic (I-N) phase transition of a number of experimental systems,\cite{Fraden, Lagerwall, Lekkerkerker-4, Zhang} and was subsequently generalised to the cholesteric phase by Straley\cite{Straley-1} in the limit of weak chirality.
\par
However, most implementations of Straley's theory have so far been limited by stringent analytical assumptions to highly simplified particle models, in which a number of further \textit{ad hoc} approximations had to be introduced for the tractable inclusion of chiral contributions in the virial formalism of the extended Onsager theory.\cite{Osipov, Odijk, Pelcovits, Evans-3, Varga-1, Varga-2, Wensink-1, Wensink-4, Wensink-3} Some numerical attempts to tackle more realistic systems --- including both steric and soft repulsion --- represented significant steps forward, but were still hindered by a cumbersome mathematical formalism and numerical procedure.\cite{Tombolato, Grelet-2, Ferrarini-1} In particular, their recourse to the partial decomposition of the particle density field in a basis of rotational invariants significantly limited both their computational efficiency and numerical precision due to potential finite-order truncation effects.\cite{Lekkerkerker-3}
\par
A notable exception to the previous pattern lies in the recent works of Belli \textit{et al.},\cite{Dijkstra-1, Dijkstra-2} in which a unified framework capable of handling arbitrarily complex particle shapes and interaction potentials independent of Straley's perturbative description was introduced. This approach, based on the use of Monte-Carlo (MC) integration for the computation of all generalised virial coefficients, however suffers from two major drawbacks. Firstly, it requires the evaluation of the full chiral free energy landscape of the molecules using a finite grid of macroscopic pitches, leading to high computational costs along with discretisation effects stemming from the limited grid resolution. Secondly, the reliable calculation of equilibrium pitches typically requires several preliminary runs in order to narrow down the pitch grid to the relevant parameter range, further increasing its computational expense.
\par
On these premises, we here aim to make progress towards an efficient, quantitatively reliable theoretical description of the cholesteric structure and underlying nematic properties of systems of arbitrary particles. Combining the numerical integration techniques of Refs.~\onlinecite{Dijkstra-1,Dijkstra-2} with Straley's perturbative framework\cite{Straley-1} enables us to circumvent the limitations of previous approaches, and allows for the expedient computation of all relevant variables with an adjustable level of precision for a wide variety of particle models. Furthermore, the relative simplicity of the corresponding mathematical expressions lends itself to a thorough examination of the different physical assumptions underpinning the DFT formalism, and of their potential shortcomings in the case of chiral nematic phases.
\par
The structure of the paper is thus organised as follows. In order that the origin of the potential shortcomings of the theory can be fully understood, we first provide in Sec.~\ref{sec:Microscopic theory of liquid-crystal elasticities} a comprehensive account of the theory and numerical implementation, focussing on the physical nature of the approximations that are being made. Specifically, we outline in Sec.~\ref{subsec:Onsager DFT for uniaxial LCs} the original Onsager DFT for nematic LCs, with a particular emphasis on phase symmetry assumptions, and focus in Sec.~\ref{subsec:Introducing finite mesogen anisotropy} on the approximations relative to the treatment of correlations between finite-size particles. We tackle in Sec.~\ref{subsec:Introducing weak nematic fluctuations} the applications of the theory to the microscopic determination of LC elasticities, which are intimately linked to the perturbative treatment of chirality in cholesterics, before detailing the computational procedure used in Sec.~\ref{subsec:Numerical procedure}. We then devote Sec.~\ref{sec:Results and discussion} to the extensive comparison of our results to previous numerical and theoretical studies of several model systems, and discuss potential deficiencies of the different density-functional assumptions stemming from both particle non-convexity and phase biaxiality. We finally conclude in Sec.~\ref{sec:Conclusion} with a few remarks on the possible consequences of these shortcomings on the general reliability of our approach, and highlight some prospective directions for future research.


\section{Microscopic theory of liquid-crystal elasticities} \label{sec:Microscopic theory of liquid-crystal elasticities}
We here choose to model cholesterics as weakly distorted uniaxial nematic phases, whose bulk free energies are typically described at the continuum level by the standard Oseen-Frank field functional\cite{Frank}
\begin{equation}
\begin{split}
\label{eq:frank}
\mathscr{F} = &\:\mathscr{F}_0 + \frac{1}{2} \int_V d\mathbf{r} \times \Big\{ K_1 \left(\nabla\cdot\mathbf{n}\right)^2 + K_2 \left(\mathbf{n} \cdot \left[\nabla\times\mathbf{n}\right]\right)^2 \\
                     &+ K_3 \left(\mathbf{n}\times\left[\nabla\times\mathbf{n}\right]\right)^2 + 2k_t\left(\mathbf{n} \cdot \left[\nabla\times\mathbf{n}\right]\right)\Big\}
\end{split}
\end{equation}
with $k_bT$ the thermal energy and $V$ the total volume of the sample. $\mathscr{F}_0$ corresponds to the free energy density of the uniformly aligned nematic phase, and $k_t$ is a chiral strength parameter which vanishes in the case of achiral mesogens. $K_1$, $K_2$ and $K_3$ are referred to as the \textit{Frank elastic constants}, and quantify the respective free energy penalties associated with splay, twist and bend deformation modes of the local director field.\cite{deGennes} It should be noted that this compact expression derives from a finite-order truncation of the free energy gradient expansion, and is therefore only valid in the limit of long-wavelength director fluctuations --- an assumption consistent with the characteristically large values of experimental cholesteric pitches.\cite{Yevdokimov, vanWinkle, Stanley, Dogic-2, Grelet-1, DuPre, Dong, Miller}
\par
Eq.~\eqref{eq:frank} thus provides a natural theoretical framework for the description of smoothly modulated nematic phases, and has been successfully used in recent years for the analytical\cite{Majumdar, Dammone-1} and numerical\cite{Majumdar, Dammone-2} modelling of a number of experimentally-relevant systems. However, its main shortcoming lies in the coarse-graining of all bulk particle properties into the three $K_i$ scalars --- understood as free parameters arising from basic mean-field symmetries, and often taken to be equal for the sake of simplicity.\cite{deGennes} In order to provide a full quantitative, bottom-up description of nematic liquid crystals, it is therefore desirable to derive microscopic expressions for these Frank elastic constants, and thus provide a link between mesogen characteristics and phase properties in such self-assembled systems.
\par
A typical starting point for this analysis, as first proposed in Straley's pioneering work,\cite{Straley-1} lies in the molecular description of the free energy $\mathscr{F}_0$ of the reference uniform nematic state. The first tractable expression for this free energy was reported by Onsager\cite{Onsager} for generic systems of highly anisotropic particles and forms the basis of his classical density functional theory, which we outline in the next section.

\subsection{The Onsager DFT for uniaxial LCs} \label{subsec:Onsager DFT for uniaxial LCs}
The original formulation of the Onsager DFT aims to quantify the onset of macroscopic nematic order in terms of competing entropic contributions at the microscopic level.\cite{Onsager} Similarly to other DFT approaches,\cite{Evans-2} its starting point lies in the derivation of the free energy of the system as a functional of the mesoscopically-averaged single-particle density $\rho(\mathbf{r}, \mathcal{R})$, defined as the number of particles at position $\mathbf{r}$ with orientation $\mathcal{R}$ per unit volume and solid angle. $\mathcal{R}$ is generally represented by a full $3\times3$ rotation matrix, which we choose to parametrise in the $z$-$y$-$z$ convention using the three Euler angles $\alpha$, $\theta$ and $\phi$, as depicted in Fig.~\ref{fig1}a. In the special case of uniaxial calamitic particles, the internal rotational degree of freedom $\alpha$ becomes irrelevant, and the symmetries of the system are fully captured by the respective polar and azimuthal angles $\theta$ and $\phi$, defining the orientation vector $\mathbf{u}$ of the particle long axis.
\par
Analogously, in the case of prolate uniaxial phases, internal particle angles $\alpha$ do not display long-range correlations, and the macroscopic structure of the sample is thus entirely characterised by the spatial distribution of the long axes of its constituent particles. The corresponding density field hence reduces to
\begin{equation}
  \label{eq:density_orientation}
  \rho(\mathbf{r}, \mathcal{R}) = \rho(\mathbf{r}, \mathbf{u}).
\end{equation}
Furthermore, owing to the lack of positional order in nematic LCs, the number of particles per unit volume is often taken to be spatially uniform within the sample, which yields the following marginal normalisation constraint:
\begin{equation}
  \label{eq:marginal_norm}
  \oint d\mathcal{R} \times \rho(\mathbf{r}, \mathbf{u}) = 2\pi \oint d\mathbf{u} \times \rho(\mathbf{r}, \mathbf{u}) = \frac{N}{V} \quad \forall \mathbf{r} \in V
\end{equation}
with $N$ the total number of particles in the system. Denoting this uniform number density by $\rho \equiv N/V$, the density field of a bulk nematic phase with director $\mathbf{n}$ then takes the general form\cite{Wensink-1}
\begin{equation}
  \label{eq:density}
  \rho(\mathbf{r}, \mathbf{u}) = \rho \times \psi\left[\mathbf{n}(\mathbf{r}) \cdot \mathbf{u} \right]
\end{equation}
with $\psi\left[\mathbf{n}(\mathbf{r}) \cdot \mathbf{u} \right]$ the so-called \textit{orientation distribution function} (ODF), which quantifies the probability of finding a particle at point $\mathbf{r}$ with its long axis borne by $\mathbf{u}$ with respect to the local director $\mathbf{n}(\mathbf{r})$. Note that ansatz \eqref{eq:density} further reduces the orientational dependence of the density field to the single scalar $\mathbf{n}(\mathbf{r}) \cdot \mathbf{u}$, and thereby assumes that the system retains local cylindrical invariance.\cite{Straley-1} This symmetry constraint thus restricts its applicability to the description of weakly distorted nematic phases, consistent with the Frank-Oseen perturbative framework introduced above. 
\par
Following in the footsteps of Onsager, let us first consider the case of a spatially uniform director field $\mathbf{n}_0$, which we define as the $z$-axis of our reference frame
\begin{equation}
  \mathbf{n}(\mathbf{r})=\mathbf{n}_0 \equiv \mathbf{e}_z \quad \forall \mathbf{r} \in V.
\end{equation}
Eq.~\eqref{eq:density} then reduces to
\begin{equation}
  \label{eq:nematic_density}
  \rho(\mathbf{r}, \mathbf{u}) = \rho(\mathbf{u}) = \rho \times \psi\left(\cos \theta \right)
\end{equation}
with $\theta$ the particle polar angle as defined in Fig.~\ref{fig1}a. The free energy of the system can then be cast in the usual form\cite{Onsager}
\begin{equation}
\label{eq:free_energy}
  \mathscr{F}_0^\infty[\rho] = \mathscr{F}_0^\infty[\psi] = \mathscr{F}_{\rm id}[\psi] + \mathscr{F}_{\rm ex}^\infty[\psi],
\end{equation}
where the subscript $0$ denotes the total free energy of the undeformed phase, and $\mathscr{F}_{\rm id}$ and $\mathscr{F}_{\rm ex}^\infty$ its respective ideal and excess components,
\begin{equation}
\begin{split}
  \label{eq:free_ideal}
  \frac{\beta\mathscr{F}_{\rm id}[\psi]}{V} = &\:\rho (\log \rho-1) + 4\pi^2 \rho \int_0^\pi d\theta \\ &\times \sin\theta \log\Big\{\psi(\cos\theta)\Big\} \psi(\cos\theta),
\end{split}
\end{equation}
\begin{equation}
\begin{split}
  \label{eq:free_excess}
  \frac{\beta\mathscr{F}_{\rm ex}^\infty[\psi]}{V} =& -\frac{\rho^2}{2V} \iint_V d\mathbf{r}_1 d\mathbf{r}_2 \oiint d\mathcal{R}_1 d\mathcal{R}_2 \\ &\times f(\mathbf{r}_{12}, \mathcal{R}_1, \mathcal{R}_2) \psi(\mathbf{n}_1 \cdot \mathbf{u}_1) \psi(\mathbf{n}_2 \cdot \mathbf{u}_2)
\end{split}
\end{equation}
with $\mathbf{r}_{1}$ and $\mathbf{r}_{2}$ the respective positions of the particle centers of mass, $\mathbf{r}_{12} \equiv \mathbf{r}_2 - \mathbf{r}_1$ and
\begin{equation}
  \label{eq:uniform}
  \mathbf{n}_i \cdot \mathbf{u}_i \equiv \mathbf{n}(\mathbf{r}_i) \cdot \mathbf{u}_i= \cos\theta_i \quad \forall i\in\llbracket 1,2 \rrbracket.
\end{equation}
The ideal free energy (Eq.~\eqref{eq:free_ideal}) corresponds to the free energy of an ideal gas of anisotropic particles with uniform density $\rho$, whose two terms can be interpreted as the respective contributions of translational and rotational entropy. The excess free energy (Eq.~\eqref{eq:free_excess}) stems from the truncation of the Mayer cluster expansion\cite{Mayer} at the second-virial level, which as discussed in the next section is only exact in the limit of infinitely thin particles,\cite{Onsager} and is hence denoted by an $\infty$ superscript. It accounts for microscopic interactions in the pairwise additive approximation through the so-called Mayer $f$-function\cite{Mayer}
\begin{equation}
\label{eq:Mayer}
  f(\mathbf{r}_{12}, \mathcal{R}_1, \mathcal{R}_2) = \exp\Big\{-\beta U(\mathbf{r}_{12}, \mathcal{R}_1, \mathcal{R}_2)\Big\} - 1
\end{equation} 
with $\beta$ the inverse thermal energy and $U$ the particle pairwise potential.
\par
It can then be proven that the equilibrium ODF $\psi^{(\rho)}_{\rm eq}$ that minimises the nematic free energy (Eq.~\eqref{eq:free_energy}) at fixed density $\rho$ satisfies the following self-consistent equation:
\begin{equation}
\begin{split}
  \label{eq:self-consistent}
  \psi^{(\rho)}_{\rm eq}(\cos\theta) = &\:\frac{1}{Z} \times \exp\bigg\{-\frac{\rho}{4\pi^2} \int_0^\pi d\theta'  \\ &\times \sin\theta' \frac{E^\infty(\theta, \theta')+E^\infty(\theta', \theta)}{2} \psi^{(\rho)}_{\rm eq}(\cos\theta') \bigg\}                                             
\end{split}
\end{equation}
with $E$ the orientation-dependent second-virial coefficient\cite{Dijkstra-1}
\begin{equation}
\begin{split}
  \label{eq:excluded}
  E^\infty(\theta_1, \theta_2) = &-\int_V d\mathbf{r}_{12}  \iint_0^{2\pi} d\alpha_1 d\alpha_2 \iint_0^{2\pi} d\phi_1 d\phi_2 \\
                   &\times f(\mathbf{r}_{12}, \mathcal{R}_1, \mathcal{R}_2)
\end{split}
\end{equation}
and $Z$ a normalisation constant ensuring that Eq.~\eqref{eq:marginal_norm} is satisfied, namely,
\begin{equation}
  Z = 4\pi^2 \int_0^\pi d\theta \times \sin\theta \times \psi^{(\rho)}_{\rm eq}(\cos\theta).
\end{equation}
Note that in the case of monodisperse systems of identical molecules, it follows from the invariance of the interaction potential with respect to particle exchange that $E^\infty(\theta_1, \theta_2) = E^\infty(\theta_2, \theta_1)$. We however choose to keep the explicit symmetric dependence of Eq.~\eqref{eq:self-consistent} on $E^\infty$ for reasons of numerical convergence, as this form provides an effective average over the non-diagonal coefficients of $E^\infty$ obtained using the computational procedure introduced in Sec.~\ref{subsec:Numerical procedure}.
\par
Eqs.~\eqref{eq:nematic_density} and \eqref{eq:self-consistent} thus fully describe the equilibrium structure of a nematic LC at any given density $\rho$. However, this original form of the Onsager DFT is restricted by geometrical assumptions to infinitely thin particles and uniform director fields, and is therefore of limited practical relevance. We hence devote the next two sections to a thorough, though non-exhaustive, overview of subsequent attempts to overcome both these limitations.

\subsection{Introducing finite mesogen anisotropy} \label{subsec:Introducing finite mesogen anisotropy}
The main difficulty in dealing with finite-size particles lies in the need to go beyond the second-virial approximation, as the relative weight of higher order contributions rapidly increases with decreasing aspect ratios;\cite{Frenkel-1} the Onsager theory has thus been found to be quantitatively inaccurate for systems with aspect ratios of up to 100.\cite{Straley-2} Therefore, several improvements have been proposed in the last decades to extend this framework to a wider class of experimentally-realistic mesogens.
\par
While some studies chose to explicitly include the third and higher order virial terms in the excess free energy expansion (Eq.~\eqref{eq:free_excess}), the quickly escalating computational complexity of such approaches with increasing truncation order largely restricts their applicability to highly simplified particle models.\cite{Masters-3} Furthermore, evidence of the decreased radius of convergence of the virial series for highly anisotropic mesogens may paradoxically limit their numerical stability range to regions of lower density,\cite{Masters-1} and even prevent the prediction of a stable nematic phase altogether for aspect ratios as low as 10 as one pushes the expansion beyond the third order in $\rho$.\cite{Masters-2}
\par
A somewhat orthogonal approach is to forego this virial formalism as a starting point, and instead exploit the more general relation between excess free energy and inter-particle correlations in the density functional framework,\cite{Hansen} which in our case takes the form
\begin{equation}
  \label{eq:DCF}
  \frac{\delta^2 \beta \mathscr{F}_{\rm ex}[\psi]}{\delta \psi(\mathbf{n}_1 \cdot \mathbf{u}_1)\delta\psi(\mathbf{n}_2 \cdot \mathbf{u}_2)}\Biggr|_{\rho\psi} = -\rho^2 \times c^{(2)}(\mathbf{r}_{12},  \mathcal{R}_1,  \mathcal{R}_2; [\rho\psi])
\end{equation}
using the notation of Sec.~\ref{subsec:Onsager DFT for uniaxial LCs}. $c^{(2)}$ is known as the \textit{two-particle direct correlation function} (DCF), and corresponds to the ``direct'' component of the usual pair correlation function (PCF), omitting indirect contributions stemming from mutual correlations with neighbouring particles. In the low-density limit, this DCF can be shown to reduce to the Mayer $f$-function (Eq.~\eqref{eq:Mayer})\cite{Lekkerkerker-1}
\begin{equation}
\label{eq:second_DCF}
  \lim_{\rho \to 0} \: c^{(2)}(\mathbf{r}_{12},  \mathcal{R}_1,  \mathcal{R}_2; [\rho\psi]) = f(\mathbf{r}_{12},  \mathcal{R}_1,  \mathcal{R}_2).
\end{equation}
However, this equality once again falters in the case of particles with a lower aspect ratio, for which the onset of liquid-crystalline order occurs at higher mesogen concentrations. In the general case, the exact relation between DCF, director profile and excess free energy at given uniform density $\rho$ can be obtained from Eq.~\eqref{eq:DCF} by double functional integration,
\begin{equation}
\begin{split}
\label{eq:excess_DCF}
  \frac{\beta\mathscr{F}_{\rm ex}[\psi]}{V} = &-\frac{\rho^2}{2V} \iint_V d\mathbf{r}_1 d\mathbf{r}_2 \oiint d\mathcal{R}_1 d\mathcal{R}_2 \\ &\times \overline{c^{(2)}}(\mathbf{r}_{12}, \mathcal{R}_1, \mathcal{R}_2; [\rho\psi]) \psi(\mathbf{n}_1 \cdot \mathbf{u}_1) \psi(\mathbf{n}_2 \cdot \mathbf{u}_2)
\end{split}
\end{equation}
with $\overline{c^{(2)}}$ the DCF integrated along a compression path from a reference state with particle density $\rho=0$,\cite{Lekkerkerker-1}
\begin{equation}
\begin{split}
\label{eq:integrated_DCF}
   \overline{c^{(2)}}(\mathbf{r}_{12}, \mathcal{R}_1, \mathcal{R}_2; [\rho\psi]) = &\:2 \int_0^1 d\lambda \int_0^\lambda d\lambda' \\ &\times c^{(2)}(\mathbf{r}_{12},  \mathcal{R}_1,  \mathcal{R}_2; [\lambda' \times \rho \psi]).
\end{split}
\end{equation}
Note that plugging Eq.~\eqref{eq:second_DCF} into Eq.~\eqref{eq:integrated_DCF} yields the density-functional equivalent of the second-virial approximation
\begin{equation}
  \label{eq:second_DFT}
  \overline{c^{(2)}}(\mathbf{r}_{12},  \mathcal{R}_1,  \mathcal{R}_2; [\rho\psi]) \simeq f(\mathbf{r}_{12},  \mathcal{R}_1,  \mathcal{R}_2)
\end{equation}
which reduces Eq.~\eqref{eq:excess_DCF} to the Onsager excess free energy (Eq.~\eqref{eq:free_excess}). Hence, Eqs.~\eqref{eq:excess_DCF} and \eqref{eq:integrated_DCF} render the accuracy of the excess free energy functional purely contingent on that of the underlying DCF for systems of arbitrarily anisotropic particles.
\par
While good approximations for the DCF of spherical particle fluids have been obtained by directly solving the Ornstein-Zernike equation using various closure relations,\cite{Wertheim, Morita} the extra orientational degrees of freedom render this process mostly intractable when particle anisotropy is introduced.\cite{Lekkerkerker-1} A number of recent efforts to derive closed-form expressions for the excess free energy and associated DCF of anisotropic systems have thus resorted to first-principle geometric considerations instead, on the basis of Rosenfeld's original fundamental measure theory (FMT) for hard spheres.\cite{Rosenfeld} Although FMT presumably provides a good description of the DCF spatial-orientational coupling, such approaches are unfortunately limited by construction to convex particle models\cite{Wittmann} and are not naturally suited for the treatment of soft interaction potentials.\cite{Schmidt}
\par
Another class of approaches revolve around the derivation of \textit{ad hoc} approximate expressions for these quantities by remapping either the relevant DCF or excess free energy to those of the well-characterised hard-sphere fluid.\cite{Lekkerkerker-1} A significant number of such methods rely on the so-called \textit{decoupling approximation}, in which inter-particle correlations are accounted for via simplified ans\"atze of the form
\begin{equation}
  \label{eq:DCF_decorrelated}
   c^{(2)}(\mathbf{r}_{12},  \mathcal{R}_1,  \mathcal{R}_2; [\rho \psi]) = c^{(2)}_{\rm ref}\left\{\frac{\lVert\mathbf{r}_{12}\rVert}{\sigma(\mathbf{u}_{12},  \mathcal{R}_1,  \mathcal{R}_2)};\rho\right\}
\end{equation}
irrespective of $\psi$, with $\lVert \cdot \rVert$ the Euclidean norm and $\mathbf{u}_{12} \equiv \mathbf{r}_{12}/\lVert\mathbf{r}_{12}\rVert$. $c^{(2)}_{\rm ref}$ then corresponds to a heuristic DCF inferred from that of a reference hard-sphere system at the same density $\rho$, either explicitly through the use of a closure relation as first proposed by Pynn\cite{Pynn} and Wulf,\cite{Wulf} or implicitly from the corresponding virial expansion as suggested independently by Parsons\cite{Parsons} and Lee.\cite{Lee-1} In this framework, particle anisotropy is introduced solely through the orientation-dependent distance of nearest approach between two particles $\sigma(\mathbf{u}_{12},  \mathcal{R}_1,  \mathcal{R}_2)$. In the case of purely hard interactions, Eqs.~\eqref{eq:excess_DCF} and \eqref{eq:integrated_DCF} then lead to effective expressions for $\overline{c^{(2)}}$ of the general form
\begin{equation}
  \label{eq:rescaled}
   \overline{c^{(2)}}(\mathbf{r}_{12},  \mathcal{R}_1,  \mathcal{R}_2; [\rho\psi]) = G(\rho v_{\rm ref}) \times f(\mathbf{r}_{12},  \mathcal{R}_1,  \mathcal{R}_2)
\end{equation}
with $v_{\rm ref}$ the molecular volume of the reference hard-sphere fluid, which may \textit{a priori} differ from that of the anisotropic particles considered, as discussed in Sec.~\ref{subsec:Numerical procedure}. In the general case, the effective volume fraction $\eta_{\rm ref}$ of the reference system is thus linked to that of the physical system with molecular volume $v$ at identical number density $\rho$ through the relation
\begin{equation}
  \label{eq:effective_eta}
  \eta_{\rm ref}^{(\rho)} \equiv \rho v_{\rm ref} = \eta^{(\rho)} \times \frac{v_{\rm ref}}{v}.
\end{equation}
The density-dependent prefactor $G$ is then related to the compressibility factor of the chosen reference framework --- namely, the Percus-Yevick free energy\cite{Percus, Wertheim} in the Pynn-Wulf (PW) approximation, and the Carnahan-Starling equation of state\cite{Carnahan} in the Parsons-Lee (PL) approach. The respective corresponding expressions for $G$ read as\cite{Lekkerkerker-1}
\begin{align}
  \label{eq:g_pw}
  G_{\rm PW}(\eta) &= \frac{3/4-3\eta/8}{(1-\eta)^2} - \frac{\ln(1-\eta)}{4\eta}, \\
  \label{eq:g_pl}
  G_{\rm PL}(\eta) &= \frac{1-3\eta/4}{(1-\eta)^2}
\end{align}
which both trivially satisfy
\begin{equation}
  \lim_{\eta \to 0} \: G(\eta) = 1.
\end{equation}
Hence, Eq.~\eqref{eq:rescaled} simply reduces to the second-virial approximation Eq.~\eqref{eq:second_DFT} in the low-density limit.
\par
In the following, we choose to account for finite particle anisostropy through the Parsons-Lee formalism (Eq.~\eqref{eq:g_pl}), due to both its simplicity and well-documented success in describing the nematic behaviour of a wide range of hard\cite{Camp, Cuetos-2} and soft\cite{Cuetos-1} particle models; however, most of the methods described in the next sections are easily generalisable to other theoretical frameworks. Plugging Eq.~\eqref{eq:rescaled} into Eq.~\eqref{eq:excess_DCF} then yields the relevant excess free energy and angle-dependent virial coefficient as simple rescaled forms of their respective Onsager expressions Eqs.~\eqref{eq:free_excess} and \eqref{eq:excluded},
\begin{align}
  \label{eq:excess_rescaled}
  \mathscr{F}_{\rm ex}[\psi] &= G_{\rm PL}\left(\eta_{\rm ref}^{(\rho)}\right) \times \mathscr{F}_{\rm ex}^\infty[\psi], \\
  \label{eq:virial_rescaled}
  E(\theta_1, \theta_2) &=G_{\rm PL}\left(\eta_{\rm ref}^{(\rho)}\right) \times E^\infty(\theta_1, \theta_2),
\end{align}
and the equilibrium properties of the system may be fully determined by substituting Eq.~\eqref{eq:excess_rescaled} for $\mathscr{F}_{\rm ex}^\infty$ in the total free energy (Eq.~\eqref{eq:free_energy}),
\begin{equation}
   \label{eq:rescaled_free_energy}
   \mathscr{F}_0[\psi] = \mathscr{F}_{\rm id}[\psi] + G_{\rm PL}\left(\eta_{\rm ref}^{(\rho)}\right) \times \mathscr{F}_{\rm ex}^\infty[\psi]
\end{equation}
and Eq.~\eqref{eq:virial_rescaled} for $E^\infty$ in the self-consistent Eq.~\eqref{eq:self-consistent},
\begin{equation}
\begin{split}
   \label{eq:rescaled_self-consistent}
   \psi^{(\rho)}_{\rm eq}(\cos\theta) = &\:\frac{1}{Z} \times \exp\bigg\{- G_{\rm PL}\left(\eta_{\rm ref}^{(\rho)}\right) \times \frac{\rho}{4\pi^2} \int_0^\pi d\theta' \\ &\times \sin\theta' \frac{E^\infty(\theta, \theta')+E^\infty(\theta', \theta)}{2} \psi^{(\rho)}_{\rm eq}(\cos\theta') \bigg\}.   
\end{split}
\end{equation}
\par
We conclude this section by remarking that computational methods for inferring accurate DCFs from the PCFs measured in direct molecular simulations have also been proposed,\cite{Schmid-1} and that Eq.~\eqref{eq:excess_DCF} thus conceptually enables one to calculate the ``exact'' excess free energy of bulk nematic phases for a wide range of particle models by simply plugging the corresponding numerical DCFs into Eq.~\eqref{eq:integrated_DCF}. Although the appeal of such a prospect is undeniable, its practical applicability is greatly hindered by the functional dependence of the DCF on the nematic density field (Eq.~\eqref{eq:density}), as $c^{(2)}$ needs to be numerically evaluated for all density distributions along the integration path (Eq.~\eqref{eq:integrated_DCF}). A useful approximation may therefore be borrowed from the DFT of simple fluids by invoking linear response theory in the limit of weakly non-uniform systems, which using our notations yields\cite{Ebner-1, Ebner-2}
\begin{equation}
  \label{eq:simpleDFT}
  \overline{c^{(2)}}(\mathbf{r}_{12},  \mathcal{R}_1,  \mathcal{R}_2; [\rho\psi]) \simeq c^{(2)}(\mathbf{r}_{12},  \mathcal{R}_1,  \mathcal{R}_2; [\rho\psi]).
\end{equation}
While the validity of the assumptions underlying Eq.~\eqref{eq:simpleDFT}, which in our case amount to treating the nematic phase as a weakly disturbed isotropic liquid, is clearly questionable, this relation provides a simple and direct way to check the effective correlations derived from the previous theories Eqs.~\eqref{eq:rescaled}--\eqref{eq:g_pl} against the numerical DCFs predicted by simulations\cite{Schmid-2} in the vicinity of the I-N transition, and is worth keeping in mind for the assessment of future theoretical developments.

\subsection{Introducing weak nematic fluctuations} \label{subsec:Introducing weak nematic fluctuations}
The inclusion of weak director fluctuations in the previous framework as first proposed by Straley\cite{Straley-1} relies on the two following related assumptions, both typical of a perturbative treatment: \textbf{(i)} that the degree of local order is identical in the distorted and undistorted phases, so that $c^{(2)}$ and $\psi$ may be taken as equal in both systems, and \textbf{(ii)} that the wavevectors $\mathbf{q}$ of the director spatial fluctuations are small, so that the Frank free energy expansion Eq.~\eqref{eq:frank} is well defined.
On the basis of these two hypotheses, a general microscopic description of the Frank elastic constants can be straightforwardly derived within the density functional framework of the previous section; the resulting set of expressions are collectively known as the Poniewierski-Stecki equations:\cite{Poniewierski-1}
\begin{equation}
\label{eq:K1}
\begin{split}
  \beta K_1[\psi] =& \:\frac{\rho^2}{2} \int_V d\mathbf{r}_{12} \oiint d\mathcal{R}_1 d\mathcal{R}_2 \\&\times c^{(2)}(\mathbf{r}_{12},  \mathcal{R}_1,  \mathcal{R}_2; [\rho \psi]) \dot{\psi}(u_{1z}) \dot{\psi}(u_{2z}) r_x^2 u_{1x} u_{2x}
\end{split}
\end{equation}
\begin{equation}
  \label{eq:K2}
  \begin{split}
  \beta K_2[\psi] =& \:\frac{\rho^2}{2} \int_V d\mathbf{r}_{12} \oiint d\mathcal{R}_1 d\mathcal{R}_2 \\&\times c^{(2)}(\mathbf{r}_{12},  \mathcal{R}_1,  \mathcal{R}_2; [\rho \psi]) \dot{\psi}(u_{1z}) \dot{\psi}(u_{2z}) r_x^2 u_{1y} u_{2y}
  \end{split}
  \end{equation}
  \begin{equation}
  \begin{split}
  \label{eq:K3}
  \beta K_3[\psi] =& \:\frac{\rho^2}{2} \int_V d\mathbf{r}_{12} \oiint d\mathcal{R}_1 d\mathcal{R}_2 \\&\times c^{(2)}(\mathbf{r}_{12},  \mathcal{R}_1,  \mathcal{R}_2; [\rho \psi]) \dot{\psi}(u_{1z}) \dot{\psi}(u_{2z}) r_z^2 u_{1x} u_{2x}
  \end{split}
  \end{equation}
which in the case of chiral particles can easily be supplemented by a fourth equation for the chiral strength parameter:
\begin{equation}
\begin{split}
  \label{eq:kt}
  \beta k_t [\psi] = &-\frac{\rho^2}{2} \int_V d\mathbf{r}_{12} \oiint d\mathcal{R}_1 d\mathcal{R}_2 \\&\times c^{(2)}(\mathbf{r}_{12},  \mathcal{R}_1,  \mathcal{R}_2; [\rho \psi]) \psi(u_{1z}) \dot{\psi}(u_{2z}) r_x u_{2y}
  \end{split}
\end{equation}
with $r_j = \mathbf{r}_{12} \cdot \mathbf{e}_j$ and $u_{ij} = \mathbf{u}_i \cdot \mathbf{e}_j$. It is worth emphasising that Eqs.~\eqref{eq:K1}--\eqref{eq:kt} can be single-handedly inferred from Eq.~\eqref{eq:DCF} using solely the two previous assumptions, and are therefore not contingent on a particular expression for the nematic free energy.
\par
In our case, this formulation is unfortunately of little use as the Parsons-Lee theory does not provide an explicit expression for the two-particle DCF $c^{(2)}$. Therefore, a more practical derivation may be obtained by following Ref.~\onlinecite{Wensink-1} and explicitly performing the Taylor expansion of the distorted free energy $\mathscr{F}$ around that of the undistorted phase (Eq.~\eqref{eq:rescaled_free_energy}). Using standard expressions for the different nematic deformation modes\cite{Straley-3} along with the two above assumptions enables one to readily recover the Poniewierski-Stecki equations from Eqs.~\eqref{eq:free_excess} and \eqref{eq:rescaled_free_energy}, with the effective DCF
\begin{equation}
  \label{eq:dcf_pl}
  c^{(2)}(\mathbf{r}_{12},  \mathcal{R}_1,  \mathcal{R}_2; [\rho \psi]) = G\left(\eta_{\rm ref}^{(\rho)}\right) \times f(\mathbf{r}_{12},  \mathcal{R}_1,  \mathcal{R}_2)
\end{equation}
which interestingly corresponds to the first-order term of the DCF obtained through different arguments by Somoza and Tarazona for hard spherocylinders,\cite{Somoza-2} as discussed further in Sec.~\ref{subsec:The Frank elastic constants}.
\par
We here briefly outline this derivation in the case of a cholesteric distortion, and refer the interested reader to Ref.~\onlinecite{Wensink-1} for the more detailed calculations. The director field then exhibits a helical modulation described by the simple expression
\begin{equation}
  \label{eq:cholesteric_director}
  \mathbf{n}(\mathbf{r}) = (0, -\sin qr_x, \cos qr_x) = (0, -qr_x, 1) + \mathcal{O}(q^2),
\end{equation}
where we used the conventions of Fig.~\ref{fig1}b along with assumption \textbf{(ii)}, ensuring that $q>0$ corresponds to a right-handed phase. Eq.~\eqref{eq:cholesteric_director} then trivially leads to
\begin{align}
  \mathbf{n} \cdot \left[\nabla\times\mathbf{n}\right] &= -q  + \mathcal{O}(q^2),\\
  \mathbf{n}\times\left[\nabla\times\mathbf{n}\right] &= \mathcal{O}(q^2), \\
  \nabla\cdot\mathbf{n} &= 0.
\end{align}
Following the previous procedure, plugging Eq.~\eqref{eq:cholesteric_director} into the Onsager excess free energy (Eq.~\eqref{eq:free_excess}) and expanding the ODF up to order 2 in $q$ allows one to recast the equilibrium free energy (Eq.~\eqref{eq:rescaled_free_energy}) in the form\cite{Wensink-1}
\begin{equation}
  \label{eq:cholesteric_free_energy}
  \frac{\mathscr{F}[\psi^{(\rho)}_{\rm eq}]}{V} = \frac{\mathscr{F}_0[\psi^{(\rho)}_{\rm eq}]}{V} -k_t[\psi^{(\rho)}_{\rm eq}] q + K_2[\psi^{(\rho)}_{\rm eq}] \frac{q^2}{2} + \mathcal{O}(q^3)
\end{equation}
which by term-to-term comparison with Eq.~\eqref{eq:frank} yields Eqs.~\eqref{eq:K2} and \eqref{eq:kt} along with the effective DCF (Eq.~\eqref{eq:dcf_pl}). Similar calculations using the corresponding distorted director fields lead to the other two elastic constants,\cite{Straley-3} with the notable difference that the linear terms of the associated free energy expansions vanish for non-polar mesogens.\cite{deGennes}
\par
Assumption \textbf{(i)} further ensures that the equilibrium ODF $\psi^{(\rho)}_{\rm eq}$ of the cholesteric phase is still given by Eq.~\eqref{eq:rescaled_self-consistent} irrespective of $q$, so that the equilibrium pitch $\mathcal{P}_{\rm eq}$ minimising the distorted free energy (Eq.~\eqref{eq:cholesteric_free_energy}) at given density $\rho$ is fully determined by
\begin{equation}
  \label{eq:q_eq}
  \mathcal{P}_{\rm eq}(\rho) = \frac{2\pi}{q_{\rm eq}(\rho)} = 2\pi \times \frac{K_2[\psi^{(\rho)}_{\rm eq}]}{k_t[\psi^{(\rho)}_{\rm eq}]}
\end{equation}
which quantifies the competition between the chiral torque $k_t$, favoring a locally-twisted phase, and the Frank twist elastic constant $K_2$, aiming to restore local alignment.\cite{Straley-1}
\par
It should be noted that this perturbative treatment of chirality --- along with most discussions of this paper --- is based on a uniaxial ansatz (Eq.~\eqref{eq:nematic_density}) for the nematic density field, which assumes a uniform distribution of particle angles $\alpha$ and $\phi$ within any mesoscopic sample of the system. This hypothesis thus precludes any long-range orientational order for the particle short axes, and therefore dictates that phase chirality arise solely from short-range correlations prescribed by the DCF (Eq.~\eqref{eq:dcf_pl}). While consistent with our weak deformation framework, this conjecture is hard to justify \textit{a priori} for systems of chiral particles and may adversely affect the reliability of the theory for cholesteric phases, as discussed in Sec.~\ref{subsec:The cholesteric pitch of hard helices}. We however leave these considerations aside for now in order to facilitate the comparison of our results with previous theoretical studies, in which similar symmetry assumptions were used.

\subsection{Numerical procedure} \label{subsec:Numerical procedure}
We now move on to more practical considerations, and outline the general numerical methods used to infer the equilibrium properties of chiral nematic phases from that of their arbitrary constituent molecules in our density functional framework. Following the considerations of Secs.~\ref{subsec:Introducing finite mesogen anisotropy} and~\ref{subsec:Introducing weak nematic fluctuations}, we hence consider from here onwards systems described by the uniaxial Onsager-Parsons-Lee equilibrium ODF $\psi_{\rm eq}^{(\rho)}$ (Eq.~\eqref{eq:rescaled_self-consistent}), and introduce the shorthand
\begin{equation}
  \mathscr{X}(\rho) \equiv \mathscr{X}[\psi_{\rm eq}^{(\rho)}]
\end{equation}
for all functionals $\mathscr{X}$ of the nematic density distribution.
\par
The working principle of our algorithm can then be summarised as follows.
\begin{enumerate}[(i), nosep]
\item \label{second-virial} We start by numerically integrating the angle-dependent virial coefficient $E(\theta_1, \theta_2)$ (Eq.~\eqref{eq:virial_rescaled}) for the chosen particle model using a discrete grid for polar angles $\theta \in [0,\pi[$.
\item \label{ODF_self-consistent} $E$ is subsequently plugged into Eq.~\eqref{eq:rescaled_self-consistent}, which is solved self-consistently using a standard iterative scheme\cite{vanRoij} to compute the equilibrium ODF $\psi^{(\rho)}_{\rm eq}$ for a given value of $\rho$.
\item \label{poniewierski} $\psi^{(\rho)}_{\rm eq}$ is then plugged into Eqs.~\eqref{eq:K1}--\eqref{eq:kt} to work out $K_i(\rho)$ and $k_t(\rho)$ for $i\in\llbracket 1,3 \rrbracket$ by further numerical integration.
\item The equilibrium pitch $\mathcal{P}_{\rm eq}$ of the system at the chosen density $\rho$ is finally given by Eq.~\eqref{eq:q_eq}.
\end{enumerate}
\par
The most computationally-intensive elements of this approach correspond to steps~\ref{second-virial} and~\ref{poniewierski}, in which functionals of the microscopic interaction potential have to be integrated over the two-particle excluded volume manifold for the chosen mesogen model. Due to the high dimensionality of this manifold, we chose to follow the lead of Ref.~\onlinecite{Dijkstra-1} and compute these integrals using stochastic sampling methods, yielding the further advantage of trivial parallelization; the details of this numerical procedure may be found in Ref.~\onlinecite{Dijkstra-2}. We further emphasise that this implementation allows for the full evaluation of the angle-dependant ODF $\psi^{(\rho)}_{\rm eq}(\theta)$, rather than its partial projection on a basis of Wigner D-matrices.\cite{Tombolato, Ferrarini-1} The current approach is thus both more efficient and not subject to truncation errors.
\par
In the following, we refer to the number of MC integration steps used in both stage~\ref{second-virial} and stage~\ref{poniewierski} by $N_{\rm MC}$, and use a finite grid of fixed size $N_\theta=250$ for polar angles $\theta$. The statistical errors of all relevant variables are hence entirely controlled by $N_{\rm MC}$, which can be tuned as required to achieve the desired numerical accuracy.
\par
We remark that this numerical framework is particularly well-suited for the study of lyotropic liquid crystals, as the second-virial coefficient $E^\infty$ defined in Eq.~\eqref{eq:excluded} is generally independent of particle density. Therefore, the full concentration dependence of the ODF in the Parsons-Lee description may be worked out from Eq.~\eqref{eq:rescaled_self-consistent} with minimal numerical effort using a single computation of the virial integral (Eq.~\eqref{eq:excluded}). Nevertheless, all of our previous discussions remain conceptually valid in the case of thermotropic liquid crystals, and our method may therefore also be applied to the investigation of thermotropic self-assembly by evaluating the temperature dependence of $E^\infty$. An important caveat however lies in the limited aspect ratios of most experimental thermotropic mesogens, which may require a more accurate description of inter-particle correlations beyond the decoupling approximation.\cite{Ping} We thus largely restrict our focus to lyotropic LC phases in the rest of this paper.
\par
Importantly, it should be noted that solving Eq.~\eqref{eq:rescaled_self-consistent} in the PL framework at step~\ref{ODF_self-consistent} requires a preliminary evaluation of the molecular volume $v_{\rm ref}$ of the reference hard-sphere system in order to work out $\eta_{\rm ref}^{(\rho)}$ from Eq.~\eqref{eq:effective_eta}. A number of previous studies obtained good agreement with simulation data for hard convex particles by using the original formulation of the PL theory, which simply amounts to setting $v_{\rm ref}=v$ in Eqs.~\eqref{eq:effective_eta}, \eqref{eq:rescaled_free_energy} and \eqref{eq:rescaled_self-consistent}. However, this assumption has been found to lead to significant underestimations of the osmotic pressure for systems of non-convex particles.\cite{Jackson} A simple explanation for this discrepancy may be provided by invoking the concept of \textit{effective molecular volume},\cite{Lago} which we introduce below. 
\par
Let $\Omega \equiv V \times SO(3)$ be the total configuration space accessible to the particles, with $SO(3)$ the group of rotations in Euclidean 3D space, and $(P,P') \in \Omega^2$ an arbitrary pair of identical hard particles. The mathematical definition of the excluded volume manifold $\Omega_{\rm exc}$ which may not be penetrated by particle $P'$ in any configuration due to the presence of $P$ reads as
\begin{equation}
  \label{eq:excluded_manifold}
  \Omega_{\rm exc} = \Big\{\mathbf{r} \in V \,\big\vert\, \forall P' \in \Omega, \, \mathbf{r} \in P' \implies P \cap P' \neq \varnothing \Big\}.
\end{equation}
The effective molecular volume $v_{\rm exc}$ of particle $P$ is then defined as
\begin{equation}
  \label{eq:exc_vol}
  v_{\rm exc} \equiv \mu(\Omega_{\rm exc}),
\end{equation}
with $\mu$ the standard Lebesgue measure for subsets of $\mathbb{R}^3$. It is easy to show that $v_{\rm exc}$ satisfies $v_{\rm exc} \geq v$, which reflects the smaller volume available to non-convex particles with respect to their convex counterparts at given density $\rho$, and therefore accounts for the higher osmotic pressures obtained in numerical simulations for the former systems. Varga and Szalai\cite{Szalai} thus proposed to assign this effective molecular volume to the volume of the reference hard-sphere particles,
\begin{equation}
  \label{eq:ref_vol}
  v_{\rm ref} = v_{\rm exc}
\end{equation}
which yields the so-called modified Parsons-Lee (MPL) approximation for hard non-convex particles.
\par
A general numerical scheme to compute this effective volume for arbitrary particle models may then be obtained by discretising $V$ into a mesh of size $N_x\times N_y \times N_z$ and evaluating the excluded volume manifold $\Omega_{\rm exc}$ by MC sampling, similarly to the integration procedure outlined above. Assuming the first particle $P$ to be fixed at the origin of the reference frame, this process amounts to generating a number $N$ of uncorrelated configurations for particle $P'$ by drawing random center-of-mass positions $\mathbf{r}' \in V$ and particle orientations $\mathcal{R}' \in SO(3)$, and checking $P$ and $P'$ for hard overlaps. In the absence of any such overlaps, all the mesh points contained within the generated configuration for particle $P'$ are guaranteed to lie outside of the excluded volume manifold, and may therefore be pruned. 
\par
This algorithm thus enables one to efficiently obtain a discretised representation of $\Omega_{\rm exc}$ for a wide range of particle models, with a numerical precision determined by both the chosen mesh resolution and the number of MC steps used. The corresponding reference volume $v_{\rm ref}$ may finally be determined from Eqs.~\eqref{eq:exc_vol} and \eqref{eq:ref_vol} as the sum of the volumes of all mesh points comprising $\Omega_{\rm exc}$ at the end of the calculation. 
\par
All effective excluded volumes used in this paper were computed employing a cuboidal integration box of size $V=L_x\times L_y\times L_z$; $L_x$, $L_y$ and $L_z$ were chosen to yield the minimal volume containing all possible overlapping configurations of two particles, prescribing the center of mass of the reference particle $P$ to the origin and its long axis to $\mathbf{e}_z$. The mesh dimensions were then set to $N_x\times N_y \times N_z = 150\times150\times1500$ and a number $N = 1\times10^8$ of random sampling steps was used, yielding a statistical dispersion of less than 1\% for all computed values of $v_{\rm ref}$ and ensuring a negligible dependence on varying mesh resolutions.


\section{Results \& discussion} \label{sec:Results and discussion}
We now illustrate the versatility of our method by determining the nematic and cholesteric phase properties of a variety of mesogenic particles, both chiral and achiral, in the (modified) Parsons-Lee approximation. An interesting first test system to assess its reliability is provided by the coarse-grained helical particle model first introduced in Ref.~\onlinecite{Ferrarini-2}, and subsequently studied extensively using both direct computer simulations\cite{Ferrarini-2, Ferrarini-5} and different theoretical approaches.\cite{Ferrarini-1,Ferrarini-2,Dijkstra-1,Dijkstra-2} Such particles are defined as chains of $N_s$ fused hard spheres of radii $\sigma$ placed along a helical backbone of microscopic pitch $p$, radius $r$ and contour length $l$, thus allowing for tunable levels of shape chirality and non-convexity, as illustrated in Fig.~\ref{fig2}. Similarly to these previous investigations, we here restrict our study to particles with a right-handed symmetry, and use the shorthand $r_ap_bl_c$ to refer to helices with microscopic radius $r=a$, pitch $p=b$ and contour length $l=c$ in units of $\sigma$.

\subsection{The isotropic-cholesteric phase transition} \label{subsec:The isotropic-cholesteric phase transition}
We start by applying our approach to the determination of the isotropic-to-cholesteric phase transition for some of the hard helices above. We first aim to characterise the onset of local nematic order in such systems, and leave the description of the longer-range cholesteric structures arising from particle chirality to Sec.~\ref{subsec:The cholesteric pitch of hard helices}.\footnote{We therefore use the words ``nematic'' and ``cholesteric'' interchangeably in this section and the next, as our perturbative framework imposes that cholesteric and uniaxial nematic phases be undistinguishable at the local scale.} We quantify the degree of local alignment through the usual nematic order parameter\cite{deGennes}
\begin{equation}
  S(\rho) = 4\pi^2 \int_0^\pi d\theta \times \sin \theta \frac{3\cos^2\theta - 1}{2} \times \psi_{\rm eq}^{(\rho)}(\cos\theta).
\end{equation}
The binodal points delimiting the stability ranges of the isotropic and nematic phases can be worked out from the osmotic pressure $\Pi$ and chemical potential $\mu$ of the system, which in our case take the form
\begin{align}
\label{eq:osmotic_pressure}
  \beta \Pi(\rho) &= -\frac{\partial \beta \mathscr{F}_0}{\partial V}\biggr|_{N,T} = \rho \times \left \{ 1 + \frac{b_2^\infty}{v} \frac{\eta_{\rm eff} - \eta_{\rm eff}^2/2}{(1-\eta_{\rm eff})^3} \right \}, \\
  \label{eq:chemical_potential}
  \beta \mu(\rho) &= \frac{\partial \beta \mathscr{F}_0}{\partial N}\biggr|_{V,T} = \log \eta - \Sigma + \rho b_2^\infty \frac{8 - 9\eta_{\rm eff} +3\eta_{\rm eff}^2}{4(1-\eta_{\rm eff})^3}
\end{align}
with $\mathscr{F}_0$ the undistorted free energy (Eq.~\eqref{eq:rescaled_free_energy}), $\Sigma$ the rotational entropy per particle,
\begin{equation}
  \Sigma(\rho) = -4\pi^2 \int_0^\pi d\theta \times \sin \theta \log \left\{ \psi_{\rm eq}^{(\rho)}(\cos\theta) \right \}  \psi_{\rm eq}^{(\rho)}(\cos\theta),
\end{equation}
and $b_2^\infty$ the angle-averaged second virial coefficient\cite{Lee-1}
\begin{equation}
\begin{split}
  b_2^\infty(\rho) = &\: \iint_0^\pi d\theta_1 d\theta_2 \times \sin \theta_1 \sin\theta_2 \times E^\infty(\theta_1, \theta_2) \\ &\times \psi_{\rm eq}^{(\rho)}(\cos\theta_1)\psi_{\rm eq}^{(\rho)}(\cos\theta_2).
\end{split}
\end{equation}
The binodal coexistence concentrations can then be obtained by equating Eqs.~\eqref{eq:osmotic_pressure} and \eqref{eq:chemical_potential} in the two phases and solving the resulting set of coupled equations via a relaxed multivariate Newton-Raphson scheme,\cite{Froberg} using the uniform limit of the ODF for the isotropic phase,
\begin{equation}
  \psi_{\rm iso}(\cos\theta) = \frac{1}{8\pi^2} \quad \forall \theta \in [0,\pi[.
\end{equation} 

\begin{figure}[htpb]
  \includegraphics[width=\columnwidth]{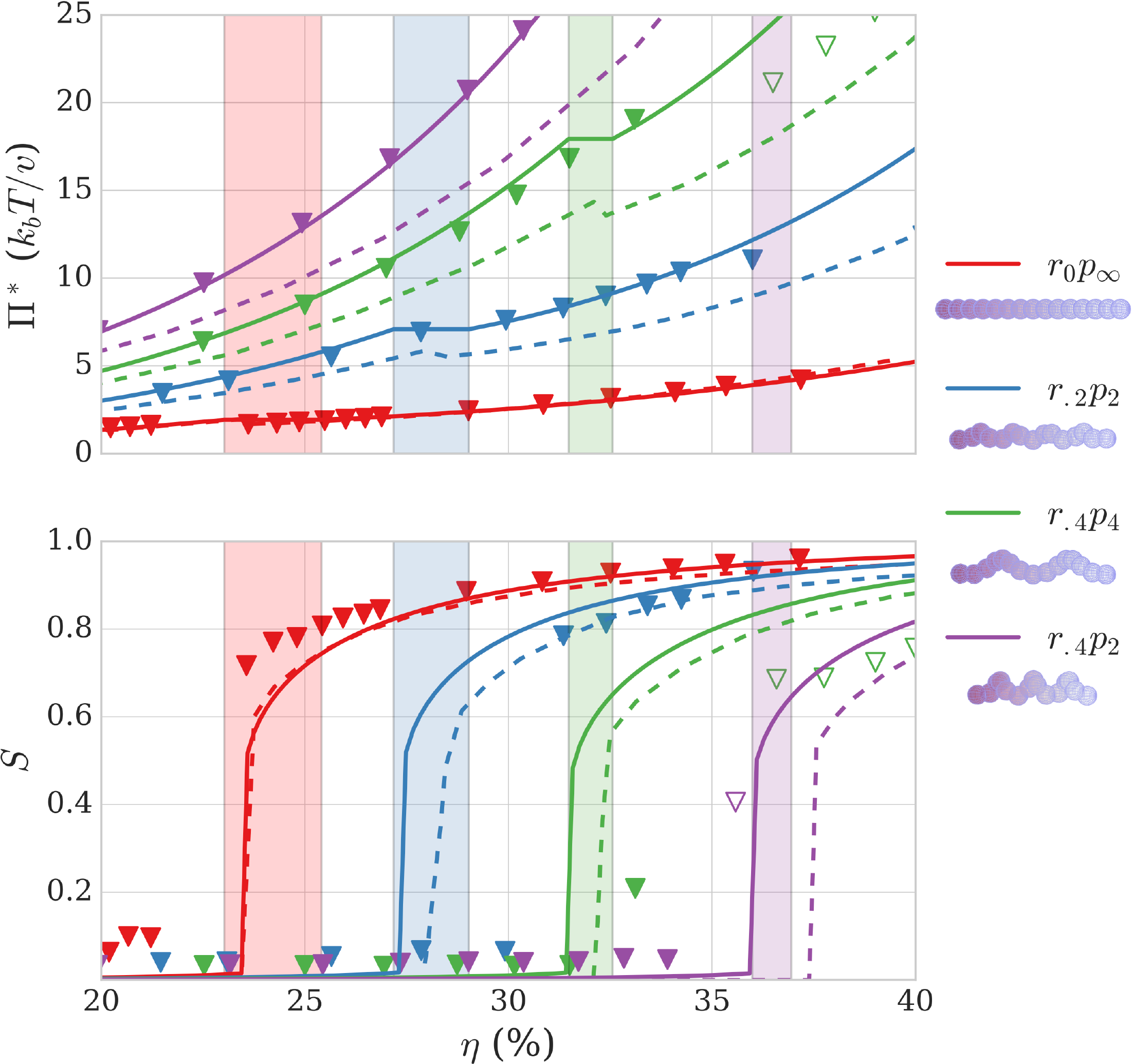}
  \caption{\label{fig2}Nematic order parameter $S$ and rescaled osmotic pressure $\Pi^\ast=\Pi \times \left[ 1 + 50r^2/(p\sigma)\right]$ as a function of particle volume fraction $\eta$ for a selection of hard helices with $N_s=15$ and $l=10\,\sigma$. Solid lines correspond to the results of this work, while symbols and dashed lines represent the simulation and theory data reported in Ref.~\onlinecite{Ferrarini-2}, respectively. Unfilled symbols denote metastable state points not fully characterised by simulations. The colored areas delimit the ranges of I-N coexistence for the different particle models as computed by our approach. Note that for consistency with Ref.~\onlinecite{Ferrarini-2}, nematic order parameters are reported for homogeneous single-phase systems, and their discontinuities thus mark the location of the nematic spinodals. All results were obtained by averaging over 32 independent runs using $N_{\rm MC}=1\times 10^{11}$ integration steps each, and numerical dispersion was found to be negligible at all densities.}
\end{figure}

Some results of our model, specifically the osmotic pressures and nematic order parameters, are summarised in Fig.~\ref{fig2} for different hard helices, including the limiting case of an achiral linear chain of hard spheres (LCHS) for $r=0$. We note that the pressures predicted by our approach are found to be in significantly better agreement with simulation data than those obtained by the theory of Ref.~\onlinecite{Ferrarini-2} in both the cholesteric and isotropic phases for all helices considered. This is because this previous approach assumed the effective molecular volumes $v_{\rm exc}$ of the particles to be given by those of the corresponding LCHS systems with equal contour lengths, for which analytical expressions have been derived.\cite{Szalai} This approximation therefore neglects the contributions of the non-linear molecular backbones to the overall shape non-convexity of the helices, and thus leads to sizeable underestimates of the associated osmotic pressures. These effects may however be accounted for by using the general procedure outlined in Sec.~\ref{subsec:Numerical procedure} to work out the full effective molecular volume of each of the chosen particles; the results of both approaches are found to be virtually identical in the case of the LCHS system, for which our numerical excluded volume converges to the analytical expression of Ref.~\onlinecite{Szalai}.
\par
Interestingly, we also remark that the quantitative agreement of the MPL theory with simulation results in terms of the nematic order parameter $S$ is somewhat less convincing, and that the inclusion of the full molecular excluded volume does conversely not lead to systematic improvements for the prediction of the I-N coexistence range. This observation seemingly mirrors the findings of Ref.~\onlinecite{Szalai} for systems of long LCHS's; however, while the MPL treatment was in that case reported to lead to a significant overestimate of the nematic transition density, our results for $r_{.2}p_2$ helices rather underestimate this quantity with respect to the simulation data of Ref.~\onlinecite{Ferrarini-2}. This would appear to suggest that the influence of particle non-convexity on nematic self-assembly is highly non-trivial, and may be too subtle to be fully captured by the simple hard-sphere analogy of the PL approximation.

\subsection{The Frank elastic constants} \label{subsec:The Frank elastic constants}
We now move on to the applications of our approach to the microscopic calculation of the Frank elastic constants, as outlined in Secs.~\ref{subsec:Introducing weak nematic fluctuations} and~\ref{subsec:Numerical procedure}. The determination of the LC elastic constants from direct molecular simulations is a notoriously challenging task, owing to the large scale of the associated nematic fluctuations in the weak deformation framework. Therefore, their reliable measurement from the spatial fluctuations of the director field requires the use of very large simulation boxes, which leads to similar technical limitations as mentioned for the direct simulation of cholesteric phases. We thus expect our method to provide a general and efficient way to work out the Frank elastic constants of a wide range of particle models, and to assess their sensitivity to microscopic mesogen structure and thermodynamic state.
\par
\begin{figure}[htpb]
  \includegraphics[width=\columnwidth]{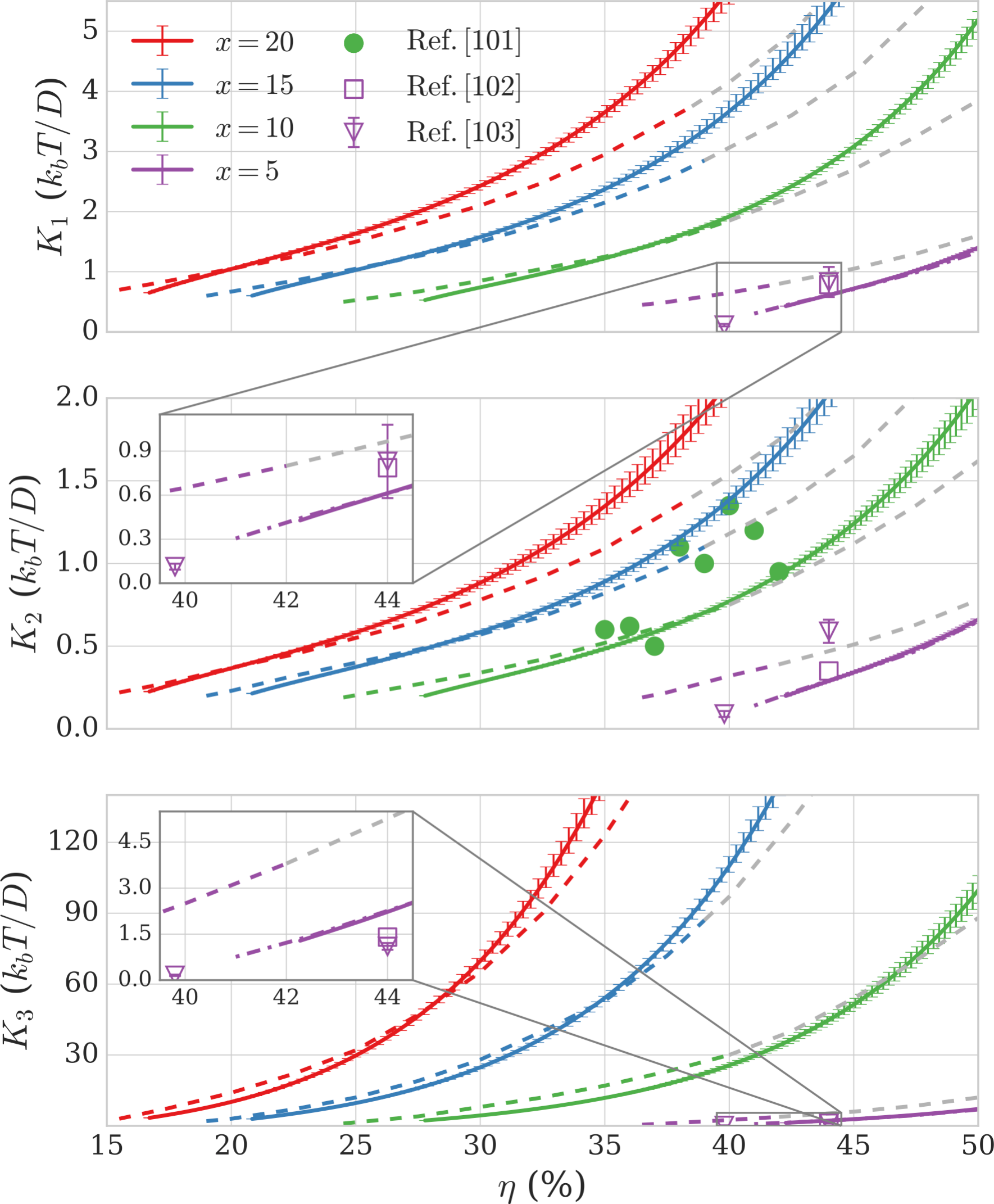}
  \caption{\label{fig3}Frank elastic constants of hard spherocylinders with different aspect ratios $x=L/D$, with $L$ and $D$ the respective length and diameter of the particles. Solid lines correspond to the results of this work and dashed lines to those of the FMT of Ref.~\onlinecite{Marechal-1}. In the latter case, the greyed-out dashed lines denote state points lying outside of the predicted region of nematic stability. The dash-dotted lines represent the DFT results of Ref.~\onlinecite{Somoza-2} for $x=5$. Symbols designate the simulation data of Refs.~\onlinecite{Marechal-2, Allen-1} for $x=5$, along with the simulated values of $K_2$ reported in Ref.~\onlinecite{Wensink-2} for $x=10$. Error bars were obtained from 16 independent runs with $N_{\rm MC}=1\times 10^{10}$.}
\end{figure}

However, a consequence of these computational difficulties is the scarcity of the available numerical data to which our results may be compared, as simulation measurements of the elastic constants have so far been limited to a few state points in the phase diagram of highly simplified particle models.\cite{Wilson} We therefore only report in Fig.~\ref{fig3} our results in the case of hard spherocylinders with aspect ratios $5 \leq x \leq 20$, for which simulation\cite{Wensink-2, Marechal-2,Allen-1} and theoretical results from both FMT\cite{Marechal-1} and DFT\cite{Somoza-2} have been reported. 

We note that our data obeys $K_3 > K_1 > K_2$ for all particles studied, as expected from general mean-field calculations for calamitic particles.\cite{Ilg} Furthermore, our results are found to be in reasonable quantitative agreement with the predictions of FMT, although increasing discrepancies are observed in the locations of the I-N transition with decreasing aspect ratios; both approaches are however in comparable agreement with the simulation values of Frenkel and Bolhuis\cite{Frenkel-2} for all reported aspect ratios. While both methods provide a good match for the limited simulation data available for the Frank elastic constants, our method predicts a significantly stronger dependence of $K_1$ and $K_2$ on particle density than estimated by FMT; this deserves to be investigated further, and more extensive simulation studies involving a larger number of state points would therefore be very useful in enabling the assessment of the validity of the different theoretical descriptions.\cite{Wensink-2}
\par
We also remark that our results are virtually identical to those of Somoza and Tarazona,\cite{Somoza-2} who made use of a more involved DFT approach based on a weighted-density approximation.\cite{Somoza-1} This is due to the fact that the full DCF obtained in Ref.~\onlinecite{Somoza-2} reduces to our effective expression Eq.~\eqref{eq:dcf_pl} in the functional subspace of Eq.~\eqref{eq:density}, and therefore leads to an equivalent description of nematic director fluctuations in the framework of Sec.~\ref{subsec:Introducing weak nematic fluctuations}. We finally note for the sake of completeness that other results using different variants of DFT have been reported for these systems,\cite{Lee-2, Poniewierski-2} but are not reproduced here due to their wide dispersion. We further expect the results of FMT, which corresponds to the most sophisticated analytical effort to date in the microscopic description of hard-particle LCs,\cite{Mederos} to provide the most reliable theoretical benchmark currently available for the validation of our approach in this case. 

\subsection{The cholesteric pitch of hard helices} \label{subsec:The cholesteric pitch of hard helices}
\begin{figure}[htpb]
  \includegraphics[width=\columnwidth]{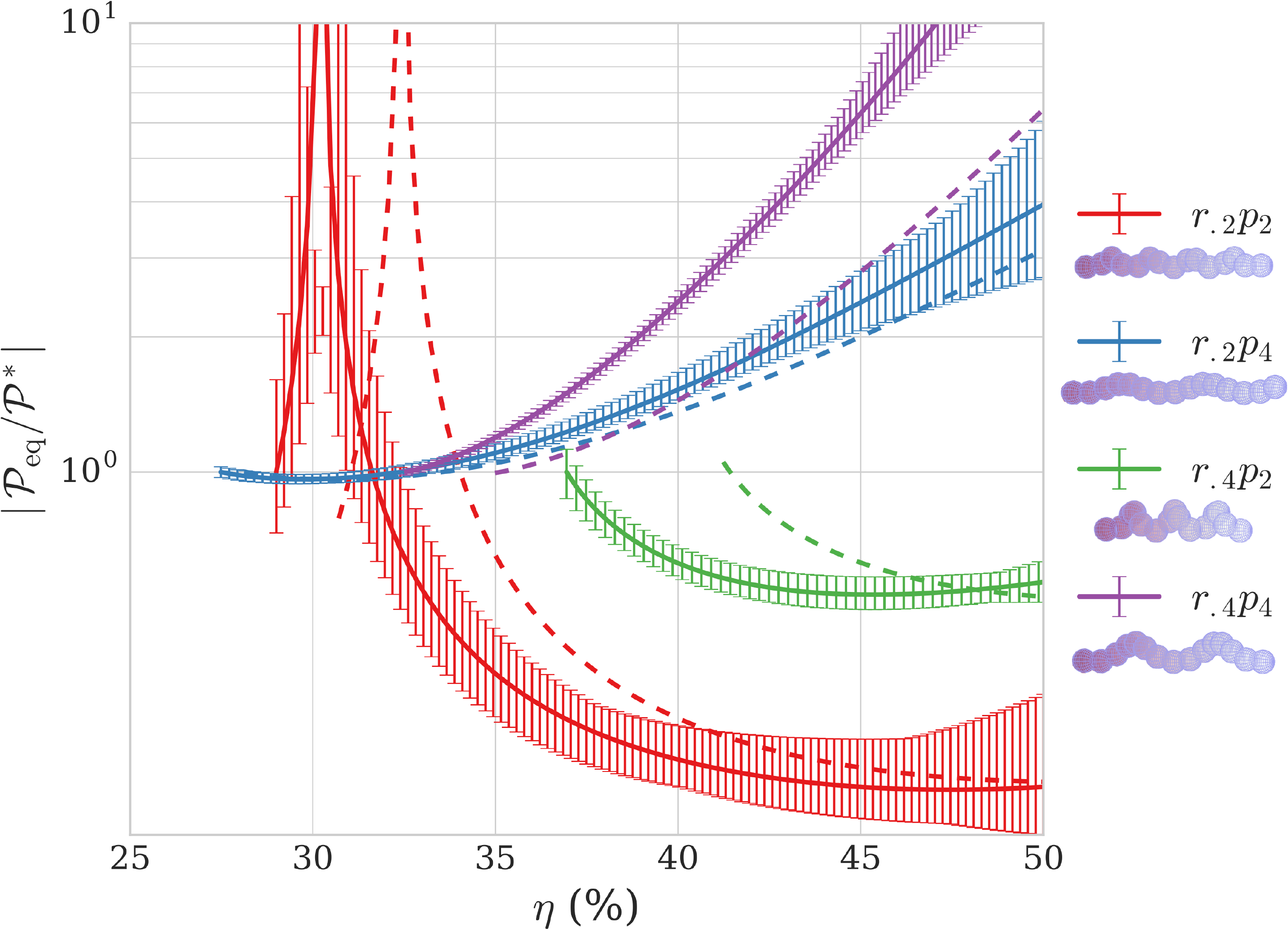}
  \caption{\label{fig4}Influence of the MPL theory on the equilibrium pitch of hard helices with $N_s=15$ and $l=10\,\sigma$. $\mathcal{P}^\ast$ corresponds to the equilibrium pitch at the isotropic-cholesteric transition as predicted by the MPL theory. The solid lines correspond to the results of our MPL approach, and the dashed lines to those of the original PL approximation. Error bars were evaluated from 32 independent runs with $N_{\rm MC} = 5\times 10^{10}$, and are only reported for the MPL results. Note that the diverging pitch for helices with $r=0.2\,\sigma$ and $p=2\,\sigma$ denotes a cholesteric inversion from left to right-handedness with increasing particle concentrations.}
\end{figure}

We finally turn our focus to larger length scales by comparing the cholesteric pitches computed using our method against those reported by previous microscopic theories\cite{Ferrarini-1,Dijkstra-2} in order to assess its reliability in predicting the emergence of cholesteric structures at the macroscopic scale. In keeping with the discussions of Sec.~\ref{subsec:The isotropic-cholesteric phase transition}, we begin by assessing the dependence of the pitch on the inclusion of particle non-convexity in the MPL approximation. We thus compare in Fig.~\ref{fig4} the results obtained using for $v_{\rm ref}$ both an analytical expression\cite{Ferrarini-2} of the real particle volume $v$ and the effective molecular volume $v_{\rm exc}$ computed using the procedure of Sec.~\ref{subsec:Numerical procedure} for different hard helices. 

We note that the MPL theory does not qualitatively alter the cholesteric behaviour of the systems, and mostly shifts the density dependence of the pitch to lower particle concentrations owing to the smaller volume available to non-convex mesogens. We however emphasise that this effect may lead to significant differences in the handedness and magnitude of the predicted pitches at given particle density, contrary to the suggestion of Ref.~\onlinecite{Ferrarini-1}; this is due to the fact that the Parsons-Lee prefactor $G_{\rm PL}$ implicitly affects the equilibrium ODF (Eq.~\eqref{eq:rescaled_self-consistent}) involved in the calculations of $K_2$ and $k_t$, even though its contribution to the effective DCF (Eq.~\eqref{eq:dcf_pl}) explicitly cancels out at all densities in the ratio of these two quantities (Eq.~\eqref{eq:q_eq}). 
\par
We further remark that the results of Sec.~\ref{subsec:The isotropic-cholesteric phase transition} point to potential deficiencies in the description of the local order parameters --- and thus the equilibrium ODFs --- worked out by both formulations of the PL approximation for hard helices, which may in turn hinder the reliability of the macroscopic pitches reported by this and previous density functional investigations of these systems; the magnitude of this effect is difficult to assess systematically in the absence of more extensive simulation studies of their nematic and cholesteric behaviour. In order to consistently compare our results with Refs.~\onlinecite{Dijkstra-1,Dijkstra-2,Ferrarini-1}, and thus critically assess the effects of our perturbative approach and numerical integration techniques, we choose in the following to restrict ourselves to the canonical form of the Parsons-Lee theory.
\par
\begin{figure}[htpb]
  \includegraphics[width=\columnwidth]{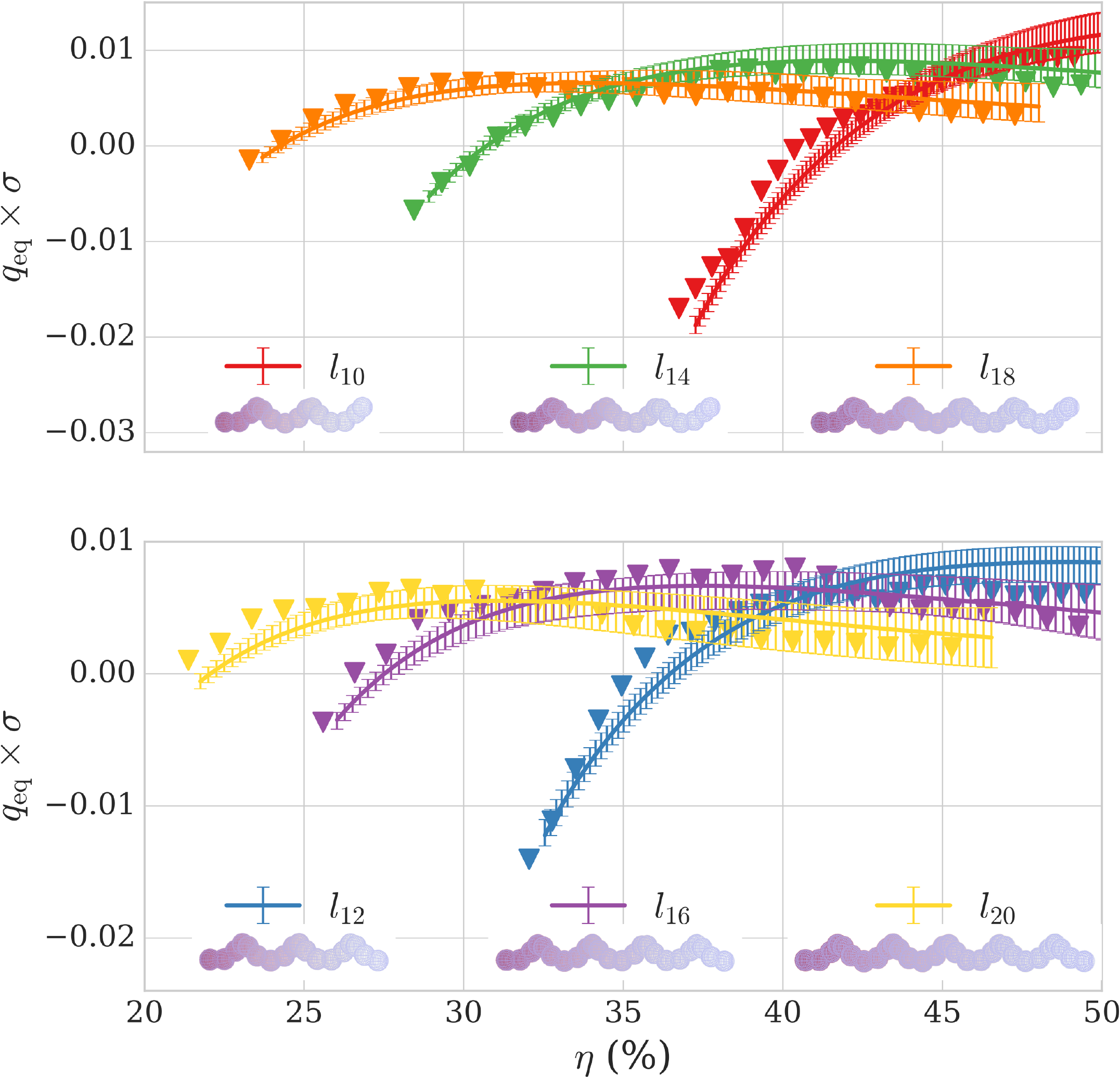}
  \caption{\label{fig5} Equilibrium cholesteric wavenumber $q_{\rm eq} = 2\pi / \mathcal{P}_{\rm eq}$ for hard helices with $r=0.4\,\sigma$, $p=3\,\sigma$ and various contour lengths $l$ and numbers of beads $N_s=3l/(2\sigma)$. Solid lines represent the results of this work, and symbols those obtained using our own implementation of Ref.~\onlinecite{Dijkstra-1}. Error bars were computed from 32 independent runs with $N_{\rm MC} = 5\times 10^{10}$, and are only reported for the results of our method.}
\end{figure}

Firstly, we compare our results against those predicted by the full-functional approach of Ref.~\onlinecite{Dijkstra-1} for weakly chiral systems --- corresponding to large equilibrium cholesteric pitches. Fig.~\ref{fig5} summarises our results in terms of the equilibrium cholesteric wavenumber $q_{\rm eq}$ in the case of hard helices with equal microscopic pitch and radius but different contour lengths, for which large equilibrium pitches $\abs{\mathcal{P}_{\rm eq}} \gtrsim 500\,\sigma$ have been reported.\cite{Dijkstra-2} We obtain excellent agreement between the two methods throughout the whole cholesteric stability range of all particles studied, providing quantitative evidence of the validity of the perturbative treatment of chirality for these systems. We further note that the computational cost of our method is roughly two orders of magnitude lower than that of our implementation of Ref.~\onlinecite{Dijkstra-1}, thanks to its use of the full angle-dependant virial coefficient (Eq.~\eqref{eq:excluded}) instead of its more expensive, pitch-dependent projections on a finite basis of Legendre polynomials.

We compare in Table~\ref{table1} the predictions of our approach to those of Refs.~\onlinecite{Dijkstra-2} and~\onlinecite{Ferrarini-1} for a variety of helices at the onset of cholesteric stability. The macroscopic pitches computed by our method are found to be within 15\% of those reported by Ref.~\onlinecite{Dijkstra-2} for all particles studied, with the larger relative discrepancies generally corresponding to the shorter pitches, i.e.~the more strongly twisted phases. This observation is fully consistent with the limitations of our weak deformation framework, and is further discussed in the next paragraphs. 
\par
The correspondence between our results and those of Ref.~\onlinecite{Ferrarini-1} is somewhat more tenuous, especially for larger values of $\mathcal{P}_{\rm eq}$ --- a fact all the more puzzling given that both our approaches are based on the same perturbative description of the cholesteric phase. These differences may most likely be attributed to finite-order truncation effects in the Wigner matrix expansions introduced in Refs.~\onlinecite{Tombolato, Ferrarini-1} for the computation of $K_2$ and $k_t$, as small discrepancies in these two coefficients may lead to significant variations of the corresponding equilibrium pitch Eq.~\eqref{eq:q_eq} in the weak chirality limit. We therefore expect our method, which allows for the calculation of all relevant quantities with a tuneable degree of precision and obviates further numerical approximations, to be more reliable in these cases.

\begin{table}
\caption{\label{table1}Comparison of this work, Ref.~\onlinecite{Dijkstra-2} and Ref.~\onlinecite{Ferrarini-1} for hard helices with $N_s=15$ and $l=10\,\sigma$ at the isotropic-cholesteric transition. $\abs{\Delta \mathcal{P}} / \mathcal{P}_{\rm eq}$ represents the relative difference between the equilibrium pitch predicted by our method and those reported in the references above, normalised by our value. Our results were obtained by averaging over 8 independent runs with $N_{\rm MC}=5\times 10^{11}$; numerical dispersion was found to be below 5\% for all computed pitches, and smaller than the last significant digit for all other variables.}
\begin{center}
\begin{tabular}{|c|c|c|c|c|c|c|c|}
\cline{3-8}
\multicolumn{2}{c|}{} & $r_{.2}p_2$ & $r_{.2}p_4$ & $r_{.2}p_8$ & $r_{.4}p_2$ & $r_{.4}p_4$ & $r_{.4}p_8$ \\
\hline
\multirow{2}{*}{\begin{tabular}{c}$\eta$ \\ ($\%$) \end{tabular}} & \textbf{here} & 30.7 & 28.1 & 26.5 & 41.2 & 35.0 & 29.2 \\
\cline{2-8}
& Ref.~\onlinecite{Dijkstra-2} & 30.0 & 27.4 & 25.8 & 40.3 & 34.0 & 28.2 \\
\hline
\hline
\multirow{3}{*}{$S$} & \textbf{here} & 0.704 & 0.684 & 0.696 & 0.629 & 0.632 & 0.621 \\
\cline{2-8}
& Ref.~\onlinecite{Dijkstra-2} & 0.699 & 0.677 & 0.690 & 0.612 & 0.622 & 0.619 \\
\cline{2-8}
& Ref.~\onlinecite{Ferrarini-1} & 0.64 & 0.66 & 0.68 & 0.60 & 0.61 & 0.61 \\
\hline
\hline
\multirow{3}{*}{\begin{tabular}{c} $-10^4 \times k_t$ \\ ($k_b T/\sigma^2$) \end{tabular}} & \textbf{here} & 4.12 & 52.1 & 46.0 & -11.6 & 130 & 139 \\
\cline{2-8}
& Ref.~\onlinecite{Dijkstra-2} & 4.27 & 47.3 & 42.3 & -10.7 & 110 & 115 \\
\cline{2-8}
& Ref.~\onlinecite{Ferrarini-1} & 4.83 & 41.36 & 29.03 & -3.83 & 98.35 & 110.13 \\
\hline
\hline
\multirow{3}{*}{\begin{tabular}{c} $K_2$ \\ ($k_b T/\sigma$) \end{tabular}} & \textbf{here} & 0.201 & 0.197 & 0.207 & 0.170 & 0.167 & 0.181 \\
\cline{2-8}
& Ref.~\onlinecite{Dijkstra-2} & 0.199 & 0.194 & 0.203 & 0.160 & 0.150 & 0.136 \\
\cline{2-8}
& Ref.~\onlinecite{Ferrarini-1} & 0.154 & 0.177 & 0.184 & 0.153 & 0.152 & 0.159 \\
\hline
\hline
\multirow{3}{*}{\begin{tabular}{c} $\mathcal{P}_{\rm eq}$ \\ ($\sigma$) \end{tabular}} & \textbf{here} & -3065 & -238 & -283 & 921 & -81 & -82 \\
\cline{2-8}
& Ref.~\onlinecite{Dijkstra-2} & -2990 & -260 & -310 & 965 & -93 & -90 \\
\cline{2-8}
& Ref.~\onlinecite{Ferrarini-1} & -2008 & -268 & -399 & 2509 & -97 & -90 \\
\hline
\hline
\multirow{2}{*}{\begin{tabular}{c}$\abs{\Delta \mathcal{P}} / \mathcal{P}_{\rm eq}$ \\ ($\%$) \end{tabular}} & Ref.~\onlinecite{Dijkstra-2} & 2.45 & 9.24 & 9.54 & 4.78 & 14.8 & 9.76 \\
\cline{2-8}
& Ref.~\onlinecite{Ferrarini-1} & 34.5 & 12.6 & 41.0 & 172 & 19.8 & 9.76 \\
\hline
\end{tabular}
\end{center}
\end{table}

\begin{figure}[htpb]
  \includegraphics[width=\columnwidth]{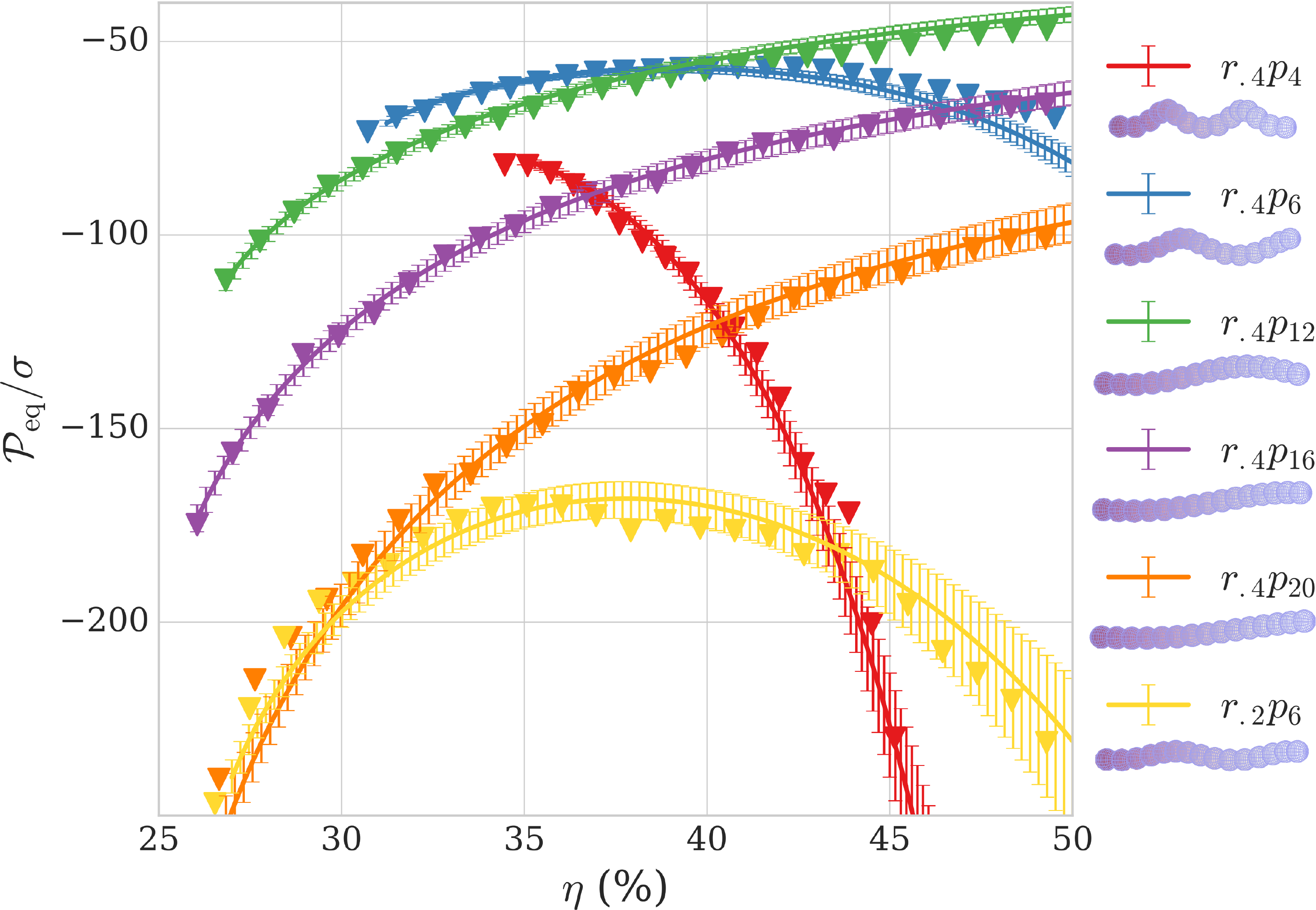}
  \caption{\label{fig6}Equilibrium pitch in the limit of strong chirality for diverse hard helices with $N_s=15$ and $l=10\,\sigma$. Lines, symbols and error bars are defined as in Fig.~\ref{fig5}.}
\end{figure}

We finally investigate the validity of our perturbative treatment in the limit of strong chirality in Fig.~\ref{fig6} by focusing on systems for which the predicted cholesteric pitches are of the order of a few dozen particle diameters, corresponding to the shortest values reported in Ref.~\onlinecite{Dijkstra-2}. Quantitative agreement with the full-functional description of Ref.~\onlinecite{Dijkstra-1} is still found to be very satisfactory, which indicates that the effects of the low-order $q$-expansion of the free energy (Eq.~\eqref{eq:cholesteric_free_energy}) and of the change in local order induced by the cholesteric twist are both of limited consequence for the systems studied. 
\par
We nevertheless note that the original framework of Ref.~\onlinecite{Dijkstra-1} also relies on the assumption of local phase uniaxiality (Eq.~\eqref{eq:density}), and therefore still qualifies as a weak-deformation account of cholesteric order. A theoretical route to incorporate the deviations from local cylindrical symmetry induced by macroscopic twist in the approach of Ref.~\onlinecite{Dijkstra-1} has subsequently been outlined by Allen,\cite{Allen-2} but has to our knowledge not yet been implemented in numerical studies. These effects are however only expected to be substantial in the limit of short pitches,\cite{Lubensky-1} and may therefore be of limited practical relevance for the study of most experimentally-realistic systems. 
\par
Conversely, it is well known that chiral mesogens are intrinsically biaxial, and that particle interactions at the microscale may thus give rise to an additional level of biaxial organisation in cholesterics beyond this simple phase-induced symmetry breaking.\cite{Lubensky-2, Lubensky-3} In this bottom-up picture, the cholesteric pitch of the system is also expected to be strongly contingent on its degree of local biaxial order, leading to a generally intricate causal relationship between phase chirality and biaxiality.\cite{Dhakal} Consequently, the use of a uniaxial ansatz for the local description of a cholesteric LC may result in a significant underestimation of its equilibrium pitch. This applies to all studies discussed in this section, and may therefore hamper the reliability of the methods described therein even in the limit of weak deformations. Extending the current density functional framework to explicitly include long-range biaxial correlations would allow for a direct and systematic investigation of the link between particle and phase biaxiality, and of its macroscopic effects on cholesteric behaviour.

\subsection{Soft threaded rods: towards more realistic particle models} \label{subsec:Soft threaded rods: towards more realistic particle models}
We finally remark that our method does not rely on any assumptions as to the nature of the microscopic interaction potential $U$ in Eq.~\eqref{eq:Mayer}, and is therefore applicable to any short-range particle potential in the pairwise additive approximation. While the previous sections focused on the case of pure steric repulsion, more experimentally-realistic particle models are likely to involve additional contributions from electrostatic and/or dispersion interactions.\cite{Kleman} Following Ref.~\onlinecite{Wensink-3}, we now consider a system of hard spherocylinders of length $L$ and diameter $D$ decorated with a helical charge distribution of microscopic pitch $p$ located at their surface. To mimic a uniform linear distribution, we discretise the charges in a number $N_p$ of identical interaction sites uniformly spread along the helical thread. The full pair interaction energy of two particles $P_1$ and $P_2$ then reads as
\begin{align*}
  U(\mathbf{r}_{12}, \mathcal{R}_1, \mathcal{R}_2) =
  \begin{dcases}
    +\infty & \text{if } P_1 \cap P_2 \neq \varnothing \\
    \left(\frac{L}{N_p} \right)^2 \sum_{i,j=1}^{N_p} u(r_{ij}) & \text{if } P_1 \cap P_2 = \varnothing
  \end{dcases},
\end{align*}
where the set intersection sign denotes the potential overlap region of the hard spherocylinders, and $u$ represents the electrostatic interaction potential per unit particle length.\cite{Wensink-3} $r_{ij}$ then corresponds to the distance between interaction sites $i \in P_1$ and $j \in P_2$, which implicitly depends on the center-of-mass separation vector $\mathbf{r}_{12}$ and respective orientations $\mathcal{R}_1, \mathcal{R}_2$ of the two particles. Following Ref.~\onlinecite{Wensink-3}, we choose to account for electrostatic repulsion in the Debye-H\"uckel formalism, introducing a finite cutoff distance $r_c$ for computational efficiency,
\begin{align*}
  \beta u(r) =
  \begin{dcases}
    u_0 \times \frac{\exp (-r/\lambda_D)}{r} & \text{if } r \leq r_c \\
    0  & \text{if } r > r_c
  \end{dcases}
\end{align*}
with $\lambda_D$ the Debye length of the system. In the following, we set the cutoff length to $r_c=10\, \lambda_D$ and number of patches to $N_p=50$, which we verified to yield negligible truncation and discretisation effects for our results, and use $u_0=1/D$ for consistency with the values of Ref.~\onlinecite{Wensink-3}.
\par
\begin{figure}[htpb]
  \includegraphics[width=\columnwidth]{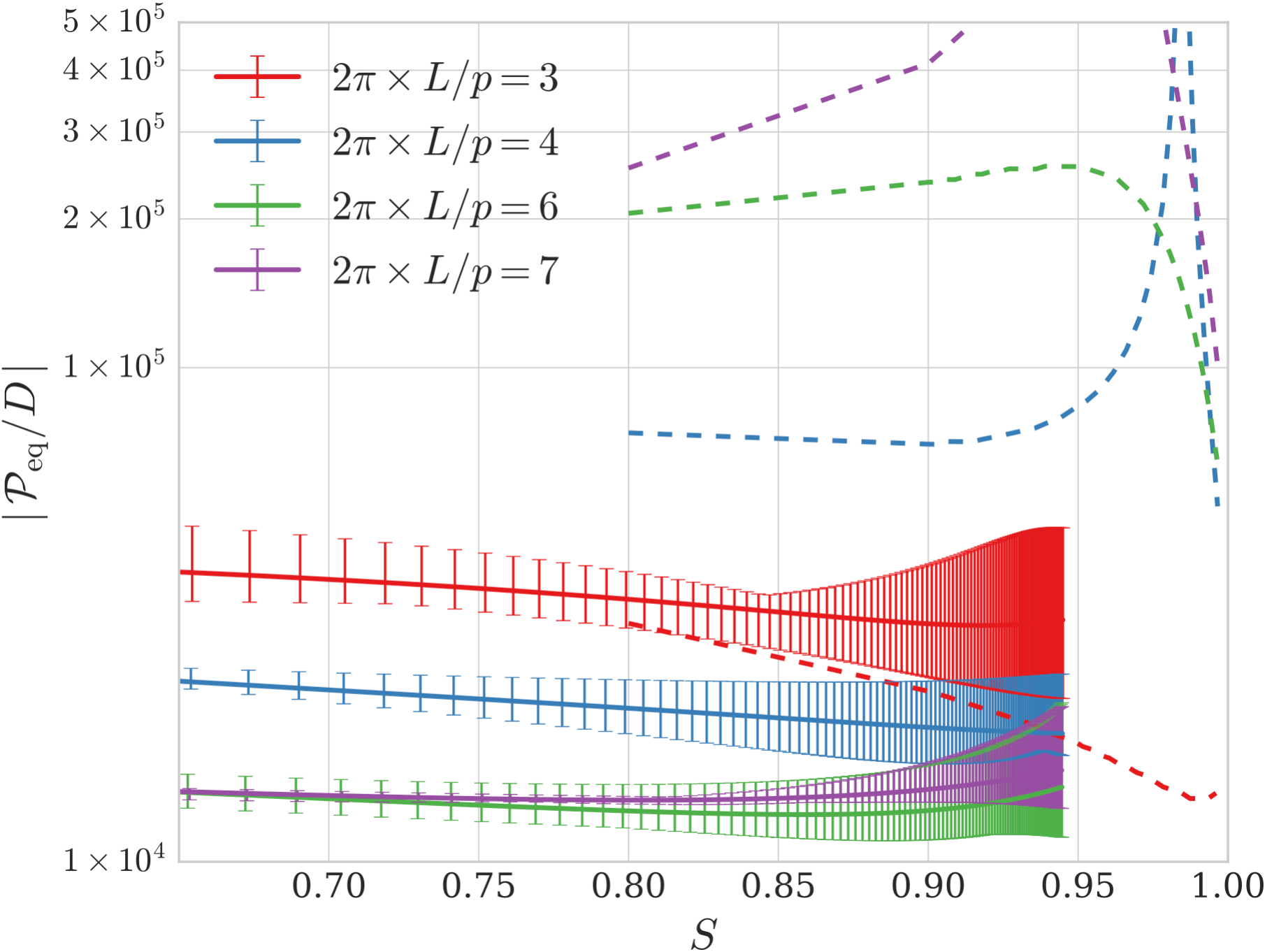}
  \caption{\label{fig7}Absolute equilibrium pitch versus nematic order parameter for spherocylinders with aspect ratio $L/D=50$, Debye length $\lambda_D=L/20$ and helical charge distributions of varying pitches $p$. Solid lines denote the results of this work, and dashed lines the data reported in Ref.~\onlinecite{Wensink-3}. All pitches predicted by our method are left-handed, while those of Ref.~\onlinecite{Wensink-3} are right-handed at lower densities with a potential handedness inversion at higher particle concentrations. Error bars obtained from 8 independent runs with $N_{\rm MC}=2.5\times 10^{11}$.}
\end{figure}

Our numerical approach then enables us to directly investigate the validity of the framework of Ref.~\onlinecite{Wensink-3}, in which a similar perturbative description was used, but where a number of approximations were introduced to work out the full pair interaction potential $U$ and Frank parameters $K_2$ and $k_t$ semi-analytically for this particle model.\cite{Wensink-4} For consistency with the approach used therein, we now revert to the original form of the Onsager theory and thus set $G_{\rm PL}=1$ for the rest of this section. The comparison of the equilibrium pitches predicted by our method to those reported in Ref.~\onlinecite{Wensink-3} is then summarised in Fig.~\ref{fig7} for helical charge distributions with a variety of microscopic pitches. 

We observe no evidence of qualitative agreement in terms of either magnitude or handedness of the predicted pitches for any of the particles studied, despite the common use of the Onsager-Straley theoretical basis in both our approaches. Furthermore, we find that the inclusion of electrostatic repulsion moves the isotropic-cholesteric transition to lower volume fractions than in the case of purely steric interactions, with the onset of cholesteric order for all systems considered occurring at $\eta \simeq 2\%$ compared to the simulation value of $\eta \simeq 7\%$ obtained for hard spherocylinders with $L/D=50$.\cite{Frenkel-2} This effect is consistent with the findings of Ref.~\onlinecite{Lekkerkerker-2} for uniformly-charged rods in the Onsager framework, but contrasts with the higher value $\eta \simeq 10\%$ reported in Ref.~\onlinecite{Wensink-3}. The origin of these discrepancies most likely lies in some of the simplifying assumptions introduced in Refs.~\onlinecite{Wensink-3,Wensink-4} to obtain tractable expressions for all microscopic integrals of the form of Eqs.~\eqref{eq:excluded} and \eqref{eq:K2}. The magnitude of the resulting disparities illustrates the difficulty of deriving reliable analytical approximations in microscopic descriptions of the cholesteric phase, due to the strong dependence of such systems on the fine structural and thermodynamic properties of their constituent particles.
\par
We also note that a number of simulation results on closely related systems have recently been reported.\cite{Wensink-2, Schilling-1, Schilling-2} However, the first of these studies\cite{Wensink-2} focused on fairly short spherocylinders in a regime of large Debye lengths, leading to effective particle aspect ratios too low to be accurately described by Onsager-like theories,\cite{Lekkerkerker-2} while the other two\cite{Schilling-1, Schilling-2} introduced an explicit dependence of the pair potential $U$ on particle density, which as discussed in Sec.~\ref{subsec:Numerical procedure} requires significantly higher computational costs to be probed using our method. We nevertheless remark that Refs.~\onlinecite{Schilling-1, Schilling-2} also found that the onset of cholesteric organisation was shifted to significantly lower volume fractions by the effects of electrostatic repulsion. 


\section{Conclusion} \label{sec:Conclusion}
We have described a novel density functional method to systematically investigate the link between microscopic structure and macroscopic liquid-crystalline properties in chiral and achiral uniaxial nematic phases. The use of numerical integration to work-out generalised virial coefficients greatly improves its applicability and reliability over that of previous analytical approaches, while the implementation of Straley's perturbative framework reduces its computational cost to a fraction of that of Ref.~\onlinecite{Dijkstra-1} for cholesteric systems. It furthermore circumvents the need for the partial decomposition of the density field in a basis of rotational invariants, which greatly increases its tractability over that of previous numerical descriptions.\cite{Tombolato, Ferrarini-1, Dijkstra-1, Dijkstra-2} 
\par
The Frank elastic constants computed by our model for hard spherocylinders are found to be consistent with previous numerical and theoretical investigations, while its predictions for the cholesteric pitch of hard helices are in excellent quantitative agreement with those obtained by Ref.~\onlinecite{Dijkstra-2} for all systems studied. The latter confirms that the additional ``weak chirality'' assumptions introduced by the perturbative approach are not particularly limiting, and still work well down to surprisingly short pitch lengths. We further reiterate that experimental lyotropic cholesterics typically have pitches that are very large compared to the particle diameters, and are therefore expected to be fully amenable to such a treatment. 
\par
However, we wish to highlight that there are a number of key approximations in most implementations of the density functional theory of cholesterics that we feel could potentially limit its quantitative accuracy, and need further consideration. Firstly, there are potential deficiencies in the Parsons-Lee rescaling in the case of strongly non-convex mesogens for the prediction of the onset of nematic order, which are not obviously improved upon by modifying the theory to account for the real effective molecular volume of the particles. This likely points to the need for a more sophisticated description of inter-particle correlations in these cases, following the considerations of Sec.~\ref{subsec:Introducing finite mesogen anisotropy}. We note that an approach involving the explicit third-order virial contribution to the excess free energy was also implemented with some success for such systems, albeit under stringent symmetry assumptions.\cite{Ferrarini-6, Ferrarini-3} It would be interesting to investigate whether its generalisation to configurations of freely-rotating particles may lead to significant improvements over the MPL approximation in our case.
\par
Secondly, there is the assumption of local uniaxiality, and we wish to further emphasise the potential importance of long-range biaxial correlations in systems of chiral mesogens. The explicit angle dependence of our excluded volume integrals provides a natural framework for the introduction of phase biaxiality, which may shed some light on the systematic underestimations of the cholesteric twist obtained by density functional methods in previous studies.\cite{Varga-1, Dijkstra-3} The explicit account of biaxial order will also represent a first step towards the description of more complex chiral nematic phases beyond the cholesteric symmetry, such as the so-called twist-bend\cite{Dozov} and screw-like\cite{Ferrarini-5} nematic phases. However, recent studies suggest that the accurate treatment of the former may require a microscopic theory of nematic elasticities beyond the perturbative Poniewierski-Stecki framework,\cite{Ferrarini-4} while the coupling between orientational and translational degrees of freedom of the latter would call for further modifications of our symmetry assumptions.\cite{Ferrarini-5} Thus, in our opinion, significant efforts are still required to achieve a unified field-theoretical description of the chiral nematic state.
\par
Hence, we stress that numerical simulations of simple cholesteric model systems along the lines of Refs.~\onlinecite{Dijkstra-3, Wensink-2, Schilling-2} are of particular importance to allow the quantitative validity of these different approximations to be probed. However, the determination of cholesteric pitches in these previous studies relies on rather involved numerical procedures, and their degree of mutual consistency remains largely unascertained due to their focus on different particle models and parameter ranges. We therefore hope that this paper will stimulate further investigations of the respective reliabilities of these different methods, and generate a reference set of data suitable for the thorough assessment of density functional descriptions of the cholesteric phase.
\par
We finally point out that the use of Monte-Carlo techniques for the computation of all microscopic integrals further allows for the simple treatment of non-rigid particle models, e.g.~by drawing mesogen configurations from an ensemble of representative structures at every integration step as part of the random sampling process. This will enable us to directly investigate the influence of particle flexibility on phase chirality, an issue so far largely unexplored despite its essential experimental relevance.\cite{Cherstvy, Dogic-1}

\begin{acknowledgments}
This project has received funding from the European Union's Horizon 2020 research and innovation programme under the Marie Sk\l{}odowska-Curie Grant Agreement No.~641839. The authors would like to acknowledge the use of the University of Oxford Advanced Research Computing (ARC) facility in carrying out this work. http://dx.doi.org/10.5281/zenodo.22558. 
\end{acknowledgments}


\bibliography{jcp}

\begin{thebibliography}{125}%
\makeatletter
\providecommand \@ifxundefined [1]{%
 \@ifx{#1\undefined}
}%
\providecommand \@ifnum [1]{%
 \ifnum #1\expandafter \@firstoftwo
 \else \expandafter \@secondoftwo
 \fi
}%
\providecommand \@ifx [1]{%
 \ifx #1\expandafter \@firstoftwo
 \else \expandafter \@secondoftwo
 \fi
}%
\providecommand \natexlab [1]{#1}%
\providecommand \enquote  [1]{``#1''}%
\providecommand \bibnamefont  [1]{#1}%
\providecommand \bibfnamefont [1]{#1}%
\providecommand \citenamefont [1]{#1}%
\providecommand \href@noop [0]{\@secondoftwo}%
\providecommand \href [0]{\begingroup \@sanitize@url \@href}%
\providecommand \@href[1]{\@@startlink{#1}\@@href}%
\providecommand \@@href[1]{\endgroup#1\@@endlink}%
\providecommand \@sanitize@url [0]{\catcode `\\12\catcode `\$12\catcode
  `\&12\catcode `\#12\catcode `\^12\catcode `\_12\catcode `\%12\relax}%
\providecommand \@@startlink[1]{}%
\providecommand \@@endlink[0]{}%
\providecommand \url  [0]{\begingroup\@sanitize@url \@url }%
\providecommand \@url [1]{\endgroup\@href {#1}{\urlprefix }}%
\providecommand \urlprefix  [0]{URL }%
\providecommand \Eprint [0]{\href }%
\providecommand \doibase [0]{http://dx.doi.org/}%
\providecommand \selectlanguage [0]{\@gobble}%
\providecommand \bibinfo  [0]{\@secondoftwo}%
\providecommand \bibfield  [0]{\@secondoftwo}%
\providecommand \translation [1]{[#1]}%
\providecommand \BibitemOpen [0]{}%
\providecommand \bibitemStop [0]{}%
\providecommand \bibitemNoStop [0]{.\EOS\space}%
\providecommand \EOS [0]{\spacefactor3000\relax}%
\providecommand \BibitemShut  [1]{\csname bibitem#1\endcsname}%
\let\auto@bib@innerbib\@empty
\bibitem [{\citenamefont {de~Gennes}\ and\ \citenamefont
  {Prost}(1993)}]{deGennes}%
  \BibitemOpen
  \bibfield  {author} {\bibinfo {author} {\bibfnamefont {P.-G.}\ \bibnamefont
  {de~Gennes}}\ and\ \bibinfo {author} {\bibfnamefont {J.}~\bibnamefont
  {Prost}},\ }\href@noop {} {\emph {\bibinfo {title} {The Physics of Liquid
  Crystals}}}\ (\bibinfo  {publisher} {Clarendon Press},\ \bibinfo {address}
  {Oxford},\ \bibinfo {year} {1993})\BibitemShut {NoStop}%
\bibitem [{\citenamefont {Goodby}\ and\ \citenamefont {Gray}(2008)}]{Goodby-1}%
  \BibitemOpen
  \bibfield  {author} {\bibinfo {author} {\bibfnamefont {J.~W.}\ \bibnamefont
  {Goodby}}\ and\ \bibinfo {author} {\bibfnamefont {G.~W.}\ \bibnamefont
  {Gray}},\ }\enquote {\bibinfo {title} {Guide to the nomenclature and
  classification of liquid crystals},}\ in\ \href {\doibase
  10.1002/9783527620760.ch2} {\emph {\bibinfo {booktitle} {Handbook of Liquid
  Crystals}}},\ \bibinfo {editor} {edited by\ \bibinfo {editor} {\bibfnamefont
  {D.}~\bibnamefont {Demus}}, \bibinfo {editor} {\bibfnamefont
  {J.}~\bibnamefont {Goodby}}, \bibinfo {editor} {\bibfnamefont {G.~W.}\
  \bibnamefont {Gray}}, \bibinfo {editor} {\bibfnamefont {H.-W.}\ \bibnamefont
  {Spiess}}, \ and\ \bibinfo {editor} {\bibfnamefont {V.}~\bibnamefont
  {Vill}}}\ (\bibinfo  {publisher} {Wiley-VCH Verlag GmbH},\ \bibinfo {address}
  {Weinheim},\ \bibinfo {year} {2008})\ pp.\ \bibinfo {pages}
  {17--23}\BibitemShut {NoStop}%
\bibitem [{\citenamefont {Wilson}(2005)}]{Wilson}%
  \BibitemOpen
  \bibfield  {author} {\bibinfo {author} {\bibfnamefont {M.~R.}\ \bibnamefont
  {Wilson}},\ }\href@noop {} {\bibfield  {journal} {\bibinfo  {journal} {Int.
  Rev. Phys. Chem.}\ }\textbf {\bibinfo {volume} {24}},\ \bibinfo {pages} {421}
  (\bibinfo {year} {2005})}\BibitemShut {NoStop}%
\bibitem [{\citenamefont {Wu}(1986)}]{Wu}%
  \BibitemOpen
  \bibfield  {author} {\bibinfo {author} {\bibfnamefont {S.-T.}\ \bibnamefont
  {Wu}},\ }\href@noop {} {\bibfield  {journal} {\bibinfo  {journal} {Phys. Rev.
  A}\ }\textbf {\bibinfo {volume} {33}},\ \bibinfo {pages} {1270} (\bibinfo
  {year} {1986})}\BibitemShut {NoStop}%
\bibitem [{\citenamefont {Fr{\'e}edericksz}\ and\ \citenamefont
  {Zolina}(1933)}]{Freedericksz}%
  \BibitemOpen
  \bibfield  {author} {\bibinfo {author} {\bibfnamefont {V.}~\bibnamefont
  {Fr{\'e}edericksz}}\ and\ \bibinfo {author} {\bibfnamefont {V.}~\bibnamefont
  {Zolina}},\ }\href@noop {} {\bibfield  {journal} {\bibinfo  {journal} {T.
  Faraday Soc.}\ }\textbf {\bibinfo {volume} {29}},\ \bibinfo {pages} {919}
  (\bibinfo {year} {1933})}\BibitemShut {NoStop}%
\bibitem [{\citenamefont {Doi}(1981)}]{Doi}%
  \BibitemOpen
  \bibfield  {author} {\bibinfo {author} {\bibfnamefont {M.}~\bibnamefont
  {Doi}},\ }\href@noop {} {\bibfield  {journal} {\bibinfo  {journal} {J. Polym.
  Sci. Polym. Phys. Ed.}\ }\textbf {\bibinfo {volume} {19}},\ \bibinfo {pages}
  {243} (\bibinfo {year} {1981})}\BibitemShut {NoStop}%
\bibitem [{\citenamefont {Reinitzer}(1888)}]{Reinitzer}%
  \BibitemOpen
  \bibfield  {author} {\bibinfo {author} {\bibfnamefont {F.}~\bibnamefont
  {Reinitzer}},\ }\href@noop {} {\bibfield  {journal} {\bibinfo  {journal}
  {Monatsh. Chem.}\ }\textbf {\bibinfo {volume} {9}},\ \bibinfo {pages} {421}
  (\bibinfo {year} {1888})}\BibitemShut {NoStop}%
\bibitem [{\citenamefont {Dunmur}\ and\ \citenamefont
  {Sluckin}(2011)}]{Dunmur}%
  \BibitemOpen
  \bibfield  {author} {\bibinfo {author} {\bibfnamefont {D.~A.}\ \bibnamefont
  {Dunmur}}\ and\ \bibinfo {author} {\bibfnamefont {T.~J.}\ \bibnamefont
  {Sluckin}},\ }\href@noop {} {\emph {\bibinfo {title} {Soap, Science, and
  Flat-screen TVs: A History of Liquid Crystals}}}\ (\bibinfo  {publisher}
  {Oxford University Press},\ \bibinfo {address} {New York, NY},\ \bibinfo
  {year} {2011})\BibitemShut {NoStop}%
\bibitem [{\citenamefont {Moreira}\ \emph {et~al.}(2004)\citenamefont
  {Moreira}, \citenamefont {Carvalho}, \citenamefont {Cao}, \citenamefont
  {Bailey}, \citenamefont {Taheri},\ and\ \citenamefont
  {Palffy-Muhoray}}]{Moreira}%
  \BibitemOpen
  \bibfield  {author} {\bibinfo {author} {\bibfnamefont {M.~F.}\ \bibnamefont
  {Moreira}}, \bibinfo {author} {\bibfnamefont {I.~C.~S.}\ \bibnamefont
  {Carvalho}}, \bibinfo {author} {\bibfnamefont {W.}~\bibnamefont {Cao}},
  \bibinfo {author} {\bibfnamefont {C.}~\bibnamefont {Bailey}}, \bibinfo
  {author} {\bibfnamefont {B.}~\bibnamefont {Taheri}}, \ and\ \bibinfo {author}
  {\bibfnamefont {P.}~\bibnamefont {Palffy-Muhoray}},\ }\href@noop {}
  {\bibfield  {journal} {\bibinfo  {journal} {Appl. Phys. Lett.}\ }\textbf
  {\bibinfo {volume} {85}},\ \bibinfo {pages} {2691} (\bibinfo {year}
  {2004})}\BibitemShut {NoStop}%
\bibitem [{\citenamefont {Mujahid}\ \emph {et~al.}(2010)\citenamefont
  {Mujahid}, \citenamefont {Stathopulos}, \citenamefont {Lieberzeit},\ and\
  \citenamefont {Dickert}}]{Mujahid}%
  \BibitemOpen
  \bibfield  {author} {\bibinfo {author} {\bibfnamefont {A.}~\bibnamefont
  {Mujahid}}, \bibinfo {author} {\bibfnamefont {H.}~\bibnamefont
  {Stathopulos}}, \bibinfo {author} {\bibfnamefont {P.~A.}\ \bibnamefont
  {Lieberzeit}}, \ and\ \bibinfo {author} {\bibfnamefont {F.~L.}\ \bibnamefont
  {Dickert}},\ }\href@noop {} {\bibfield  {journal} {\bibinfo  {journal}
  {Sensors}\ }\textbf {\bibinfo {volume} {10}},\ \bibinfo {pages} {4887}
  (\bibinfo {year} {2010})}\BibitemShut {NoStop}%
\bibitem [{\citenamefont {Liu}\ and\ \citenamefont {Broer}(2015)}]{Liu}%
  \BibitemOpen
  \bibfield  {author} {\bibinfo {author} {\bibfnamefont {D.}~\bibnamefont
  {Liu}}\ and\ \bibinfo {author} {\bibfnamefont {D.~J.}\ \bibnamefont
  {Broer}},\ }\href@noop {} {\bibfield  {journal} {\bibinfo  {journal} {Nat.
  Commun.}\ }\textbf {\bibinfo {volume} {6}},\ \bibinfo {pages} {8334}
  (\bibinfo {year} {2015})}\BibitemShut {NoStop}%
\bibitem [{\citenamefont {Finkelmann}\ \emph {et~al.}(2001)\citenamefont
  {Finkelmann}, \citenamefont {Kim}, \citenamefont {Munoz}, \citenamefont
  {Palffy-Muhoray}, \citenamefont {Taheri} \emph {et~al.}}]{Finkelmann}%
  \BibitemOpen
  \bibfield  {author} {\bibinfo {author} {\bibfnamefont {H.}~\bibnamefont
  {Finkelmann}}, \bibinfo {author} {\bibfnamefont {S.~T.}\ \bibnamefont {Kim}},
  \bibinfo {author} {\bibfnamefont {A.}~\bibnamefont {Munoz}}, \bibinfo
  {author} {\bibfnamefont {P.}~\bibnamefont {Palffy-Muhoray}}, \bibinfo
  {author} {\bibfnamefont {B.}~\bibnamefont {Taheri}},  \emph {et~al.},\
  }\href@noop {} {\bibfield  {journal} {\bibinfo  {journal} {Adv. Mater.}\
  }\textbf {\bibinfo {volume} {13}},\ \bibinfo {pages} {1069} (\bibinfo {year}
  {2001})}\BibitemShut {NoStop}%
\bibitem [{\citenamefont {Enz}\ and\ \citenamefont {Lagerwall}(2010)}]{Enz}%
  \BibitemOpen
  \bibfield  {author} {\bibinfo {author} {\bibfnamefont {E.}~\bibnamefont
  {Enz}}\ and\ \bibinfo {author} {\bibfnamefont {J.~P.~F.}\ \bibnamefont
  {Lagerwall}},\ }\href@noop {} {\bibfield  {journal} {\bibinfo  {journal} {J.
  Mater. Chem.}\ }\textbf {\bibinfo {volume} {20}},\ \bibinfo {pages} {6866}
  (\bibinfo {year} {2010})}\BibitemShut {NoStop}%
\bibitem [{\citenamefont {Robinson}(1961)}]{Robinson}%
  \BibitemOpen
  \bibfield  {author} {\bibinfo {author} {\bibfnamefont {C.}~\bibnamefont
  {Robinson}},\ }\href@noop {} {\bibfield  {journal} {\bibinfo  {journal}
  {Tetrahedron}\ }\textbf {\bibinfo {volume} {13}},\ \bibinfo {pages} {219}
  (\bibinfo {year} {1961})}\BibitemShut {NoStop}%
\bibitem [{\citenamefont {Livolant}\ and\ \citenamefont
  {Leforestier}(1996)}]{Livolant}%
  \BibitemOpen
  \bibfield  {author} {\bibinfo {author} {\bibfnamefont {F.}~\bibnamefont
  {Livolant}}\ and\ \bibinfo {author} {\bibfnamefont {A.}~\bibnamefont
  {Leforestier}},\ }\href@noop {} {\bibfield  {journal} {\bibinfo  {journal}
  {Prog. Polym. Sci.}\ }\textbf {\bibinfo {volume} {21}},\ \bibinfo {pages}
  {1115} (\bibinfo {year} {1996})}\BibitemShut {NoStop}%
\bibitem [{\citenamefont {Dogic}\ and\ \citenamefont {Fraden}(2006)}]{Dogic-1}%
  \BibitemOpen
  \bibfield  {author} {\bibinfo {author} {\bibfnamefont {Z.}~\bibnamefont
  {Dogic}}\ and\ \bibinfo {author} {\bibfnamefont {S.}~\bibnamefont {Fraden}},\
  }\href@noop {} {\bibfield  {journal} {\bibinfo  {journal} {Curr. Opin.
  Colloid In.}\ }\textbf {\bibinfo {volume} {11}},\ \bibinfo {pages} {47}
  (\bibinfo {year} {2006})}\BibitemShut {NoStop}%
\bibitem [{\citenamefont {Uematsu}\ and\ \citenamefont
  {Uematsu}(1984)}]{Uematsu}%
  \BibitemOpen
  \bibfield  {author} {\bibinfo {author} {\bibfnamefont {I.}~\bibnamefont
  {Uematsu}}\ and\ \bibinfo {author} {\bibfnamefont {Y.}~\bibnamefont
  {Uematsu}},\ }\enquote {\bibinfo {title} {Polypeptide liquid crystals},}\ in\
  \href {\doibase 10.1007/3-540-12818-2_6} {\emph {\bibinfo {booktitle} {Liquid
  Crystal Polymers I}}},\ \bibinfo {editor} {edited by\ \bibinfo {editor}
  {\bibfnamefont {N.~A.}\ \bibnamefont {Plat{\'e}}}}\ (\bibinfo  {publisher}
  {Springer Berlin Heidelberg},\ \bibinfo {address} {Berlin, Heidelberg},\
  \bibinfo {year} {1984})\ pp.\ \bibinfo {pages} {37--73}\BibitemShut {NoStop}%
\bibitem [{\citenamefont {Hiltrop}(2001)}]{Hiltrop}%
  \BibitemOpen
  \bibfield  {author} {\bibinfo {author} {\bibfnamefont {K.}~\bibnamefont
  {Hiltrop}},\ }\enquote {\bibinfo {title} {Phase chirality of micellar
  lyotropic liquid crystals},}\ in\ \href {\doibase 10.1007/0-387-21642-1_14}
  {\emph {\bibinfo {booktitle} {Chirality in Liquid Crystals}}},\ \bibinfo
  {editor} {edited by\ \bibinfo {editor} {\bibfnamefont {H.-S.}\ \bibnamefont
  {Kitzerow}}\ and\ \bibinfo {editor} {\bibfnamefont {C.}~\bibnamefont
  {Bahr}}}\ (\bibinfo  {publisher} {Springer New York},\ \bibinfo {address}
  {New York, NY},\ \bibinfo {year} {2001})\ pp.\ \bibinfo {pages}
  {447--480}\BibitemShut {NoStop}%
\bibitem [{\citenamefont {Sato}\ and\ \citenamefont {Teramoto}(1996)}]{Sato-2}%
  \BibitemOpen
  \bibfield  {author} {\bibinfo {author} {\bibfnamefont {T.}~\bibnamefont
  {Sato}}\ and\ \bibinfo {author} {\bibfnamefont {A.}~\bibnamefont
  {Teramoto}},\ }\enquote {\bibinfo {title} {Concentrated solutions of
  liquid-crystalline polymers},}\ in\ \href {\doibase 10.1007/3-540-60484-7_3}
  {\emph {\bibinfo {booktitle} {Biopolymers Liquid Crystalline Polymers Phase
  Emulsion}}}\ (\bibinfo  {publisher} {Springer Berlin Heidelberg},\ \bibinfo
  {address} {Berlin, Heidelberg},\ \bibinfo {year} {1996})\ pp.\ \bibinfo
  {pages} {85--161}\BibitemShut {NoStop}%
\bibitem [{\citenamefont {Werbowyj}\ and\ \citenamefont
  {Gray}(1976)}]{Werbowyj}%
  \BibitemOpen
  \bibfield  {author} {\bibinfo {author} {\bibfnamefont {R.~S.}\ \bibnamefont
  {Werbowyj}}\ and\ \bibinfo {author} {\bibfnamefont {D.~G.}\ \bibnamefont
  {Gray}},\ }\href@noop {} {\bibfield  {journal} {\bibinfo  {journal} {Mol.
  Cryst. Liq. Cryst.}\ }\textbf {\bibinfo {volume} {34}},\ \bibinfo {pages}
  {97} (\bibinfo {year} {1976})}\BibitemShut {NoStop}%
\bibitem [{\citenamefont {Revol}\ and\ \citenamefont
  {Marchessault}(1993)}]{Revol}%
  \BibitemOpen
  \bibfield  {author} {\bibinfo {author} {\bibfnamefont {J.-F.}\ \bibnamefont
  {Revol}}\ and\ \bibinfo {author} {\bibfnamefont {R.~H.}\ \bibnamefont
  {Marchessault}},\ }\href@noop {} {\bibfield  {journal} {\bibinfo  {journal}
  {Int. J. Biol. Macromol.}\ }\textbf {\bibinfo {volume} {15}},\ \bibinfo
  {pages} {329} (\bibinfo {year} {1993})}\BibitemShut {NoStop}%
\bibitem [{\citenamefont {Aharoni}(1979)}]{Aharoni}%
  \BibitemOpen
  \bibfield  {author} {\bibinfo {author} {\bibfnamefont {S.~M.}\ \bibnamefont
  {Aharoni}},\ }\href@noop {} {\bibfield  {journal} {\bibinfo  {journal}
  {Macromolecules}\ }\textbf {\bibinfo {volume} {12}},\ \bibinfo {pages} {94}
  (\bibinfo {year} {1979})}\BibitemShut {NoStop}%
\bibitem [{\citenamefont {Sato}\ \emph {et~al.}(1993)\citenamefont {Sato},
  \citenamefont {Sato}, \citenamefont {Umemura}, \citenamefont {Teramoto},
  \citenamefont {Nagamura}, \citenamefont {Wagner}, \citenamefont {Weng},
  \citenamefont {Okamoto}, \citenamefont {Hatada},\ and\ \citenamefont
  {Green}}]{Sato-1}%
  \BibitemOpen
  \bibfield  {author} {\bibinfo {author} {\bibfnamefont {T.}~\bibnamefont
  {Sato}}, \bibinfo {author} {\bibfnamefont {Y.}~\bibnamefont {Sato}}, \bibinfo
  {author} {\bibfnamefont {Y.}~\bibnamefont {Umemura}}, \bibinfo {author}
  {\bibfnamefont {A.}~\bibnamefont {Teramoto}}, \bibinfo {author}
  {\bibfnamefont {Y.}~\bibnamefont {Nagamura}}, \bibinfo {author}
  {\bibfnamefont {J.}~\bibnamefont {Wagner}}, \bibinfo {author} {\bibfnamefont
  {D.}~\bibnamefont {Weng}}, \bibinfo {author} {\bibfnamefont {Y.}~\bibnamefont
  {Okamoto}}, \bibinfo {author} {\bibfnamefont {K.}~\bibnamefont {Hatada}}, \
  and\ \bibinfo {author} {\bibfnamefont {M.~M.}\ \bibnamefont {Green}},\
  }\href@noop {} {\bibfield  {journal} {\bibinfo  {journal} {Macromolecules}\
  }\textbf {\bibinfo {volume} {26}},\ \bibinfo {pages} {4551} (\bibinfo {year}
  {1993})}\BibitemShut {NoStop}%
\bibitem [{\citenamefont {Watanabe}, \citenamefont {Kamee},\ and\ \citenamefont
  {Fujiki}(2001)}]{Watanabe}%
  \BibitemOpen
  \bibfield  {author} {\bibinfo {author} {\bibfnamefont {J.}~\bibnamefont
  {Watanabe}}, \bibinfo {author} {\bibfnamefont {H.}~\bibnamefont {Kamee}}, \
  and\ \bibinfo {author} {\bibfnamefont {M.}~\bibnamefont {Fujiki}},\
  }\href@noop {} {\bibfield  {journal} {\bibinfo  {journal} {Polym. J.}\
  }\textbf {\bibinfo {volume} {33}},\ \bibinfo {pages} {495} (\bibinfo {year}
  {2001})}\BibitemShut {NoStop}%
\bibitem [{\citenamefont {Yevdokimov}, \citenamefont {Skuridin},\ and\
  \citenamefont {Salyanov}(1988)}]{Yevdokimov}%
  \BibitemOpen
  \bibfield  {author} {\bibinfo {author} {\bibfnamefont {Y.~M.}\ \bibnamefont
  {Yevdokimov}}, \bibinfo {author} {\bibfnamefont {S.~G.}\ \bibnamefont
  {Skuridin}}, \ and\ \bibinfo {author} {\bibfnamefont {V.~I.}\ \bibnamefont
  {Salyanov}},\ }\href@noop {} {\bibfield  {journal} {\bibinfo  {journal} {Liq.
  Cryst.}\ }\textbf {\bibinfo {volume} {3}},\ \bibinfo {pages} {1443} (\bibinfo
  {year} {1988})}\BibitemShut {NoStop}%
\bibitem [{\citenamefont {Van~Winkle}\ \emph {et~al.}(1990)\citenamefont
  {Van~Winkle}, \citenamefont {Davidson}, \citenamefont {Chen},\ and\
  \citenamefont {Rill}}]{vanWinkle}%
  \BibitemOpen
  \bibfield  {author} {\bibinfo {author} {\bibfnamefont {D.~H.}\ \bibnamefont
  {Van~Winkle}}, \bibinfo {author} {\bibfnamefont {M.~W.}\ \bibnamefont
  {Davidson}}, \bibinfo {author} {\bibfnamefont {W.~X.}\ \bibnamefont {Chen}},
  \ and\ \bibinfo {author} {\bibfnamefont {R.~L.}\ \bibnamefont {Rill}},\
  }\href@noop {} {\bibfield  {journal} {\bibinfo  {journal} {Macromolecules}\
  }\textbf {\bibinfo {volume} {23}},\ \bibinfo {pages} {4140} (\bibinfo {year}
  {1990})}\BibitemShut {NoStop}%
\bibitem [{\citenamefont {Stanley}, \citenamefont {Hong},\ and\ \citenamefont
  {Strey}(2005)}]{Stanley}%
  \BibitemOpen
  \bibfield  {author} {\bibinfo {author} {\bibfnamefont {C.~B.}\ \bibnamefont
  {Stanley}}, \bibinfo {author} {\bibfnamefont {H.}~\bibnamefont {Hong}}, \
  and\ \bibinfo {author} {\bibfnamefont {H.~H.}\ \bibnamefont {Strey}},\
  }\href@noop {} {\bibfield  {journal} {\bibinfo  {journal} {Biophys. J.}\
  }\textbf {\bibinfo {volume} {89}},\ \bibinfo {pages} {2552} (\bibinfo {year}
  {2005})}\BibitemShut {NoStop}%
\bibitem [{\citenamefont {Dogic}\ and\ \citenamefont {Fraden}(2000)}]{Dogic-2}%
  \BibitemOpen
  \bibfield  {author} {\bibinfo {author} {\bibfnamefont {Z.}~\bibnamefont
  {Dogic}}\ and\ \bibinfo {author} {\bibfnamefont {S.}~\bibnamefont {Fraden}},\
  }\href@noop {} {\bibfield  {journal} {\bibinfo  {journal} {Langmuir}\
  }\textbf {\bibinfo {volume} {16}},\ \bibinfo {pages} {7820} (\bibinfo {year}
  {2000})}\BibitemShut {NoStop}%
\bibitem [{\citenamefont {Grelet}\ and\ \citenamefont
  {Fraden}(2003)}]{Grelet-1}%
  \BibitemOpen
  \bibfield  {author} {\bibinfo {author} {\bibfnamefont {E.}~\bibnamefont
  {Grelet}}\ and\ \bibinfo {author} {\bibfnamefont {S.}~\bibnamefont
  {Fraden}},\ }\href@noop {} {\bibfield  {journal} {\bibinfo  {journal} {Phys.
  Rev. Lett.}\ }\textbf {\bibinfo {volume} {90}},\ \bibinfo {pages} {198302}
  (\bibinfo {year} {2003})}\BibitemShut {NoStop}%
\bibitem [{\citenamefont {DuPr{\'e}}\ and\ \citenamefont {Duke}(1975)}]{DuPre}%
  \BibitemOpen
  \bibfield  {author} {\bibinfo {author} {\bibfnamefont {D.~B.}\ \bibnamefont
  {DuPr{\'e}}}\ and\ \bibinfo {author} {\bibfnamefont {R.~W.}\ \bibnamefont
  {Duke}},\ }\href@noop {} {\bibfield  {journal} {\bibinfo  {journal} {J. Chem.
  Phys.}\ }\textbf {\bibinfo {volume} {63}},\ \bibinfo {pages} {143} (\bibinfo
  {year} {1975})}\BibitemShut {NoStop}%
\bibitem [{\citenamefont {Dong}\ and\ \citenamefont {Gray}(1997)}]{Dong}%
  \BibitemOpen
  \bibfield  {author} {\bibinfo {author} {\bibfnamefont {X.~M.}\ \bibnamefont
  {Dong}}\ and\ \bibinfo {author} {\bibfnamefont {D.~G.}\ \bibnamefont
  {Gray}},\ }\href@noop {} {\bibfield  {journal} {\bibinfo  {journal}
  {Langmuir}\ }\textbf {\bibinfo {volume} {13}},\ \bibinfo {pages} {2404}
  (\bibinfo {year} {1997})}\BibitemShut {NoStop}%
\bibitem [{\citenamefont {Miller}\ and\ \citenamefont {Donald}(2003)}]{Miller}%
  \BibitemOpen
  \bibfield  {author} {\bibinfo {author} {\bibfnamefont {A.~F.}\ \bibnamefont
  {Miller}}\ and\ \bibinfo {author} {\bibfnamefont {A.~M.}\ \bibnamefont
  {Donald}},\ }\href@noop {} {\bibfield  {journal} {\bibinfo  {journal}
  {Biomacromolecules}\ }\textbf {\bibinfo {volume} {4}},\ \bibinfo {pages}
  {510} (\bibinfo {year} {2003})}\BibitemShut {NoStop}%
\bibitem [{\citenamefont {Frenkel}(2013)}]{Frenkel-3}%
  \BibitemOpen
  \bibfield  {author} {\bibinfo {author} {\bibfnamefont {D.}~\bibnamefont
  {Frenkel}},\ }\href@noop {} {\bibfield  {journal} {\bibinfo  {journal} {Eur.
  Phys. J. Plus}\ }\textbf {\bibinfo {volume} {128}},\ \bibinfo {pages} {10}
  (\bibinfo {year} {2013})}\BibitemShut {NoStop}%
\bibitem [{\citenamefont {Usov}\ \emph {et~al.}(2015)\citenamefont {Usov},
  \citenamefont {Nystr{\"o}m}, \citenamefont {Adamcik}, \citenamefont
  {Handschin}, \citenamefont {Sch{\"u}tz}, \citenamefont {Fall}, \citenamefont
  {Bergstr{\"o}m},\ and\ \citenamefont {Mezzenga}}]{Usov}%
  \BibitemOpen
  \bibfield  {author} {\bibinfo {author} {\bibfnamefont {I.}~\bibnamefont
  {Usov}}, \bibinfo {author} {\bibfnamefont {G.}~\bibnamefont {Nystr{\"o}m}},
  \bibinfo {author} {\bibfnamefont {J.}~\bibnamefont {Adamcik}}, \bibinfo
  {author} {\bibfnamefont {S.}~\bibnamefont {Handschin}}, \bibinfo {author}
  {\bibfnamefont {C.}~\bibnamefont {Sch{\"u}tz}}, \bibinfo {author}
  {\bibfnamefont {A.}~\bibnamefont {Fall}}, \bibinfo {author} {\bibfnamefont
  {L.}~\bibnamefont {Bergstr{\"o}m}}, \ and\ \bibinfo {author} {\bibfnamefont
  {R.}~\bibnamefont {Mezzenga}},\ }\href@noop {} {\bibfield  {journal}
  {\bibinfo  {journal} {Nat. Commun.}\ }\textbf {\bibinfo {volume} {6}},\
  \bibinfo {pages} {7564} (\bibinfo {year} {2015})}\BibitemShut {NoStop}%
\bibitem [{\citenamefont {Wang}, \citenamefont {Hamad},\ and\ \citenamefont
  {MacLachlan}(2016)}]{Wang}%
  \BibitemOpen
  \bibfield  {author} {\bibinfo {author} {\bibfnamefont {P.-X.}\ \bibnamefont
  {Wang}}, \bibinfo {author} {\bibfnamefont {W.~Y.}\ \bibnamefont {Hamad}}, \
  and\ \bibinfo {author} {\bibfnamefont {M.~J.}\ \bibnamefont {MacLachlan}},\
  }\href@noop {} {\bibfield  {journal} {\bibinfo  {journal} {Nat. Commun.}\
  }\textbf {\bibinfo {volume} {7}},\ \bibinfo {pages} {11515} (\bibinfo {year}
  {2016})}\BibitemShut {NoStop}%
\bibitem [{\citenamefont {Onsager}(1949)}]{Onsager}%
  \BibitemOpen
  \bibfield  {author} {\bibinfo {author} {\bibfnamefont {L.}~\bibnamefont
  {Onsager}},\ }\href@noop {} {\bibfield  {journal} {\bibinfo  {journal} {Ann.
  N.Y. Acad. Sci.}\ }\textbf {\bibinfo {volume} {51}},\ \bibinfo {pages} {627}
  (\bibinfo {year} {1949})}\BibitemShut {NoStop}%
\bibitem [{\citenamefont {Evans}(1992{\natexlab{a}})}]{Evans-1}%
  \BibitemOpen
  \bibfield  {author} {\bibinfo {author} {\bibfnamefont {R.}~\bibnamefont
  {Evans}},\ }\enquote {\bibinfo {title} {Density functionals in the theory of
  nonuniform fluids},}\ in\ \href@noop {} {\emph {\bibinfo {booktitle}
  {Fundamentals of Inhomogeneous Fluids}}},\ \bibinfo {editor} {edited by\
  \bibinfo {editor} {\bibfnamefont {D.}~\bibnamefont {Henderson}}}\ (\bibinfo
  {publisher} {Marcel Dekker},\ \bibinfo {address} {New York, NY},\ \bibinfo
  {year} {1992})\ pp.\ \bibinfo {pages} {85--175}\BibitemShut {NoStop}%
\bibitem [{\citenamefont {Fraden}\ \emph {et~al.}(1989)\citenamefont {Fraden},
  \citenamefont {Maret}, \citenamefont {Caspar},\ and\ \citenamefont
  {Meyer}}]{Fraden}%
  \BibitemOpen
  \bibfield  {author} {\bibinfo {author} {\bibfnamefont {S.}~\bibnamefont
  {Fraden}}, \bibinfo {author} {\bibfnamefont {G.}~\bibnamefont {Maret}},
  \bibinfo {author} {\bibfnamefont {D.~L.~D.}\ \bibnamefont {Caspar}}, \ and\
  \bibinfo {author} {\bibfnamefont {R.~B.}\ \bibnamefont {Meyer}},\ }\href@noop
  {} {\bibfield  {journal} {\bibinfo  {journal} {Phys. Rev. Lett.}\ }\textbf
  {\bibinfo {volume} {63}},\ \bibinfo {pages} {2068} (\bibinfo {year}
  {1989})}\BibitemShut {NoStop}%
\bibitem [{\citenamefont {Lagerwall}\ and\ \citenamefont
  {Scalia}(2008)}]{Lagerwall}%
  \BibitemOpen
  \bibfield  {author} {\bibinfo {author} {\bibfnamefont {J.~P.~F.}\
  \bibnamefont {Lagerwall}}\ and\ \bibinfo {author} {\bibfnamefont
  {G.}~\bibnamefont {Scalia}},\ }\href@noop {} {\bibfield  {journal} {\bibinfo
  {journal} {J. Mater. Chem.}\ }\textbf {\bibinfo {volume} {18}},\ \bibinfo
  {pages} {2890} (\bibinfo {year} {2008})}\BibitemShut {NoStop}%
\bibitem [{\citenamefont {Buining}\ and\ \citenamefont
  {Lekkerkerker}(1993)}]{Lekkerkerker-4}%
  \BibitemOpen
  \bibfield  {author} {\bibinfo {author} {\bibfnamefont {P.~A.}\ \bibnamefont
  {Buining}}\ and\ \bibinfo {author} {\bibfnamefont {H.~N.~W.}\ \bibnamefont
  {Lekkerkerker}},\ }\href@noop {} {\bibfield  {journal} {\bibinfo  {journal}
  {J. Phys. Chem.}\ }\textbf {\bibinfo {volume} {97}},\ \bibinfo {pages}
  {11510} (\bibinfo {year} {1993})}\BibitemShut {NoStop}%
\bibitem [{\citenamefont {Zhang}\ and\ \citenamefont {van
  Duijneveldt}(2006)}]{Zhang}%
  \BibitemOpen
  \bibfield  {author} {\bibinfo {author} {\bibfnamefont {Z.~X.}\ \bibnamefont
  {Zhang}}\ and\ \bibinfo {author} {\bibfnamefont {J.~S.}\ \bibnamefont {van
  Duijneveldt}},\ }\href@noop {} {\bibfield  {journal} {\bibinfo  {journal} {J.
  Chem. Phys.}\ }\textbf {\bibinfo {volume} {124}},\ \bibinfo {pages} {154910}
  (\bibinfo {year} {2006})}\BibitemShut {NoStop}%
\bibitem [{\citenamefont {Straley}(1976)}]{Straley-1}%
  \BibitemOpen
  \bibfield  {author} {\bibinfo {author} {\bibfnamefont {J.~P.}\ \bibnamefont
  {Straley}},\ }\href@noop {} {\bibfield  {journal} {\bibinfo  {journal} {Phys.
  Rev. A}\ }\textbf {\bibinfo {volume} {14}},\ \bibinfo {pages} {1835}
  (\bibinfo {year} {1976})}\BibitemShut {NoStop}%
\bibitem [{\citenamefont {Osipov}(1985)}]{Osipov}%
  \BibitemOpen
  \bibfield  {author} {\bibinfo {author} {\bibfnamefont {M.~A.}\ \bibnamefont
  {Osipov}},\ }\href@noop {} {\bibfield  {journal} {\bibinfo  {journal} {Chem.
  Phys.}\ }\textbf {\bibinfo {volume} {96}},\ \bibinfo {pages} {259} (\bibinfo
  {year} {1985})}\BibitemShut {NoStop}%
\bibitem [{\citenamefont {Odijk}(1987)}]{Odijk}%
  \BibitemOpen
  \bibfield  {author} {\bibinfo {author} {\bibfnamefont {T.}~\bibnamefont
  {Odijk}},\ }\href@noop {} {\bibfield  {journal} {\bibinfo  {journal} {J.
  Phys. Chem.}\ }\textbf {\bibinfo {volume} {91}},\ \bibinfo {pages} {6060}
  (\bibinfo {year} {1987})}\BibitemShut {NoStop}%
\bibitem [{\citenamefont {Pelcovits}(1996)}]{Pelcovits}%
  \BibitemOpen
  \bibfield  {author} {\bibinfo {author} {\bibfnamefont {R.~A.}\ \bibnamefont
  {Pelcovits}},\ }\href@noop {} {\bibfield  {journal} {\bibinfo  {journal}
  {Liq. Cryst.}\ }\textbf {\bibinfo {volume} {21}},\ \bibinfo {pages} {361}
  (\bibinfo {year} {1996})}\BibitemShut {NoStop}%
\bibitem [{\citenamefont {Evans}(1992{\natexlab{b}})}]{Evans-3}%
  \BibitemOpen
  \bibfield  {author} {\bibinfo {author} {\bibfnamefont {G.~T.}\ \bibnamefont
  {Evans}},\ }\href@noop {} {\bibfield  {journal} {\bibinfo  {journal} {Mol.
  Phys.}\ }\textbf {\bibinfo {volume} {77}},\ \bibinfo {pages} {969} (\bibinfo
  {year} {1992}{\natexlab{b}})}\BibitemShut {NoStop}%
\bibitem [{\citenamefont {Varga}\ and\ \citenamefont
  {Jackson}(2006)}]{Varga-1}%
  \BibitemOpen
  \bibfield  {author} {\bibinfo {author} {\bibfnamefont {S.}~\bibnamefont
  {Varga}}\ and\ \bibinfo {author} {\bibfnamefont {G.}~\bibnamefont
  {Jackson}},\ }\href@noop {} {\bibfield  {journal} {\bibinfo  {journal} {Mol.
  Phys.}\ }\textbf {\bibinfo {volume} {104}},\ \bibinfo {pages} {3681}
  (\bibinfo {year} {2006})}\BibitemShut {NoStop}%
\bibitem [{\citenamefont {Varga}\ and\ \citenamefont
  {Jackson}(2011)}]{Varga-2}%
  \BibitemOpen
  \bibfield  {author} {\bibinfo {author} {\bibfnamefont {S.}~\bibnamefont
  {Varga}}\ and\ \bibinfo {author} {\bibfnamefont {G.}~\bibnamefont
  {Jackson}},\ }\href@noop {} {\bibfield  {journal} {\bibinfo  {journal} {Mol.
  Phys.}\ }\textbf {\bibinfo {volume} {109}},\ \bibinfo {pages} {1313}
  (\bibinfo {year} {2011})}\BibitemShut {NoStop}%
\bibitem [{\citenamefont {Wensink}\ and\ \citenamefont
  {Jackson}(2009)}]{Wensink-1}%
  \BibitemOpen
  \bibfield  {author} {\bibinfo {author} {\bibfnamefont {H.~H.}\ \bibnamefont
  {Wensink}}\ and\ \bibinfo {author} {\bibfnamefont {G.}~\bibnamefont
  {Jackson}},\ }\href@noop {} {\bibfield  {journal} {\bibinfo  {journal} {J.
  Chem. Phys.}\ }\textbf {\bibinfo {volume} {130}},\ \bibinfo {pages} {234911}
  (\bibinfo {year} {2009})}\BibitemShut {NoStop}%
\bibitem [{\citenamefont {Wensink}\ and\ \citenamefont
  {Jackson}(2011)}]{Wensink-4}%
  \BibitemOpen
  \bibfield  {author} {\bibinfo {author} {\bibfnamefont {H.~H.}\ \bibnamefont
  {Wensink}}\ and\ \bibinfo {author} {\bibfnamefont {G.}~\bibnamefont
  {Jackson}},\ }\href@noop {} {\bibfield  {journal} {\bibinfo  {journal} {J.
  Phys.: Condens. Matter}\ }\textbf {\bibinfo {volume} {23}},\ \bibinfo {pages}
  {194107} (\bibinfo {year} {2011})}\BibitemShut {NoStop}%
\bibitem [{\citenamefont {Wensink}(2014)}]{Wensink-3}%
  \BibitemOpen
  \bibfield  {author} {\bibinfo {author} {\bibfnamefont {H.~H.}\ \bibnamefont
  {Wensink}},\ }\href@noop {} {\bibfield  {journal} {\bibinfo  {journal}
  {Europhys. Lett.}\ }\textbf {\bibinfo {volume} {107}},\ \bibinfo {pages}
  {36001} (\bibinfo {year} {2014})}\BibitemShut {NoStop}%
\bibitem [{\citenamefont {Tombolato}\ and\ \citenamefont
  {Ferrarini}(2005)}]{Tombolato}%
  \BibitemOpen
  \bibfield  {author} {\bibinfo {author} {\bibfnamefont {F.}~\bibnamefont
  {Tombolato}}\ and\ \bibinfo {author} {\bibfnamefont {A.}~\bibnamefont
  {Ferrarini}},\ }\href@noop {} {\bibfield  {journal} {\bibinfo  {journal} {J.
  Chem. Phys.}\ }\textbf {\bibinfo {volume} {122}},\ \bibinfo {pages} {054908}
  (\bibinfo {year} {2005})}\BibitemShut {NoStop}%
\bibitem [{\citenamefont {Tombolato}, \citenamefont {Ferrarini},\ and\
  \citenamefont {Grelet}(2006)}]{Grelet-2}%
  \BibitemOpen
  \bibfield  {author} {\bibinfo {author} {\bibfnamefont {F.}~\bibnamefont
  {Tombolato}}, \bibinfo {author} {\bibfnamefont {A.}~\bibnamefont
  {Ferrarini}}, \ and\ \bibinfo {author} {\bibfnamefont {E.}~\bibnamefont
  {Grelet}},\ }\href@noop {} {\bibfield  {journal} {\bibinfo  {journal} {Phys.
  Rev. Lett.}\ }\textbf {\bibinfo {volume} {96}},\ \bibinfo {pages} {258302}
  (\bibinfo {year} {2006})}\BibitemShut {NoStop}%
\bibitem [{\citenamefont {Frezza}\ \emph {et~al.}(2014)\citenamefont {Frezza},
  \citenamefont {Ferrarini}, \citenamefont {Kolli}, \citenamefont
  {Giacometti},\ and\ \citenamefont {Cinacchi}}]{Ferrarini-1}%
  \BibitemOpen
  \bibfield  {author} {\bibinfo {author} {\bibfnamefont {E.}~\bibnamefont
  {Frezza}}, \bibinfo {author} {\bibfnamefont {A.}~\bibnamefont {Ferrarini}},
  \bibinfo {author} {\bibfnamefont {H.~B.}\ \bibnamefont {Kolli}}, \bibinfo
  {author} {\bibfnamefont {A.}~\bibnamefont {Giacometti}}, \ and\ \bibinfo
  {author} {\bibfnamefont {G.}~\bibnamefont {Cinacchi}},\ }\href@noop {}
  {\bibfield  {journal} {\bibinfo  {journal} {Phys. Chem. Chem. Phys.}\
  }\textbf {\bibinfo {volume} {16}},\ \bibinfo {pages} {16225} (\bibinfo {year}
  {2014})}\BibitemShut {NoStop}%
\bibitem [{\citenamefont {Lekkerkerker}\ \emph {et~al.}(1984)\citenamefont
  {Lekkerkerker}, \citenamefont {Coulon}, \citenamefont {Van Der~Haegen},\ and\
  \citenamefont {Deblieck}}]{Lekkerkerker-3}%
  \BibitemOpen
  \bibfield  {author} {\bibinfo {author} {\bibfnamefont {H.~N.~W.}\
  \bibnamefont {Lekkerkerker}}, \bibinfo {author} {\bibfnamefont
  {P.}~\bibnamefont {Coulon}}, \bibinfo {author} {\bibfnamefont
  {R.}~\bibnamefont {Van Der~Haegen}}, \ and\ \bibinfo {author} {\bibfnamefont
  {R.}~\bibnamefont {Deblieck}},\ }\href@noop {} {\bibfield  {journal}
  {\bibinfo  {journal} {J. Chem. Phys.}\ }\textbf {\bibinfo {volume} {80}},\
  \bibinfo {pages} {3427} (\bibinfo {year} {1984})}\BibitemShut {NoStop}%
\bibitem [{\citenamefont {Belli}\ \emph {et~al.}(2014)\citenamefont {Belli},
  \citenamefont {Dussi}, \citenamefont {Dijkstra},\ and\ \citenamefont {van
  Roij}}]{Dijkstra-1}%
  \BibitemOpen
  \bibfield  {author} {\bibinfo {author} {\bibfnamefont {S.}~\bibnamefont
  {Belli}}, \bibinfo {author} {\bibfnamefont {S.}~\bibnamefont {Dussi}},
  \bibinfo {author} {\bibfnamefont {M.}~\bibnamefont {Dijkstra}}, \ and\
  \bibinfo {author} {\bibfnamefont {R.}~\bibnamefont {van Roij}},\ }\href@noop
  {} {\bibfield  {journal} {\bibinfo  {journal} {Phys. Rev. E}\ }\textbf
  {\bibinfo {volume} {90}},\ \bibinfo {pages} {020503} (\bibinfo {year}
  {2014})}\BibitemShut {NoStop}%
\bibitem [{\citenamefont {Dussi}\ \emph {et~al.}(2015)\citenamefont {Dussi},
  \citenamefont {Belli}, \citenamefont {van Roij},\ and\ \citenamefont
  {Dijkstra}}]{Dijkstra-2}%
  \BibitemOpen
  \bibfield  {author} {\bibinfo {author} {\bibfnamefont {S.}~\bibnamefont
  {Dussi}}, \bibinfo {author} {\bibfnamefont {S.}~\bibnamefont {Belli}},
  \bibinfo {author} {\bibfnamefont {R.}~\bibnamefont {van Roij}}, \ and\
  \bibinfo {author} {\bibfnamefont {M.}~\bibnamefont {Dijkstra}},\ }\href@noop
  {} {\bibfield  {journal} {\bibinfo  {journal} {J. Chem. Phys.}\ }\textbf
  {\bibinfo {volume} {142}},\ \bibinfo {pages} {074905} (\bibinfo {year}
  {2015})}\BibitemShut {NoStop}%
\bibitem [{\citenamefont {Frank}(1958)}]{Frank}%
  \BibitemOpen
  \bibfield  {author} {\bibinfo {author} {\bibfnamefont {F.~C.}\ \bibnamefont
  {Frank}},\ }\href {\doibase 10.1039/DF9582500019} {\bibfield  {journal}
  {\bibinfo  {journal} {Discuss. Faraday Soc.}\ }\textbf {\bibinfo {volume}
  {25}},\ \bibinfo {pages} {19} (\bibinfo {year} {1958})}\BibitemShut {NoStop}%
\bibitem [{\citenamefont {Majumdar}\ and\ \citenamefont
  {Lewis}(2017)}]{Majumdar}%
  \BibitemOpen
  \bibfield  {author} {\bibinfo {author} {\bibfnamefont {A.}~\bibnamefont
  {Majumdar}}\ and\ \bibinfo {author} {\bibfnamefont {A.~H.}\ \bibnamefont
  {Lewis}},\ }\enquote {\bibinfo {title} {A theoretician's approach to nematic
  liquid crystals and their applications},}\ in\ \href {\doibase
  10.1007/978-981-10-2502-0_8} {\emph {\bibinfo {booktitle} {Variational
  Methods in Molecular Modeling}}},\ \bibinfo {editor} {edited by\ \bibinfo
  {editor} {\bibfnamefont {J.}~\bibnamefont {Wu}}}\ (\bibinfo  {publisher}
  {Springer Singapore},\ \bibinfo {address} {Singapore},\ \bibinfo {year}
  {2017})\ pp.\ \bibinfo {pages} {223--254}\BibitemShut {NoStop}%
\bibitem [{\citenamefont {Lewis}\ \emph {et~al.}(2014)\citenamefont {Lewis},
  \citenamefont {Garlea}, \citenamefont {Alvarado}, \citenamefont {Dammone},
  \citenamefont {Howell}, \citenamefont {Majumdar}, \citenamefont {Mulder},
  \citenamefont {Lettinga}, \citenamefont {Koenderink},\ and\ \citenamefont
  {Aarts}}]{Dammone-1}%
  \BibitemOpen
  \bibfield  {author} {\bibinfo {author} {\bibfnamefont {A.~H.}\ \bibnamefont
  {Lewis}}, \bibinfo {author} {\bibfnamefont {I.}~\bibnamefont {Garlea}},
  \bibinfo {author} {\bibfnamefont {J.}~\bibnamefont {Alvarado}}, \bibinfo
  {author} {\bibfnamefont {O.~J.}\ \bibnamefont {Dammone}}, \bibinfo {author}
  {\bibfnamefont {P.~D.}\ \bibnamefont {Howell}}, \bibinfo {author}
  {\bibfnamefont {A.}~\bibnamefont {Majumdar}}, \bibinfo {author}
  {\bibfnamefont {B.~M.}\ \bibnamefont {Mulder}}, \bibinfo {author}
  {\bibfnamefont {M.~P.}\ \bibnamefont {Lettinga}}, \bibinfo {author}
  {\bibfnamefont {G.~H.}\ \bibnamefont {Koenderink}}, \ and\ \bibinfo {author}
  {\bibfnamefont {D.~G. A.~L.}\ \bibnamefont {Aarts}},\ }\href@noop {}
  {\bibfield  {journal} {\bibinfo  {journal} {Soft Matter}\ }\textbf {\bibinfo
  {volume} {10}},\ \bibinfo {pages} {7865} (\bibinfo {year}
  {2014})}\BibitemShut {NoStop}%
\bibitem [{\citenamefont {Dammone}\ \emph {et~al.}(2012)\citenamefont
  {Dammone}, \citenamefont {Zacharoudiou}, \citenamefont {Dullens},
  \citenamefont {Yeomans}, \citenamefont {Lettinga},\ and\ \citenamefont
  {Aarts}}]{Dammone-2}%
  \BibitemOpen
  \bibfield  {author} {\bibinfo {author} {\bibfnamefont {O.~J.}\ \bibnamefont
  {Dammone}}, \bibinfo {author} {\bibfnamefont {I.}~\bibnamefont
  {Zacharoudiou}}, \bibinfo {author} {\bibfnamefont {R.~P.~A.}\ \bibnamefont
  {Dullens}}, \bibinfo {author} {\bibfnamefont {J.~M.}\ \bibnamefont
  {Yeomans}}, \bibinfo {author} {\bibfnamefont {M.~P.}\ \bibnamefont
  {Lettinga}}, \ and\ \bibinfo {author} {\bibfnamefont {D.~G. A.~L.}\
  \bibnamefont {Aarts}},\ }\href {\doibase 10.1103/PhysRevLett.109.108303}
  {\bibfield  {journal} {\bibinfo  {journal} {Phys. Rev. Lett.}\ }\textbf
  {\bibinfo {volume} {109}},\ \bibinfo {pages} {108303} (\bibinfo {year}
  {2012})}\BibitemShut {NoStop}%
\bibitem [{\citenamefont {Evans}\ \emph {et~al.}(2016)\citenamefont {Evans},
  \citenamefont {Oettel}, \citenamefont {Roth},\ and\ \citenamefont
  {Kahl}}]{Evans-2}%
  \BibitemOpen
  \bibfield  {author} {\bibinfo {author} {\bibfnamefont {R.}~\bibnamefont
  {Evans}}, \bibinfo {author} {\bibfnamefont {M.}~\bibnamefont {Oettel}},
  \bibinfo {author} {\bibfnamefont {R.}~\bibnamefont {Roth}}, \ and\ \bibinfo
  {author} {\bibfnamefont {G.}~\bibnamefont {Kahl}},\ }\href
  {http://stacks.iop.org/0953-8984/28/i=24/a=240401} {\bibfield  {journal}
  {\bibinfo  {journal} {J. Phys.: Condens. Matter}\ }\textbf {\bibinfo {volume}
  {28}},\ \bibinfo {pages} {240401} (\bibinfo {year} {2016})}\BibitemShut
  {NoStop}%
\bibitem [{\citenamefont {Mayer}\ and\ \citenamefont {Mayer}(1940)}]{Mayer}%
  \BibitemOpen
  \bibfield  {author} {\bibinfo {author} {\bibfnamefont {J.~E.}\ \bibnamefont
  {Mayer}}\ and\ \bibinfo {author} {\bibfnamefont {M.~G.}\ \bibnamefont
  {Mayer}},\ }\href@noop {} {\emph {\bibinfo {title} {Statistical Mechanics}}}\
  (\bibinfo  {publisher} {John Wiley \& Sons, Ltd},\ \bibinfo {address} {New
  York, NY},\ \bibinfo {year} {1940})\BibitemShut {NoStop}%
\bibitem [{\citenamefont {Frenkel}(1987)}]{Frenkel-1}%
  \BibitemOpen
  \bibfield  {author} {\bibinfo {author} {\bibfnamefont {D.}~\bibnamefont
  {Frenkel}},\ }\href@noop {} {\bibfield  {journal} {\bibinfo  {journal} {J.
  Phys. Chem.}\ }\textbf {\bibinfo {volume} {91}},\ \bibinfo {pages} {4912}
  (\bibinfo {year} {1987})}\BibitemShut {NoStop}%
\bibitem [{\citenamefont {Straley}(1973{\natexlab{a}})}]{Straley-2}%
  \BibitemOpen
  \bibfield  {author} {\bibinfo {author} {\bibfnamefont {J.~P.}\ \bibnamefont
  {Straley}},\ }\href@noop {} {\bibfield  {journal} {\bibinfo  {journal} {Mol.
  Cryst. Liq. Cryst.}\ }\textbf {\bibinfo {volume} {24}},\ \bibinfo {pages} {7}
  (\bibinfo {year} {1973}{\natexlab{a}})}\BibitemShut {NoStop}%
\bibitem [{\citenamefont {Masters}(2008)}]{Masters-3}%
  \BibitemOpen
  \bibfield  {author} {\bibinfo {author} {\bibfnamefont {A.~J.}\ \bibnamefont
  {Masters}},\ }\href {http://stacks.iop.org/0953-8984/20/i=28/a=283102}
  {\bibfield  {journal} {\bibinfo  {journal} {J. Phys.: Condens. Matter}\
  }\textbf {\bibinfo {volume} {20}},\ \bibinfo {pages} {283102} (\bibinfo
  {year} {2008})}\BibitemShut {NoStop}%
\bibitem [{\citenamefont {You}, \citenamefont {Vlasov},\ and\ \citenamefont
  {Masters}(2005)}]{Masters-1}%
  \BibitemOpen
  \bibfield  {author} {\bibinfo {author} {\bibfnamefont {X.-M.}\ \bibnamefont
  {You}}, \bibinfo {author} {\bibfnamefont {A.~Y.}\ \bibnamefont {Vlasov}}, \
  and\ \bibinfo {author} {\bibfnamefont {A.~J.}\ \bibnamefont {Masters}},\
  }\href@noop {} {\bibfield  {journal} {\bibinfo  {journal} {J. Chem. Phys.}\
  }\textbf {\bibinfo {volume} {123}},\ \bibinfo {pages} {034510} (\bibinfo
  {year} {2005})}\BibitemShut {NoStop}%
\bibitem [{\citenamefont {You}\ \emph {et~al.}(2012)\citenamefont {You},
  \citenamefont {Vlasov}, \citenamefont {Anton},\ and\ \citenamefont
  {Masters}}]{Masters-2}%
  \BibitemOpen
  \bibfield  {author} {\bibinfo {author} {\bibfnamefont {X.-M.}\ \bibnamefont
  {You}}, \bibinfo {author} {\bibfnamefont {A.~Y.}\ \bibnamefont {Vlasov}},
  \bibinfo {author} {\bibfnamefont {L.}~\bibnamefont {Anton}}, \ and\ \bibinfo
  {author} {\bibfnamefont {A.~J.}\ \bibnamefont {Masters}},\ }\href@noop {}
  {\bibfield  {journal} {\bibinfo  {journal} {Phys. Rev. E}\ }\textbf {\bibinfo
  {volume} {85}},\ \bibinfo {pages} {061706} (\bibinfo {year}
  {2012})}\BibitemShut {NoStop}%
\bibitem [{\citenamefont {Hansen}\ and\ \citenamefont
  {McDonald}(1976)}]{Hansen}%
  \BibitemOpen
  \bibfield  {author} {\bibinfo {author} {\bibfnamefont {J.-P.}\ \bibnamefont
  {Hansen}}\ and\ \bibinfo {author} {\bibfnamefont {I.~R.}\ \bibnamefont
  {McDonald}},\ }\href@noop {} {\emph {\bibinfo {title} {Theory of Simple
  Liquids}}}\ (\bibinfo  {publisher} {Academic Press},\ \bibinfo {address}
  {London},\ \bibinfo {year} {1976})\BibitemShut {NoStop}%
\bibitem [{\citenamefont {Vroege}\ and\ \citenamefont
  {Lekkerkerker}(1992)}]{Lekkerkerker-1}%
  \BibitemOpen
  \bibfield  {author} {\bibinfo {author} {\bibfnamefont {G.~J.}\ \bibnamefont
  {Vroege}}\ and\ \bibinfo {author} {\bibfnamefont {H.~N.~W.}\ \bibnamefont
  {Lekkerkerker}},\ }\href@noop {} {\bibfield  {journal} {\bibinfo  {journal}
  {Rep. Prog. Phys.}\ }\textbf {\bibinfo {volume} {55}},\ \bibinfo {pages}
  {1241} (\bibinfo {year} {1992})}\BibitemShut {NoStop}%
\bibitem [{\citenamefont {Wertheim}(1963)}]{Wertheim}%
  \BibitemOpen
  \bibfield  {author} {\bibinfo {author} {\bibfnamefont {M.~S.}\ \bibnamefont
  {Wertheim}},\ }\href@noop {} {\bibfield  {journal} {\bibinfo  {journal}
  {Phys. Rev. Lett.}\ }\textbf {\bibinfo {volume} {10}},\ \bibinfo {pages}
  {321} (\bibinfo {year} {1963})}\BibitemShut {NoStop}%
\bibitem [{\citenamefont {Morita}\ and\ \citenamefont
  {Hiroike}(1960)}]{Morita}%
  \BibitemOpen
  \bibfield  {author} {\bibinfo {author} {\bibfnamefont {T.}~\bibnamefont
  {Morita}}\ and\ \bibinfo {author} {\bibfnamefont {K.}~\bibnamefont
  {Hiroike}},\ }\href@noop {} {\bibfield  {journal} {\bibinfo  {journal} {Prog.
  Theor. Phys.}\ }\textbf {\bibinfo {volume} {23}},\ \bibinfo {pages} {385}
  (\bibinfo {year} {1960})}\BibitemShut {NoStop}%
\bibitem [{\citenamefont {Rosenfeld}(1988)}]{Rosenfeld}%
  \BibitemOpen
  \bibfield  {author} {\bibinfo {author} {\bibfnamefont {Y.}~\bibnamefont
  {Rosenfeld}},\ }\href@noop {} {\bibfield  {journal} {\bibinfo  {journal} {J.
  Chem. Phys.}\ }\textbf {\bibinfo {volume} {89}},\ \bibinfo {pages} {4272}
  (\bibinfo {year} {1988})}\BibitemShut {NoStop}%
\bibitem [{\citenamefont {Wittmann}, \citenamefont {Marechal},\ and\
  \citenamefont {Mecke}(2016)}]{Wittmann}%
  \BibitemOpen
  \bibfield  {author} {\bibinfo {author} {\bibfnamefont {R.}~\bibnamefont
  {Wittmann}}, \bibinfo {author} {\bibfnamefont {M.}~\bibnamefont {Marechal}},
  \ and\ \bibinfo {author} {\bibfnamefont {K.}~\bibnamefont {Mecke}},\
  }\href@noop {} {\bibfield  {journal} {\bibinfo  {journal} {J. Phys.: Condens.
  Matter}\ }\textbf {\bibinfo {volume} {28}},\ \bibinfo {pages} {244003}
  (\bibinfo {year} {2016})}\BibitemShut {NoStop}%
\bibitem [{\citenamefont {Schmidt}(1999)}]{Schmidt}%
  \BibitemOpen
  \bibfield  {author} {\bibinfo {author} {\bibfnamefont {M.}~\bibnamefont
  {Schmidt}},\ }\href@noop {} {\bibfield  {journal} {\bibinfo  {journal} {Phys.
  Rev. E}\ }\textbf {\bibinfo {volume} {60}},\ \bibinfo {pages} {R6291}
  (\bibinfo {year} {1999})}\BibitemShut {NoStop}%
\bibitem [{\citenamefont {Pynn}(1974)}]{Pynn}%
  \BibitemOpen
  \bibfield  {author} {\bibinfo {author} {\bibfnamefont {R.}~\bibnamefont
  {Pynn}},\ }\href@noop {} {\bibfield  {journal} {\bibinfo  {journal} {Solid
  State Commun.}\ }\textbf {\bibinfo {volume} {14}},\ \bibinfo {pages} {29}
  (\bibinfo {year} {1974})}\BibitemShut {NoStop}%
\bibitem [{\citenamefont {Wulf}(1977)}]{Wulf}%
  \BibitemOpen
  \bibfield  {author} {\bibinfo {author} {\bibfnamefont {A.}~\bibnamefont
  {Wulf}},\ }\href@noop {} {\bibfield  {journal} {\bibinfo  {journal} {J. Chem.
  Phys.}\ }\textbf {\bibinfo {volume} {67}},\ \bibinfo {pages} {2254} (\bibinfo
  {year} {1977})}\BibitemShut {NoStop}%
\bibitem [{\citenamefont {Parsons}(1979)}]{Parsons}%
  \BibitemOpen
  \bibfield  {author} {\bibinfo {author} {\bibfnamefont {J.~D.}\ \bibnamefont
  {Parsons}},\ }\href@noop {} {\bibfield  {journal} {\bibinfo  {journal} {Phys.
  Rev. A}\ }\textbf {\bibinfo {volume} {19}},\ \bibinfo {pages} {1225}
  (\bibinfo {year} {1979})}\BibitemShut {NoStop}%
\bibitem [{\citenamefont {Lee}(1987)}]{Lee-1}%
  \BibitemOpen
  \bibfield  {author} {\bibinfo {author} {\bibfnamefont {S.-D.}\ \bibnamefont
  {Lee}},\ }\href@noop {} {\bibfield  {journal} {\bibinfo  {journal} {J. Chem.
  Phys.}\ }\textbf {\bibinfo {volume} {87}},\ \bibinfo {pages} {4972} (\bibinfo
  {year} {1987})}\BibitemShut {NoStop}%
\bibitem [{\citenamefont {Percus}\ and\ \citenamefont {Yevick}(1958)}]{Percus}%
  \BibitemOpen
  \bibfield  {author} {\bibinfo {author} {\bibfnamefont {J.~K.}\ \bibnamefont
  {Percus}}\ and\ \bibinfo {author} {\bibfnamefont {G.~J.}\ \bibnamefont
  {Yevick}},\ }\href@noop {} {\bibfield  {journal} {\bibinfo  {journal} {Phys.
  Rev.}\ }\textbf {\bibinfo {volume} {110}},\ \bibinfo {pages} {1} (\bibinfo
  {year} {1958})}\BibitemShut {NoStop}%
\bibitem [{\citenamefont {Carnahan}\ and\ \citenamefont
  {Starling}(1969)}]{Carnahan}%
  \BibitemOpen
  \bibfield  {author} {\bibinfo {author} {\bibfnamefont {N.~F.}\ \bibnamefont
  {Carnahan}}\ and\ \bibinfo {author} {\bibfnamefont {K.~E.}\ \bibnamefont
  {Starling}},\ }\href@noop {} {\bibfield  {journal} {\bibinfo  {journal} {J.
  Chem. Phys.}\ }\textbf {\bibinfo {volume} {51}},\ \bibinfo {pages} {635}
  (\bibinfo {year} {1969})}\BibitemShut {NoStop}%
\bibitem [{\citenamefont {Camp}\ \emph {et~al.}(1996)\citenamefont {Camp},
  \citenamefont {Mason}, \citenamefont {Allen}, \citenamefont {Khare},\ and\
  \citenamefont {Kofke}}]{Camp}%
  \BibitemOpen
  \bibfield  {author} {\bibinfo {author} {\bibfnamefont {P.~J.}\ \bibnamefont
  {Camp}}, \bibinfo {author} {\bibfnamefont {C.~P.}\ \bibnamefont {Mason}},
  \bibinfo {author} {\bibfnamefont {M.~P.}\ \bibnamefont {Allen}}, \bibinfo
  {author} {\bibfnamefont {A.~A.}\ \bibnamefont {Khare}}, \ and\ \bibinfo
  {author} {\bibfnamefont {D.~A.}\ \bibnamefont {Kofke}},\ }\href@noop {}
  {\bibfield  {journal} {\bibinfo  {journal} {J. Chem. Phys.}\ }\textbf
  {\bibinfo {volume} {105}},\ \bibinfo {pages} {2837} (\bibinfo {year}
  {1996})}\BibitemShut {NoStop}%
\bibitem [{\citenamefont {Cuetos}\ \emph {et~al.}(2007)\citenamefont {Cuetos},
  \citenamefont {Mart{\'\i}nez-Haya}, \citenamefont {Lago},\ and\ \citenamefont
  {Rull}}]{Cuetos-2}%
  \BibitemOpen
  \bibfield  {author} {\bibinfo {author} {\bibfnamefont {A.}~\bibnamefont
  {Cuetos}}, \bibinfo {author} {\bibfnamefont {B.}~\bibnamefont
  {Mart{\'\i}nez-Haya}}, \bibinfo {author} {\bibfnamefont {S.}~\bibnamefont
  {Lago}}, \ and\ \bibinfo {author} {\bibfnamefont {L.~F.}\ \bibnamefont
  {Rull}},\ }\href@noop {} {\bibfield  {journal} {\bibinfo  {journal} {Phys.
  Rev. E}\ }\textbf {\bibinfo {volume} {75}},\ \bibinfo {pages} {061701}
  (\bibinfo {year} {2007})}\BibitemShut {NoStop}%
\bibitem [{\citenamefont {Cuetos}\ \emph {et~al.}(2005)\citenamefont {Cuetos},
  \citenamefont {Martinez-Haya}, \citenamefont {Lago},\ and\ \citenamefont
  {Rull}}]{Cuetos-1}%
  \BibitemOpen
  \bibfield  {author} {\bibinfo {author} {\bibfnamefont {A.}~\bibnamefont
  {Cuetos}}, \bibinfo {author} {\bibfnamefont {B.}~\bibnamefont
  {Martinez-Haya}}, \bibinfo {author} {\bibfnamefont {S.}~\bibnamefont {Lago}},
  \ and\ \bibinfo {author} {\bibfnamefont {L.~F.}\ \bibnamefont {Rull}},\
  }\href@noop {} {\bibfield  {journal} {\bibinfo  {journal} {J. Phys. Chem. B}\
  }\textbf {\bibinfo {volume} {109}},\ \bibinfo {pages} {13729} (\bibinfo
  {year} {2005})}\BibitemShut {NoStop}%
\bibitem [{\citenamefont {Phuong}, \citenamefont {Germano},\ and\ \citenamefont
  {Schmid}(2001)}]{Schmid-1}%
  \BibitemOpen
  \bibfield  {author} {\bibinfo {author} {\bibfnamefont {N.~H.}\ \bibnamefont
  {Phuong}}, \bibinfo {author} {\bibfnamefont {G.}~\bibnamefont {Germano}}, \
  and\ \bibinfo {author} {\bibfnamefont {F.}~\bibnamefont {Schmid}},\
  }\href@noop {} {\bibfield  {journal} {\bibinfo  {journal} {J. Chem. Phys.}\
  }\textbf {\bibinfo {volume} {115}},\ \bibinfo {pages} {7227} (\bibinfo {year}
  {2001})}\BibitemShut {NoStop}%
\bibitem [{\citenamefont {Ebner}, \citenamefont {Saam},\ and\ \citenamefont
  {Stroud}(1976)}]{Ebner-1}%
  \BibitemOpen
  \bibfield  {author} {\bibinfo {author} {\bibfnamefont {C.}~\bibnamefont
  {Ebner}}, \bibinfo {author} {\bibfnamefont {W.~F.}\ \bibnamefont {Saam}}, \
  and\ \bibinfo {author} {\bibfnamefont {D.}~\bibnamefont {Stroud}},\
  }\href@noop {} {\bibfield  {journal} {\bibinfo  {journal} {Phys. Rev. A}\
  }\textbf {\bibinfo {volume} {14}},\ \bibinfo {pages} {2264} (\bibinfo {year}
  {1976})}\BibitemShut {NoStop}%
\bibitem [{\citenamefont {Saam}\ and\ \citenamefont {Ebner}(1977)}]{Ebner-2}%
  \BibitemOpen
  \bibfield  {author} {\bibinfo {author} {\bibfnamefont {W.~F.}\ \bibnamefont
  {Saam}}\ and\ \bibinfo {author} {\bibfnamefont {C.}~\bibnamefont {Ebner}},\
  }\href@noop {} {\bibfield  {journal} {\bibinfo  {journal} {Phys. Rev. A}\
  }\textbf {\bibinfo {volume} {15}},\ \bibinfo {pages} {2566} (\bibinfo {year}
  {1977})}\BibitemShut {NoStop}%
\bibitem [{\citenamefont {Phuong}, \citenamefont {Germano},\ and\ \citenamefont
  {Schmid}(2002)}]{Schmid-2}%
  \BibitemOpen
  \bibfield  {author} {\bibinfo {author} {\bibfnamefont {N.~H.}\ \bibnamefont
  {Phuong}}, \bibinfo {author} {\bibfnamefont {G.}~\bibnamefont {Germano}}, \
  and\ \bibinfo {author} {\bibfnamefont {F.}~\bibnamefont {Schmid}},\
  }\href@noop {} {\bibfield  {journal} {\bibinfo  {journal} {Comput. Phys.
  Commun.}\ }\textbf {\bibinfo {volume} {147}},\ \bibinfo {pages} {350}
  (\bibinfo {year} {2002})}\BibitemShut {NoStop}%
\bibitem [{\citenamefont {Poniewierski}\ and\ \citenamefont
  {Stecki}(1979)}]{Poniewierski-1}%
  \BibitemOpen
  \bibfield  {author} {\bibinfo {author} {\bibfnamefont {A.}~\bibnamefont
  {Poniewierski}}\ and\ \bibinfo {author} {\bibfnamefont {J.}~\bibnamefont
  {Stecki}},\ }\href@noop {} {\bibfield  {journal} {\bibinfo  {journal} {Mol.
  Phys.}\ }\textbf {\bibinfo {volume} {38}},\ \bibinfo {pages} {1931} (\bibinfo
  {year} {1979})}\BibitemShut {NoStop}%
\bibitem [{\citenamefont {Straley}(1973{\natexlab{b}})}]{Straley-3}%
  \BibitemOpen
  \bibfield  {author} {\bibinfo {author} {\bibfnamefont {J.~P.}\ \bibnamefont
  {Straley}},\ }\href@noop {} {\bibfield  {journal} {\bibinfo  {journal} {Phys.
  Rev. A}\ }\textbf {\bibinfo {volume} {8}},\ \bibinfo {pages} {2181} (\bibinfo
  {year} {1973}{\natexlab{b}})}\BibitemShut {NoStop}%
\bibitem [{\citenamefont {Somoza}\ and\ \citenamefont
  {Tarazona}(1989{\natexlab{a}})}]{Somoza-2}%
  \BibitemOpen
  \bibfield  {author} {\bibinfo {author} {\bibfnamefont {A.~M.}\ \bibnamefont
  {Somoza}}\ and\ \bibinfo {author} {\bibfnamefont {P.}~\bibnamefont
  {Tarazona}},\ }\href@noop {} {\bibfield  {journal} {\bibinfo  {journal}
  {Phys. Rev. A}\ }\textbf {\bibinfo {volume} {40}},\ \bibinfo {pages} {6069}
  (\bibinfo {year} {1989}{\natexlab{a}})}\BibitemShut {NoStop}%
\bibitem [{\citenamefont {van Roij}(2005)}]{vanRoij}%
  \BibitemOpen
  \bibfield  {author} {\bibinfo {author} {\bibfnamefont {R.}~\bibnamefont {van
  Roij}},\ }\href@noop {} {\bibfield  {journal} {\bibinfo  {journal} {Eur. J.
  Phys.}\ }\textbf {\bibinfo {volume} {26}},\ \bibinfo {pages} {S57} (\bibinfo
  {year} {2005})}\BibitemShut {NoStop}%
\bibitem [{\citenamefont {Xiao}\ and\ \citenamefont {Sheng}(2013)}]{Ping}%
  \BibitemOpen
  \bibfield  {author} {\bibinfo {author} {\bibfnamefont {X.}~\bibnamefont
  {Xiao}}\ and\ \bibinfo {author} {\bibfnamefont {P.}~\bibnamefont {Sheng}},\
  }\href {\doibase 10.1103/PhysRevE.88.062501} {\bibfield  {journal} {\bibinfo
  {journal} {Phys. Rev. E}\ }\textbf {\bibinfo {volume} {88}},\ \bibinfo
  {pages} {062501} (\bibinfo {year} {2013})}\BibitemShut {NoStop}%
\bibitem [{\citenamefont {Williamson}\ and\ \citenamefont
  {Jackson}(1998)}]{Jackson}%
  \BibitemOpen
  \bibfield  {author} {\bibinfo {author} {\bibfnamefont {D.~C.}\ \bibnamefont
  {Williamson}}\ and\ \bibinfo {author} {\bibfnamefont {G.}~\bibnamefont
  {Jackson}},\ }\href@noop {} {\bibfield  {journal} {\bibinfo  {journal} {J.
  Chem. Phys.}\ }\textbf {\bibinfo {volume} {108}},\ \bibinfo {pages} {10294}
  (\bibinfo {year} {1998})}\BibitemShut {NoStop}%
\bibitem [{\citenamefont {Abascal}\ and\ \citenamefont {Lago}(1985)}]{Lago}%
  \BibitemOpen
  \bibfield  {author} {\bibinfo {author} {\bibfnamefont {J.~L.~F.}\
  \bibnamefont {Abascal}}\ and\ \bibinfo {author} {\bibfnamefont
  {S.}~\bibnamefont {Lago}},\ }\href@noop {} {\bibfield  {journal} {\bibinfo
  {journal} {J. Mol. Liq.}\ }\textbf {\bibinfo {volume} {30}},\ \bibinfo
  {pages} {133} (\bibinfo {year} {1985})}\BibitemShut {NoStop}%
\bibitem [{\citenamefont {Varga}\ and\ \citenamefont {Szalai}(2000)}]{Szalai}%
  \BibitemOpen
  \bibfield  {author} {\bibinfo {author} {\bibfnamefont {S.}~\bibnamefont
  {Varga}}\ and\ \bibinfo {author} {\bibfnamefont {I.}~\bibnamefont {Szalai}},\
  }\href@noop {} {\bibfield  {journal} {\bibinfo  {journal} {Mol. Phys.}\
  }\textbf {\bibinfo {volume} {98}},\ \bibinfo {pages} {693} (\bibinfo {year}
  {2000})}\BibitemShut {NoStop}%
\bibitem [{\citenamefont {Frezza}\ \emph {et~al.}(2013)\citenamefont {Frezza},
  \citenamefont {Ferrarini}, \citenamefont {Kolli}, \citenamefont
  {Giacometti},\ and\ \citenamefont {Cinacchi}}]{Ferrarini-2}%
  \BibitemOpen
  \bibfield  {author} {\bibinfo {author} {\bibfnamefont {E.}~\bibnamefont
  {Frezza}}, \bibinfo {author} {\bibfnamefont {A.}~\bibnamefont {Ferrarini}},
  \bibinfo {author} {\bibfnamefont {H.~B.}\ \bibnamefont {Kolli}}, \bibinfo
  {author} {\bibfnamefont {A.}~\bibnamefont {Giacometti}}, \ and\ \bibinfo
  {author} {\bibfnamefont {G.}~\bibnamefont {Cinacchi}},\ }\href@noop {}
  {\bibfield  {journal} {\bibinfo  {journal} {J. Chem. Phys.}\ }\textbf
  {\bibinfo {volume} {138}},\ \bibinfo {pages} {164906} (\bibinfo {year}
  {2013})}\BibitemShut {NoStop}%
\bibitem [{\citenamefont {Kolli}\ \emph
  {et~al.}(2014{\natexlab{a}})\citenamefont {Kolli}, \citenamefont {Frezza},
  \citenamefont {Cinacchi}, \citenamefont {Ferrarini}, \citenamefont
  {Giacometti},\ and\ \citenamefont {Hudson}}]{Ferrarini-5}%
  \BibitemOpen
  \bibfield  {author} {\bibinfo {author} {\bibfnamefont {H.~B.}\ \bibnamefont
  {Kolli}}, \bibinfo {author} {\bibfnamefont {E.}~\bibnamefont {Frezza}},
  \bibinfo {author} {\bibfnamefont {G.}~\bibnamefont {Cinacchi}}, \bibinfo
  {author} {\bibfnamefont {A.}~\bibnamefont {Ferrarini}}, \bibinfo {author}
  {\bibfnamefont {A.}~\bibnamefont {Giacometti}}, \ and\ \bibinfo {author}
  {\bibfnamefont {T.~S.}\ \bibnamefont {Hudson}},\ }\href@noop {} {\bibfield
  {journal} {\bibinfo  {journal} {J. Chem. Phys.}\ }\textbf {\bibinfo {volume}
  {140}},\ \bibinfo {pages} {081101} (\bibinfo {year}
  {2014}{\natexlab{a}})}\BibitemShut {NoStop}%
\bibitem [{Note1()}]{Note1}%
  \BibitemOpen
  \bibinfo {note} {We therefore use the words ``nematic'' and ``cholesteric''
  interchangeably in this section and the next, as our perturbative framework
  imposes that cholesteric and uniaxial nematic phases be undistinguishable at
  the local scale.}\BibitemShut {Stop}%
\bibitem [{\citenamefont {Froberg}(1969)}]{Froberg}%
  \BibitemOpen
  \bibfield  {author} {\bibinfo {author} {\bibfnamefont {C.-E.}\ \bibnamefont
  {Froberg}},\ }\href@noop {} {\emph {\bibinfo {title} {Introduction to
  Numerical Analysis}}}\ (\bibinfo  {publisher} {Addison-Wesley Reading,
  Massachusetts},\ \bibinfo {year} {1969})\BibitemShut {NoStop}%
\bibitem [{\citenamefont {Ruzicka}\ and\ \citenamefont
  {Wensink}(2016)}]{Wensink-2}%
  \BibitemOpen
  \bibfield  {author} {\bibinfo {author} {\bibfnamefont {S.}~\bibnamefont
  {Ruzicka}}\ and\ \bibinfo {author} {\bibfnamefont {H.~H.}\ \bibnamefont
  {Wensink}},\ }\href@noop {} {\bibfield  {journal} {\bibinfo  {journal} {Soft
  Matter}\ }\textbf {\bibinfo {volume} {12}},\ \bibinfo {pages} {5205}
  (\bibinfo {year} {2016})}\BibitemShut {NoStop}%
\bibitem [{\citenamefont {Fischermeier}\ \emph {et~al.}(2014)\citenamefont
  {Fischermeier}, \citenamefont {Bartuschat}, \citenamefont {Preclik},
  \citenamefont {Marechal},\ and\ \citenamefont {Mecke}}]{Marechal-2}%
  \BibitemOpen
  \bibfield  {author} {\bibinfo {author} {\bibfnamefont {E.}~\bibnamefont
  {Fischermeier}}, \bibinfo {author} {\bibfnamefont {D.}~\bibnamefont
  {Bartuschat}}, \bibinfo {author} {\bibfnamefont {T.}~\bibnamefont {Preclik}},
  \bibinfo {author} {\bibfnamefont {M.}~\bibnamefont {Marechal}}, \ and\
  \bibinfo {author} {\bibfnamefont {K.}~\bibnamefont {Mecke}},\ }\href@noop {}
  {\bibfield  {journal} {\bibinfo  {journal} {Comput. Phys. Commun.}\ }\textbf
  {\bibinfo {volume} {185}},\ \bibinfo {pages} {3156} (\bibinfo {year}
  {2014})}\BibitemShut {NoStop}%
\bibitem [{\citenamefont {Allen}\ and\ \citenamefont
  {Frenkel}(1988)}]{Allen-1}%
  \BibitemOpen
  \bibfield  {author} {\bibinfo {author} {\bibfnamefont {M.~P.}\ \bibnamefont
  {Allen}}\ and\ \bibinfo {author} {\bibfnamefont {D.}~\bibnamefont
  {Frenkel}},\ }\href@noop {} {\bibfield  {journal} {\bibinfo  {journal} {Phys.
  Rev. A}\ }\textbf {\bibinfo {volume} {37}},\ \bibinfo {pages} {1813}
  (\bibinfo {year} {1988})}\BibitemShut {NoStop}%
\bibitem [{\citenamefont {Wittmann}, \citenamefont {Marechal},\ and\
  \citenamefont {Mecke}(2015)}]{Marechal-1}%
  \BibitemOpen
  \bibfield  {author} {\bibinfo {author} {\bibfnamefont {R.}~\bibnamefont
  {Wittmann}}, \bibinfo {author} {\bibfnamefont {M.}~\bibnamefont {Marechal}},
  \ and\ \bibinfo {author} {\bibfnamefont {K.}~\bibnamefont {Mecke}},\
  }\href@noop {} {\bibfield  {journal} {\bibinfo  {journal} {Phys. Rev. E}\
  }\textbf {\bibinfo {volume} {91}},\ \bibinfo {pages} {052501} (\bibinfo
  {year} {2015})}\BibitemShut {NoStop}%
\bibitem [{\citenamefont {Kr{\"o}ger}\ and\ \citenamefont {Ilg}(2007)}]{Ilg}%
  \BibitemOpen
  \bibfield  {author} {\bibinfo {author} {\bibfnamefont {M.}~\bibnamefont
  {Kr{\"o}ger}}\ and\ \bibinfo {author} {\bibfnamefont {P.}~\bibnamefont
  {Ilg}},\ }\href@noop {} {\bibfield  {journal} {\bibinfo  {journal} {J. Chem.
  Phys.}\ }\textbf {\bibinfo {volume} {127}},\ \bibinfo {pages} {034903}
  (\bibinfo {year} {2007})}\BibitemShut {NoStop}%
\bibitem [{\citenamefont {Bolhuis}\ and\ \citenamefont
  {Frenkel}(1997)}]{Frenkel-2}%
  \BibitemOpen
  \bibfield  {author} {\bibinfo {author} {\bibfnamefont {P.}~\bibnamefont
  {Bolhuis}}\ and\ \bibinfo {author} {\bibfnamefont {D.}~\bibnamefont
  {Frenkel}},\ }\href@noop {} {\bibfield  {journal} {\bibinfo  {journal} {J.
  Chem. Phys.}\ }\textbf {\bibinfo {volume} {106}},\ \bibinfo {pages} {666}
  (\bibinfo {year} {1997})}\BibitemShut {NoStop}%
\bibitem [{\citenamefont {Somoza}\ and\ \citenamefont
  {Tarazona}(1989{\natexlab{b}})}]{Somoza-1}%
  \BibitemOpen
  \bibfield  {author} {\bibinfo {author} {\bibfnamefont {A.~M.}\ \bibnamefont
  {Somoza}}\ and\ \bibinfo {author} {\bibfnamefont {P.}~\bibnamefont
  {Tarazona}},\ }\href@noop {} {\bibfield  {journal} {\bibinfo  {journal} {J.
  Chem. Phys.}\ }\textbf {\bibinfo {volume} {91}},\ \bibinfo {pages} {517}
  (\bibinfo {year} {1989}{\natexlab{b}})}\BibitemShut {NoStop}%
\bibitem [{\citenamefont {Lee}(1989)}]{Lee-2}%
  \BibitemOpen
  \bibfield  {author} {\bibinfo {author} {\bibfnamefont {S.-D.}\ \bibnamefont
  {Lee}},\ }\href@noop {} {\bibfield  {journal} {\bibinfo  {journal} {Phys.
  Rev. A}\ }\textbf {\bibinfo {volume} {39}},\ \bibinfo {pages} {3631}
  (\bibinfo {year} {1989})}\BibitemShut {NoStop}%
\bibitem [{\citenamefont {Poniewierski}\ and\ \citenamefont
  {Ho\l{}yst}(1990)}]{Poniewierski-2}%
  \BibitemOpen
  \bibfield  {author} {\bibinfo {author} {\bibfnamefont {A.}~\bibnamefont
  {Poniewierski}}\ and\ \bibinfo {author} {\bibfnamefont {R.}~\bibnamefont
  {Ho\l{}yst}},\ }\href {\doibase 10.1103/PhysRevA.41.6871} {\bibfield
  {journal} {\bibinfo  {journal} {Phys. Rev. A}\ }\textbf {\bibinfo {volume}
  {41}},\ \bibinfo {pages} {6871} (\bibinfo {year} {1990})}\BibitemShut
  {NoStop}%
\bibitem [{\citenamefont {Mederos}, \citenamefont {Velasco},\ and\
  \citenamefont {Mart{\'\i}nez-Rat{\'o}n}(2014)}]{Mederos}%
  \BibitemOpen
  \bibfield  {author} {\bibinfo {author} {\bibfnamefont {L.}~\bibnamefont
  {Mederos}}, \bibinfo {author} {\bibfnamefont {E.}~\bibnamefont {Velasco}}, \
  and\ \bibinfo {author} {\bibfnamefont {Y.}~\bibnamefont
  {Mart{\'\i}nez-Rat{\'o}n}},\ }\href@noop {} {\bibfield  {journal} {\bibinfo
  {journal} {J. Phys.: Condens. Matter}\ }\textbf {\bibinfo {volume} {26}},\
  \bibinfo {pages} {463101} (\bibinfo {year} {2014})}\BibitemShut {NoStop}%
\bibitem [{\citenamefont {Allen}(2016)}]{Allen-2}%
  \BibitemOpen
  \bibfield  {author} {\bibinfo {author} {\bibfnamefont {M.~P.}\ \bibnamefont
  {Allen}},\ }\href@noop {} {\bibfield  {journal} {\bibinfo  {journal} {Mol.
  Phys.}\ }\textbf {\bibinfo {volume} {114}},\ \bibinfo {pages} {2574}
  (\bibinfo {year} {2016})}\BibitemShut {NoStop}%
\bibitem [{\citenamefont {Priest}\ and\ \citenamefont
  {Lubensky}(1974)}]{Lubensky-1}%
  \BibitemOpen
  \bibfield  {author} {\bibinfo {author} {\bibfnamefont {R.~G.}\ \bibnamefont
  {Priest}}\ and\ \bibinfo {author} {\bibfnamefont {T.~C.}\ \bibnamefont
  {Lubensky}},\ }\href@noop {} {\bibfield  {journal} {\bibinfo  {journal}
  {Phys. Rev. A}\ }\textbf {\bibinfo {volume} {9}},\ \bibinfo {pages} {893}
  (\bibinfo {year} {1974})}\BibitemShut {NoStop}%
\bibitem [{\citenamefont {Harris}, \citenamefont {Kamien},\ and\ \citenamefont
  {Lubensky}(1997)}]{Lubensky-2}%
  \BibitemOpen
  \bibfield  {author} {\bibinfo {author} {\bibfnamefont {A.~B.}\ \bibnamefont
  {Harris}}, \bibinfo {author} {\bibfnamefont {R.~D.}\ \bibnamefont {Kamien}},
  \ and\ \bibinfo {author} {\bibfnamefont {T.~C.}\ \bibnamefont {Lubensky}},\
  }\href@noop {} {\bibfield  {journal} {\bibinfo  {journal} {Phys. Rev. Lett.}\
  }\textbf {\bibinfo {volume} {78}},\ \bibinfo {pages} {1476} (\bibinfo {year}
  {1997})}\BibitemShut {NoStop}%
\bibitem [{\citenamefont {Harris}, \citenamefont {Kamien},\ and\ \citenamefont
  {Lubensky}(1999)}]{Lubensky-3}%
  \BibitemOpen
  \bibfield  {author} {\bibinfo {author} {\bibfnamefont {A.~B.}\ \bibnamefont
  {Harris}}, \bibinfo {author} {\bibfnamefont {R.~D.}\ \bibnamefont {Kamien}},
  \ and\ \bibinfo {author} {\bibfnamefont {T.~C.}\ \bibnamefont {Lubensky}},\
  }\href@noop {} {\bibfield  {journal} {\bibinfo  {journal} {Rev. Mod. Phys.}\
  }\textbf {\bibinfo {volume} {71}},\ \bibinfo {pages} {1745} (\bibinfo {year}
  {1999})}\BibitemShut {NoStop}%
\bibitem [{\citenamefont {Dhakal}\ and\ \citenamefont
  {Selinger}(2011)}]{Dhakal}%
  \BibitemOpen
  \bibfield  {author} {\bibinfo {author} {\bibfnamefont {S.}~\bibnamefont
  {Dhakal}}\ and\ \bibinfo {author} {\bibfnamefont {J.~V.}\ \bibnamefont
  {Selinger}},\ }\href@noop {} {\bibfield  {journal} {\bibinfo  {journal}
  {Phys. Rev. E}\ }\textbf {\bibinfo {volume} {83}},\ \bibinfo {pages} {020702}
  (\bibinfo {year} {2011})}\BibitemShut {NoStop}%
\bibitem [{\citenamefont {Kleman}\ and\ \citenamefont
  {Lavrentovich}(2007)}]{Kleman}%
  \BibitemOpen
  \bibfield  {author} {\bibinfo {author} {\bibfnamefont {M.}~\bibnamefont
  {Kleman}}\ and\ \bibinfo {author} {\bibfnamefont {O.~D.}\ \bibnamefont
  {Lavrentovich}},\ }\href@noop {} {\emph {\bibinfo {title} {Soft Matter
  Physics: An Introduction}}}\ (\bibinfo  {publisher} {Springer Science \&
  Business Media},\ \bibinfo {address} {New York, NY},\ \bibinfo {year}
  {2007})\BibitemShut {NoStop}%
\bibitem [{\citenamefont {Stroobants}, \citenamefont {Lekkerkerker},\ and\
  \citenamefont {Odijk}(1986)}]{Lekkerkerker-2}%
  \BibitemOpen
  \bibfield  {author} {\bibinfo {author} {\bibfnamefont {A.}~\bibnamefont
  {Stroobants}}, \bibinfo {author} {\bibfnamefont {H.~N.~W.}\ \bibnamefont
  {Lekkerkerker}}, \ and\ \bibinfo {author} {\bibfnamefont {T.}~\bibnamefont
  {Odijk}},\ }\href@noop {} {\bibfield  {journal} {\bibinfo  {journal}
  {Macromolecules}\ }\textbf {\bibinfo {volume} {19}},\ \bibinfo {pages} {2232}
  (\bibinfo {year} {1986})}\BibitemShut {NoStop}%
\bibitem [{\citenamefont {Honorato-Rios}\ \emph {et~al.}(2016)\citenamefont
  {Honorato-Rios}, \citenamefont {Kuhnhold}, \citenamefont {Bruckner},
  \citenamefont {Dannert}, \citenamefont {Schilling},\ and\ \citenamefont
  {Lagerwall}}]{Schilling-1}%
  \BibitemOpen
  \bibfield  {author} {\bibinfo {author} {\bibfnamefont {C.}~\bibnamefont
  {Honorato-Rios}}, \bibinfo {author} {\bibfnamefont {A.}~\bibnamefont
  {Kuhnhold}}, \bibinfo {author} {\bibfnamefont {J.~R.}\ \bibnamefont
  {Bruckner}}, \bibinfo {author} {\bibfnamefont {R.}~\bibnamefont {Dannert}},
  \bibinfo {author} {\bibfnamefont {T.}~\bibnamefont {Schilling}}, \ and\
  \bibinfo {author} {\bibfnamefont {J.~P.~F.}\ \bibnamefont {Lagerwall}},\
  }\href@noop {} {\bibfield  {journal} {\bibinfo  {journal} {Front. Mater.}\
  }\textbf {\bibinfo {volume} {3}},\ \bibinfo {pages} {21} (\bibinfo {year}
  {2016})}\BibitemShut {NoStop}%
\bibitem [{\citenamefont {Kuhnhold}\ and\ \citenamefont
  {Schilling}(2016)}]{Schilling-2}%
  \BibitemOpen
  \bibfield  {author} {\bibinfo {author} {\bibfnamefont {A.}~\bibnamefont
  {Kuhnhold}}\ and\ \bibinfo {author} {\bibfnamefont {T.}~\bibnamefont
  {Schilling}},\ }\href@noop {} {\bibfield  {journal} {\bibinfo  {journal} {J.
  Chem. Phys.}\ }\textbf {\bibinfo {volume} {145}},\ \bibinfo {pages} {194904}
  (\bibinfo {year} {2016})}\BibitemShut {NoStop}%
\bibitem [{\citenamefont {Kolli}\ \emph
  {et~al.}(2014{\natexlab{b}})\citenamefont {Kolli}, \citenamefont {Frezza},
  \citenamefont {Cinacchi}, \citenamefont {Ferrarini}, \citenamefont
  {Giacometti}, \citenamefont {Hudson}, \citenamefont {De~Michele},\ and\
  \citenamefont {Sciortino}}]{Ferrarini-6}%
  \BibitemOpen
  \bibfield  {author} {\bibinfo {author} {\bibfnamefont {H.~B.}\ \bibnamefont
  {Kolli}}, \bibinfo {author} {\bibfnamefont {E.}~\bibnamefont {Frezza}},
  \bibinfo {author} {\bibfnamefont {G.}~\bibnamefont {Cinacchi}}, \bibinfo
  {author} {\bibfnamefont {A.}~\bibnamefont {Ferrarini}}, \bibinfo {author}
  {\bibfnamefont {A.}~\bibnamefont {Giacometti}}, \bibinfo {author}
  {\bibfnamefont {T.~S.}\ \bibnamefont {Hudson}}, \bibinfo {author}
  {\bibfnamefont {C.}~\bibnamefont {De~Michele}}, \ and\ \bibinfo {author}
  {\bibfnamefont {F.}~\bibnamefont {Sciortino}},\ }\href@noop {} {\bibfield
  {journal} {\bibinfo  {journal} {Soft Matter}\ }\textbf {\bibinfo {volume}
  {10}},\ \bibinfo {pages} {8171} (\bibinfo {year}
  {2014}{\natexlab{b}})}\BibitemShut {NoStop}%
\bibitem [{\citenamefont {Cinacchi}\ \emph {et~al.}(2016)\citenamefont
  {Cinacchi}, \citenamefont {Ferrarini}, \citenamefont {Frezza}, \citenamefont
  {Giacometti},\ and\ \citenamefont {Kolli}}]{Ferrarini-3}%
  \BibitemOpen
  \bibfield  {author} {\bibinfo {author} {\bibfnamefont {G.}~\bibnamefont
  {Cinacchi}}, \bibinfo {author} {\bibfnamefont {A.}~\bibnamefont {Ferrarini}},
  \bibinfo {author} {\bibfnamefont {E.}~\bibnamefont {Frezza}}, \bibinfo
  {author} {\bibfnamefont {A.}~\bibnamefont {Giacometti}}, \ and\ \bibinfo
  {author} {\bibfnamefont {H.~B.}\ \bibnamefont {Kolli}},\ }\enquote {\bibinfo
  {title} {Theory and simulation studies of self-assembly of helical
  particles},}\ in\ \href {\doibase 10.1002/9781119113171.ch3} {\emph {\bibinfo
  {booktitle} {Self-Assembling Systems: Theory and Simulation}}},\ \bibinfo
  {editor} {edited by\ \bibinfo {editor} {\bibfnamefont {L.-T.}\ \bibnamefont
  {Yan}}}\ (\bibinfo  {publisher} {John Wiley \& Sons, Ltd},\ \bibinfo
  {address} {Hoboken, NJ},\ \bibinfo {year} {2016})\ pp.\ \bibinfo {pages}
  {53--84}\BibitemShut {NoStop}%
\bibitem [{\citenamefont {Dussi}\ and\ \citenamefont
  {Dijkstra}(2016)}]{Dijkstra-3}%
  \BibitemOpen
  \bibfield  {author} {\bibinfo {author} {\bibfnamefont {S.}~\bibnamefont
  {Dussi}}\ and\ \bibinfo {author} {\bibfnamefont {M.}~\bibnamefont
  {Dijkstra}},\ }\href@noop {} {\bibfield  {journal} {\bibinfo  {journal} {Nat.
  Commun.}\ }\textbf {\bibinfo {volume} {7}},\ \bibinfo {pages} {11175}
  (\bibinfo {year} {2016})}\BibitemShut {NoStop}%
\bibitem [{\citenamefont {Dozov}(2001)}]{Dozov}%
  \BibitemOpen
  \bibfield  {author} {\bibinfo {author} {\bibfnamefont {I.}~\bibnamefont
  {Dozov}},\ }\href@noop {} {\bibfield  {journal} {\bibinfo  {journal}
  {Europhys. Lett.}\ }\textbf {\bibinfo {volume} {56}},\ \bibinfo {pages} {247}
  (\bibinfo {year} {2001})}\BibitemShut {NoStop}%
\bibitem [{\citenamefont {De~Gregorio}\ \emph {et~al.}(2016)\citenamefont
  {De~Gregorio}, \citenamefont {Frezza}, \citenamefont {Greco},\ and\
  \citenamefont {Ferrarini}}]{Ferrarini-4}%
  \BibitemOpen
  \bibfield  {author} {\bibinfo {author} {\bibfnamefont {P.}~\bibnamefont
  {De~Gregorio}}, \bibinfo {author} {\bibfnamefont {E.}~\bibnamefont {Frezza}},
  \bibinfo {author} {\bibfnamefont {C.}~\bibnamefont {Greco}}, \ and\ \bibinfo
  {author} {\bibfnamefont {A.}~\bibnamefont {Ferrarini}},\ }\href@noop {}
  {\bibfield  {journal} {\bibinfo  {journal} {Soft Matter}\ }\textbf {\bibinfo
  {volume} {12}},\ \bibinfo {pages} {5188} (\bibinfo {year}
  {2016})}\BibitemShut {NoStop}%
\bibitem [{\citenamefont {Cherstvy}(2008)}]{Cherstvy}%
  \BibitemOpen
  \bibfield  {author} {\bibinfo {author} {\bibfnamefont {A.~G.}\ \bibnamefont
  {Cherstvy}},\ }\href@noop {} {\bibfield  {journal} {\bibinfo  {journal} {J.
  Phys. Chem. B}\ }\textbf {\bibinfo {volume} {112}},\ \bibinfo {pages} {12585}
  (\bibinfo {year} {2008})}\BibitemShut {NoStop}%
\end{thebibliography}%


\end{document}